\documentclass[acmsmall]{acmart}

\usepackage{tabularx} % Add this to the preamble
\usepackage{graphicx}
\usepackage{subcaption}
\usepackage{soul}
\usepackage{multirow}
\usepackage{wrapfig}
\usepackage{microtype}  % Load the microtype package
\usepackage{algorithm}
\usepackage{algorithmic}

\usepackage{nth}
\usepackage{circledsteps}
\usepackage[nolist]{acronym}
\usepackage{booktabs}
\usepackage{xspace}
\usepackage{comment}
\usepackage{bm} % bold math
\usepackage{todonotes}
\usepackage{arydshln} % Draw dash-lines in array/tabular
\usepackage{graphicx}
\usepackage{pifont} % Wingdings font stuff.
\usepackage{tikz}
\usepackage[acronym]{glossaries}
\makeglossaries
\usepackage{ifthen}
\usepackage{csquotes}
\usepackage{hyperref}
\usepackage[inline]{enumitem}
\usepackage{mathtools}
\usepackage{multicol}
\usepackage{appendix}
\usepackage[table,dvipsnames]{xcolor} 
\usepackage[framemethod=tikz]{mdframed}  
\usepackage[most]{tcolorbox}
\definecolor{+green}{HTML}{1a9e18}
\definecolor{-red}{HTML}{ea0000}

\usepackage{placeins} % Leo: For float barriers in the appendix. Not whitelisted by the ACM

\usepackage{ragged2e}
\usepackage{textgreek}

\usepackage{cleveref} % Cref etc.

\usepackage{natbib}
\usepackage{caption}
\usepackage{amsmath}
\usepackage{makecell}
\usepackage{siunitx}
\usepackage{textcase} % for \NoCaseChange

\usepackage[T1]{fontenc}
\usepackage[utf8]{inputenc}
\usepackage[polish,english]{babel}
\hypersetup{unicode=true}

\AtBeginDocument{%
  }

\usepackage{booktabs}
\usepackage{xspace}
\usepackage{comment}
\usepackage{todonotes}
\usepackage{graphicx}
\usepackage{pifont} % Wingdings font stuff.
\usepackage{tikz}
\usepackage{soul} % Hyphenation for letterspacing, underlining, and more
\usepackage{ifthen}
\usepackage{csquotes}
\usepackage[inline]{enumitem}
\usepackage{mathtools}
\usepackage{cleveref} % Cref etc.

\newboolean{showcomments}
\setboolean{showcomments}{true}
\newcommand\important[1]{\todo[inline]{\textbf{Important:} #1}}
\newcommand\gareth[1]{\textbf{\textcolor{blue}{GT: #1}}}
\newcommand\leo[1]{\todo[color=orange,inline]{\textbf{Leo:} #1}}
\newcommand\michal[1]{\textbf{\textcolor{magenta}{Michal: #1}}}
\newcommand\onur[1]{\textbf{\textcolor{red}{Onur: #1}}}
\newcommand\saidu[1]{\todo[color=yellow,inline]{\textbf{Saidu:} #1}}

\newcommand\wen[1]{\textbf{\textcolor{purple}{wen: #1}}}
\newcommand\qiming[1]{\textbf{\textcolor{pink}{Qiming: #1}}}
\ifthenelse{\boolean{showcomments}} { }
{
	\renewcommand\important[1]{}
	\renewcommand\leo[1]{}
	\renewcommand\michal[1]{}
	\renewcommand\onur[1]{}
	\renewcommand\saidu[1]{}
        \renewcommand\qiming[1]{}
	\renewcommand\gareth[1]{}
        \renewcommand\wen[1]{}
}

\ifthenelse{\boolean{showcomments}}
{ \newcommand{\mynote}[3]{
		\protect\fbox{\bfseries\sffamily\scriptsize#1}
		{\small\textsf{\emph{\color{#3}{#2}}}}}}
{ \newcommand{\mynote}[3]{}}

\newcommand{\vs}{vs.\@\xspace}

\newcommand{\eg}{\textit{e.g.}\@\xspace}

\newcommand{\ie}{\textit{i.e.}\@\xspace}

\definecolor{verylightgray}{gray}{0.8}

\crefname{section}{Sec.}{Sec.}
\Crefname{section}{Section}{Sections}
\crefname{equation}{eq.}{eq.}
\crefname{figure}{Fig.}{Fig.s}
\Crefname{figure}{Figure}{Figures}

\floatstyle{plain}
\newfloat{lstfloat}{htbp}{lop}
\floatname{lstfloat}{Listing}
\crefalias{lstfloat}{listing}

\setcopyright{acmlicensed}
\acmJournal{POMACS}
\acmYear{2025} \acmVolume{9} \acmNumber{3}
\acmArticle{50} \acmMonth{12}\acmDOI{10.1145/3771565}
\begin{document}
\begin{acronym}[Derp]
    \acro{did}[DID]{Decentralized Identifier}
    \acro{nsfw}[NSFW]{not safe for work}
    \acro{atp}[ATProto]{Authenticated Transfer Protocol}
    \acro{pds}[PDS]{Personal Data Server}
    \acroplural{pds}[PDSes]{Personal Data Servers}
    \acro{uri}[URI]{Uniform Resource Identifier}
    \acro{cbor}[CBOR]{Concise Binary Object Representation}
    \acro{crdt}[CRDT]{Conflict-free Replicated Data Type}
    \acro{ens}[ENS]{Ethereum Name Service}
    \acro{dfs}[DFS]{depth-first search}
    \acro{bfs}[BFS]{breadth-first search}
    \acro{as}[AS]{Autonomous System}
    \acro{qps}[QPS]{queries per second}
    \acro{cv}[CV]{coefficient of variation}
    \acro{ttl}[TTL]{time to live}
    \acro{rtt}[RTT]{round trip time}
    \acro{p2p}[P2P]{peer-to-peer}
    \acro{dosn}[DOSN]{Decentralized Online Social Network}
    \acro{dht}[DHT]{Distributed Hash Table}
    \acro{acc}[ACC]{Average Clustering Coefficient}
    \acro{var}[VAR]{Vector Autoregression}
    \acro{irf}[IRF]{Impulse Response Function}
    \acro{te}[TE]{Transfer Entropy}
    \acro{psm}[PSM]{Propensity Score Matching}
    \acro{did}[DID]{Difference-in-Differences}
    \acro{att}[ATT]{Average Treatment Effect on the Treated}
    \acro{ols}[OLS]{Ordinary Least Squares}
    \acro{ecdf}[ECDF]{Empirical Cumulative Distribution Function}
    \acro{fid}[FID]{Farcaster Identifier}
    \acro{fname}[Fname]{Farcaster User Name}
    
\end{acronym}

\title{Beyond Single-Tokenomics: How Farcaster’s Pluralistic Incentives Reshape Social Networking}

\author{\NoCaseChange{Wen Yang}}
\affiliation{%
  \institution{Hong Kong University of Science and Technology (Guangzhou)}
  \city{Guangzhou}
  % \state{Guangdong}
  \country{China}
}
\email{wyang330@connect.hkust-gz.edu.cn}
\orcid{0009-0009-8481-6405}

\author{\NoCaseChange{Qiming Ye}}
\affiliation{%
  \institution{Hong Kong University of Science and Technology (Guangzhou)}
  \city{Guangzhou}
  % \state{Guangdong}
  \country{China}
}
\email{qiming@connect.hkust-gz.edu.cn}
\orcid{0009-0009-7106-9404}

\author{\NoCaseChange{Onur Ascigil}}
\affiliation{%
  \institution{Lancaster University}
  \city{Lancaster}
  % \state{England}
  \country{UK}
}
\email{o.ascigil@lancaster.ac.uk}
\orcid{0000-0002-3023-6431}

\author{\NoCaseChange{Saidu Sokoto}}
\affiliation{%
  \institution{City St George’s, University of London}
  \city{London}
  % \state{England}
  \country{UK}
}
\email{saidu.sokoto@city.ac.uk}
\orcid{0000-0001-7152-546X}

\author{\NoCaseChange{Leonhard Balduf}}
\affiliation{%
  \institution{TU Darmstadt}
  \city{Darmstadt}
  % \state{Hessen}
  \country{Germany}
}
\email{leonhard.balduf@tu-darmstadt.de}
\orcid{0000-0002-3519-7160}

% \author{Michał Król}
\author{\NoCaseChange{\foreignlanguage{polish}{Michał Król}}}
%\author{\foreignlanguage{polish}{MICHA\L{} KR\'{O}L}}
% \author{Micha\l Kr\'{o}l}
\affiliation{%
  \institution{City St George’s, University of London}
  \city{London}
  % \state{England}
  \country{UK}
}
\email{michal.krol@city.ac.uk}
\orcid{0000-0002-3437-8621}

\author{\NoCaseChange{Gareth Tyson}}
% \authornote{Corresponding author}
\affiliation{%
  \institution{Hong Kong University of Science and Technology (Guangzhou)}
  \city{Guangzhou}
  % \state{Guangdong}
  \country{China}
}
\email{gtyson@ust.hk}
\orcid{0000-0003-3010-791X}

\renewcommand{\shortauthors}{Yang et al.}

\begin{abstract}
This paper presents the first empirical analysis of how diverse token-based reward mechanisms impact platform dynamics and user behaviors.
For this, we gather a unique, large-scale dataset from Farcaster. This blockchain-based, decentralized social network incorporates multiple incentive mechanisms spanning platform-native rewards, third-party token programs, and peer-to-peer tipping.
Our dataset captures token transactions and social interactions from 574,829 wallet-linked users, representing 64.25\% of the platform's user base. 
Our socioeconomic analyses reveal how different tokenomics design shape varying participation rates (7.6\%--70\%) and wealth concentration patterns (Gini 0.72--0.94), 
whereas inter-community tipping is 1.3--2x more frequent among non-following pairs, thereby mitigating echo chambers.
Our causal analyses further uncover several critical trade-offs:
\begin{enumerate*}
\item while most token rewards boost content creation, they often fail to enhance---sometimes undermining---content quality;
\item token rewards increase follower acquisition but show neutral or negative effects on outbound following, suggesting potential asymmetric network growth;
\item repeated algorithmic rewards demonstrate strong cumulative effects that may encourage strategic optimization.
\end{enumerate*}
Our findings advance understanding of cryptocurrency integration in social platforms and highlight challenges in aligning economic incentives with authentic social value.
\end{abstract}


% \ccsdesc[300]{Applied computing~Economics}
%%%%%%%% XML only version %%%%%%%%%%%%
% <concept_id>10002951.10003260.10003282.10003292</concept_id>
% <concept_desc>Information systems~Social networks</concept_desc>
% <concept_id>10003120.10003130.10011762</concept_id>
% <concept_desc>Human-centered computing~Empirical studies in collaborative and social computing</concept_desc>
% <concept_id>10003033.10003068.10003078</concept_id>
% <concept_desc>Networks~Network economics</concept_desc>
% <concept_significance>100</concept_significance>


\maketitle
%%%%%%%%%%%%%%%%%%%%%%%%%%%%%%%%%%%%%%%%%%%
\section{Introduction} 
\label{sec:intro}
%%%%%%%%%%%%%%%%%%%%%%%%%%%%%%%%%%%%%%%%%%%
The emergence of \acp{dosn} marks a shift in social networking, emphasizing user autonomy, data sovereignty, and censorship resistance~\cite{survey_on_perf, jeong2025navigating}.
Despite this, most \acp{dosn} have struggled to incentivize high-quality content, large-scale user uptake, and sustained engagement~\cite{raman2019challenges,Wei2024ExploringTN}. 
Most notably, their emphasis on user sovereignty has limited the adoption of commonly used monetization models~\cite{liang2022end, Cai2024InnovationOM, mastodon_challenges, Wei2024ExploringTN}, often resulting in insufficient funding being available to compete with larger players. 

Consequently, some have attempted to integrate cryptocurrency-based token incentives to encourage participation by both content creators and infrastructure operators~\cite{steemit_doc,deso_official_doc,lens_official_doc,zora_doc,memo_set_in_stone}.
This, however, comes with key challenges, most notably the reliance on a single, platform-issued token incentive mechanism.
For instance, Steemit~\cite{steemit_doc}, a token-based \ac{dosn} launched in 2016, utilizes its self-issued token for content interaction incentives. 
However, the failure of such a token renders the rewards worthless.
Furthermore, research has revealed that Steemit's single designated token incentive mechanism is susceptible to token price fluctuations~\cite{ba2022role}, and has suffered from gaming and farming (\ie strategic interactions between colluding users designed to exploit reward systems),~\cite{li2019incentivized} and bot-driven adversarial manipulation~\cite{delkhosh2023impact}. 
This led to reward concentration among a small group of colluding users, increasing centralization and economic inequality while losing its effectiveness in promoting social engagement~\cite{steemit2024richgetricher}.

In response to this, a new \ac{dosn} called Farcaster was publicly launched in 2023 to support multiple incentive mechanisms~\cite{farcaster_official_doc}. 
Functionally similar to X (Twitter), Farcaster stands out from current \acp{dosn} in two key aspects.
First, Farcaster supports \enquote{\emph{modular}} wallet binding --- unlike platforms constrained by primary account-bound blockchain addresses~\cite{steemit_doc,memocash_official_doc,lens_official_doc,zora_doc}, Farcaster enables users to link any external Ethereum-compatible addresses~\cite{evm_addresses}, functioning as on-chain transaction wallets, alongside their user accounts (termed \acp{fid})~\cite{farcaster_wallet_connection}, providing greater economic flexibility and autonomy. 
Second, Farcaster is the first to implement a \enquote{\emph{pluralistic}} token incentive ecosystem. 
We refer to it as pluralistic because, unlike existing \acp{dosn}, Farcaster does not have an officially issued token or a centrally designated incentive mechanism. 
Instead, Farcaster allows any token or incentive mechanism to coexist within the ecosystem, regardless of the token used (medium of reward) or the eligibility criteria designed. This opens up incentive design to users, third-party developers, or the platform’s administrators themselves.

Thus, Farcaster enables users and developers to easily create and distribute their own tokens, creating an entirely decentralized reward ecosystem rather than a fixed incentive paradigm managed centrally. 
Such tokens can be used for any purpose deemed appropriate, including tipping content creators and operators who manage the infrastructure.
Moreover, third-party developers can create custom applications (\emph{mini-apps}) with algorithmic token reward distribution mechanisms~\cite{farcaster_miniapp_doc}, supporting a more community-driven incentive paradigm.
We believe this presents a unique use case for studying the feasibility of a system where multiple tokens and diverse incentive mechanisms coexist to incentivize positive user behavior within social networks.

To understand its broader implications, this paper empirically examines how Farcaster's pluralistic incentive paradigm shapes platform dynamics and user behaviors.
We gather both on-chain token transactions and off-chain social interactions relevant to Farcaster. 
As of April 27, 2025, our dataset covers 574,829 (64.25\% of the user base) users who have at least one Ethereum-compatible wallet bound to their \acp{fid}, with 5,878 unique tokens traded between users (far surpassing other \acp{dosn})~\cite{base_farconomy, lens_vs_farcaster, zora_dune_dashboard}.
Exploiting this data, we study the impact of multiple incentive mechanisms within the ecosystem.

Specifically, we explore the following three research questions:

\textbf{RQ1:} How widespread and diverse is the token economy within Farcaster's ecosystem, specifically regarding: 
\begin{enumerate*}
    \item the temporal dynamics of people binding their external cryptocurrency wallets to their Farcaster accounts,
    \item how prevalent the various available tokens are,
    and
    \item how these tokens serve different social functions through their incentive mechanisms?
\end{enumerate*}

\textbf{RQ2:} What socioeconomic risks are inherent in Farcaster's incentive system, specifically concerning:
\begin{enumerate*}
    \item disparities in new user participation rates across different token rewards,
    \item inequity in reward distribution, alongside
    \item echo chamber effects in tipping?
\end{enumerate*}

\textbf{RQ3:} What causal relationships exist between token incentives and subsequent social activities, and how do these dynamics vary across: 
\begin{enumerate*}
    \item different token categories (volatile tokens vs. stablecoins),
    \item distinct incentive mechanisms (user-to-user tipping vs. algorithmic rewards)?
\end{enumerate*}

To the best of our knowledge, we are the first to empirically study Farcaster's pluralistic incentive ecosystem. 
Our contributions are as follows: 
\begin{itemize}
    \item We reveal how specific eligibility criteria designs (\eg nomination-based vs. behavioral scoring) and reward distribution structure (\eg bot-driven tipping, redistribution mechanism) significantly impact both user inclusion (70\% vs. 7.6\% new participants) and income equality (Gini coefficients 0.72-0.94) (see \Cref{sec:rq2_incentive_inclusion,sec:rq2_rich_get_richer}).
    
    \item We demonstrate that, while user-to-user tipping represents the most flexible incentive mechanism, it is predominantly unidirectional (with less than 10\% of users acting as both tip receivers and senders) (see \Cref{sec:rq2_incentive_inclusion}). 
    Additionally, tips that occur across community boundaries are 1.3--2x more frequent among non-following pairs. This suggests that tipping incentives can facilitate value exchange beyond established social community structures (see \Cref{sec:rq2_echo_chamber}).
    
    \item We reveal trade-offs in incentivised social activities: while algorithmic rewards leveraging volatile tokens as the medium effectively increase content quantity, they show limited or negative effects on content quality (see \Cref{sec:rq3_results_findings}). 
    
    \item We uncover that repeated algorithmic rewards correlate with asymmetric social network growth (increased follower acquisition but decreased outbound following) and strategic engagement optimization (prioritizing immediate reactions over share-worthy content creation), highlighting risks in token-incentivized social platforms (see \Cref{sec:rq3_results_findings}).
\end{itemize}

These findings advance both the theoretical understanding of token-based incentive design and provide practical guidance for implementing sustainable reward mechanisms in social platforms.
%%%%%%%%%%%%%%%%%%%%%%%%%%%%%%%%%%%%%%%%
\section{A Primer on Farcaster}  
\label{sec:background}
%%%%%%%%%%%%%%%%%%%%%%%%%%%%%%%%%%%%%%%%
We begin by outlining the core design of \emph{Farcaster}.
Below, we provide brief descriptions of:
\begin{enumerate*}
\item social interactions; and
\item token transactions.
\end{enumerate*}
For full technical details, we refer readers to the official documentation~\cite{farcaster_official_doc}.

\paragraph{Social Interactions.}
Upon registration, Farcaster users receive an on-chain identifier (an Ethereum \emph{custody address}) anchored on the Optimism Layer-2 chain\footnote{
A Layer-2 (L2) is a scaling solution atop a Layer-1 (L1) blockchain (\eg Ethereum), enabling faster and cheaper transactions while inheriting its security.
}~\cite{optimism_doc} and managed through Farcaster's smart contracts~\cite{farcaster_contract_addresses}.
Users must pay an annual storage fee~\cite{storage_registry_contract_address} to rent network storage capacity during registration.\footnote{
Farcaster’s storage fee has been reduced three times since launching in October 2023, from \$7 to the current \$2~\cite{storage_fee_changes}} 
Users maintain exclusive control over their account's private key.
To facilitate network interaction, each address is associated with both a unique numeric identifier (\ac{fid}) and a human-readable username (\ac{fname}, \eg, \texttt{@vitalik}).

The off-chain social interactions --- referred to using Farcaster-specific terminology as \enquote{casts} (posts and replies), \enquote{reactions} (likes and re-posts), and \enquote{links} (follow actions) --- are exchanged through a \ac{p2p} network of independently operated servers called \emph{hubs}~\cite{farcaster_hub_tutorial}.
Each hub maintains a complete copy of the interaction data and synchronizes with peers using the GossipSub~\cite{GossipSub} and Diff Sync protocol~\cite{diff_sync_project}.\footnote{Note, Farcaster has transitioned to a new \ac{p2p} coordination layer called \emph{Snapchain} since May 2025. Built upon GossipSub, Snapchain replaces full replication with a partitioned model, where each hub stores only a subset of data based on user \acp{fid}~\cite{snapchain_doc}.}
The system demonstrates robust fault tolerance: network functionality remains intact as long as a single hub remains operational~\cite{hackmd_farcaster_hub_costs}.

All social interactions (\eg casts, links, and reactions) require a digital signature using the private key corresponding to the custody address.
These signed actions are broadcast across the network, where participants (\ie hubs, clients, and third-party applications) verify message authenticity by checking the digital signature against the on-chain registered public key for that \ac{fid}.
This hybrid (\ie on-chain/off-chain) architecture preserves user ownership and interoperability while circumventing the scalability and cost constraints inherent in fully on-chain systems~\cite{steemit_doc,memo_blog,deso_official_doc}.

\paragraph{Token Transactions.}
Custody addresses linked to \acp{fid} are primarily intended for account management (\eg signing social actions) rather than token transactions~\cite{farcaster_official_doc}. Farcaster enables users to bind \emph{external} Ethereum-compatible addresses to their \ac{fid} as transaction wallets~\cite{farcaster_wallet_connection}, allowing for trading, rewarding, or payment activities.
We refer to this flexibility as a \emph{modular} wallet architecture. This architecture facilitates broader token interoperability and economic autonomy. By isolating user accounts from token transactions, it also enhances security and reduces risks associated with private key exposure (\eg phishing/scam attacks~\cite{torres2019art}). 

Since February 22, 2025, Farcaster has implemented a phased roll-out of official Ethereum-compatible wallets.
This provides users with both optionally bound and officially issued Farcaster transaction wallets, along with the flexibility to designate any wallet as their primary wallet~\cite{farcaster_wallet_connection}. 

Note, while Farcaster allows users to bind 
% \onur{may not be clear what binding a wallet means} 
both Ethereum~\cite{eth_whitepaper} and Solana wallets~\cite{solana_doc} to their \acp{fid}, Ethereum addresses significantly outnumber Solana addresses ($794,386$ vs. $186,434$ as of April 27, 2025).
Moreover, Farcaster only introduced Solana Wallet Standard integration on May 21, 2025~\cite{farcaster_solana}, beyond our study period. 
Therefore, all subsequent references to \enquote{wallets} in this paper denote optionally bound and officially issued Ethereum-compatible wallets for the purpose of token transaction, distinct from both custody addresses and Solana-compatible wallets.

Moreover, Farcaster users can exchange tokens across over 50 Ethereum-compatible L1 and L2 chains (\eg Base, Optimism, Polygon, and BSC). However, we find that:
\begin{enumerate*}
\item all Farcaster's top-ten tokens by daily transaction volume originate from Base chain deployments~\cite{base_farconomy};
\item Base chain transactions constitute nearly 90\% of total activity among Farcaster users~\cite{base_chain_90_percentage}; and
\item Farcaster's native reward mechanism exclusively employs Base-chain USDC for weekly distributions to qualified accounts~\cite{farcaster_official_usdc_reward}.
\end{enumerate*}
We thus focus our analysis on the Base chain alone~\cite{base_doc}.

\section{Data Collection Methodology}
\label{sec:data_methodology}

%%%%%%%%%%%%%%%%%%%%%%%%%%%%%%%%%%%%%%%%%%
Farcaster's hybrid data architecture necessitates both on-chain and off-chain data collection:
\begin{enumerate*}
\item \emph{Off-chain Data:} we gather a complete snapshot of Farcaster's hub data as of April 27, 2025, including all user profiles (\ie \acp{fid}, user names, \ac{fid}-bound wallet addresses) and social interactions (\ie following, posting, liking, replying, and re-posting.) with their creation timestamps.
\item \emph{On-chain Data:} We use Alchemy APIs\footnote{\url{https://www.alchemy.com/}} to collect Farcaster’s token transaction data from the Base chain and construct transaction graphs that capture interactions between users' wallets, as well as those between user wallets and non-user addresses (\eg smart contracts). 
\end{enumerate*}

\subsection{Off-chain Data (User Profiles and Social Interactions)}
\label{sec:measure:offchain_data}
%%%%%%%%%%%%%%%%%%%%%%%%%%%%%%%%%%%%%%%%%%
Following Farcaster's official documentation~\cite{farcaster_official_doc_hub_install} and code-base~\cite{farcaster_official_github}, we deploy two hub server instances (one in Asia and one in Europe) to synchronize the off-chain data.

\paragraph{\ac{fid} Registration.}
As of April 27, 2025, our dataset encompasses 1,059,655 registered accounts, among which we identify 894,678 valid \acp{fid}.\footnote{We exclude invalid accounts by identifying and removing \acp{fid} without historical storage units. See \url{https://farcaster.xyz/wayne24/0x24410400} for more details.}
Note, we discover in the hub data that all timestamps of \acp{fid} registered before November 7, 2023 are aggregated to November 7, 2023 by Farcaster's official team. 
Therefore, in our analyses requiring \ac{fid} registration timestamps, we set November 7, 2023 as the starting point.

\paragraph{Wallet Binding Records.}
While associations between Farcaster-issued wallets and \acp{fid} are recorded both in the \emph{KeyRegistry} smart contract's transaction logs~\cite{farcaster_contract_addresses} and hub data, users' optionally bound external wallets are recorded solely in the hubs and not on-chain~\cite{farcaster_wallet_connection}.
However, hubs periodically purge old data~\cite{farcaster_official_github}, resulting in the loss of information about wallets that had been previously associated with an \ac{fid} but were later unbound.
To recover a complete list of external wallets bound to each \ac{fid}, we query Neynar's API~\cite{neynar_official_dash}.\footnote{Neynar is an independent 3rd party provider that offers API services for Farcaster data.}
Since Neynar only provides mappings of historical bound wallets and \acp{fid}, without any binding and unbinding timestamps, we must rely on the incomplete wallet records in hubs with timestamps for data analyses where binding time is necessary.

For wallets recorded in hubs, we discover that 574,829 (64.25\%) of \acp{fid} have at least one transaction wallet, whether optionally bound or officially issued, totaling 794,386 Ethereum-compatible wallets.
After retrieving the complete historical bound wallets, we identify a total of 1,282,783 external wallets bound to 606,827 (64.5\%) \acp{fid}.
We find that 488,397 (38\%) wallets were unbound as of May 2025 after their initial binding.

\paragraph{Social Interactions.} The social interaction data provided by our hub contains 159,539,953 unique following relationships, 164,984,116 casts (comprising 36,646,412 posts (22.21\%) and 128,337,704 replies (77.79\%)), and 299,079,720 reactions (consisting of 252,771,162 likes (84.52\%) and 46,308,558 re-posts (15.48\%)). 
For clarity and consistency with conventional terminology in the literature~\cite{trunfio2021conceptualising}, we use standard terms such as \enquote{follow}, \enquote{post}, \enquote{reply}, \enquote{like}, and \enquote{re-post} to denote these social interactions throughout the remainder of this paper.

\subsection{On-chain Data (Token Transactions)}
\label{sec:data_onchain_data}
%%%%%%%%%%%%%%%%%%%%%%%%%%%%%%%%%%%%%%%%%%
Recall, we identify that Farcaster's token transactions predominantly happen on the Base chain (see \Cref{sec:background}). 
Therefore, we extract all token transfer records on the Base chain involving Farcaster users' \ac{fid}-linked wallets. 
To do so, we use the Alchemy APIs~\cite{alchemy_transfer_api} to retrieve historical transfer data for all 1,282,783 wallet addresses, as of April 27, 2025.

Additionally, to capture transactions involving smart contracts and other non-user wallet interactions, we include transactions where at least one party (either sender or recipient) is a user wallet.
We collect a total of 87,687,791 transaction records, encompassing 5,878 distinct tokens (1.34\% of all 440,274 tokens that have appeared in all user wallets but may not necessarily have been traded between users) transferred between users' \ac{fid}-linked wallets.

%%%%%%%%%%%%%%%%%%%%
\subsection{Dataset Availability and Reproducibility}
\label{sec:data_availability}
%%%%%%%%%%%%%%%%%%%%
All data used in this paper is derived from publicly accessible sources. 
The off-chain data (including user profiles, \ac{fid}-linked wallet addresses, and social engagement) can be reproduced by participating in the Farcaster node network following the official documentation.
Note that our study utilize the now-deprecated hub architecture. 
Following May 2025, Farcaster has transitioned to a new blockchain-style architecture called Snapchain (see more details in~\Cref{sec:background}); subsequent researchers will need to deploy a Snapchain node to access the data that is previously available via hubs.
The on-chain token transaction data is retrievable via Ethereum-compatible node API providers (\eg Alchemy) using the wallet lists linked to \acp{fid}.
We release the code to perform this dataset reconstruction.

In light of ethical considerations (user profiles and content) and the substantial footprint of the raw datasets (Farcaster social engagement $\approx172$\,GB; token transactions $\approx37$\,GB), we do not mirror or redistribute the raw data. 
Instead, we publicly release weekly aggregated metrics (social engagement and token receipt) derived from the raw data that underpin our causal analyses in \Cref{sec:rq3_token_social_causality}. 
These aggregated datasets provide sufficient granularity for independent replication of all results reported in \Cref{sec:rq3_token_social_causality}, while substantially reducing the risk of re-identification.

In summary, we release: (i) the weekly aggregated metrics dataset, and (ii) the scripts and configuration templates to reproduce the Base-chain transfer retrieval pipeline. 
These materials are accessible at: \url{https://zenodo.org/records/17317121}.

%%%%%%%%%%%%%%%%%%%%
\subsection{Potential Biases from Hub Data Collection}
\label{sec:hub_data_bias}
%%%%%%%%%%%%%%%%%%%%
Farcaster's hub architecture performs \emph{full} replication of off-chain data across each hub. 
In our setup, two hubs (geographically distributed) are bootstrapped from an official snapshot and then continuously synchronized via GossipSub and Diff Sync (see~\Cref{sec:background}). 
While transient discrepancies can occur across hubs, our empirical checks confirm these are trivial: over 99.9\% of records converge across our two hubs within the observation window. 
For on-chain data, we retrieve all historical transfers for all \ac{fid}-linked wallets on the Base chain; due to the immutability of blockchain ledgers, this component is complete by definition. 
Taken together, we do not find evidence of sampling biases that would alter our findings or their interpretation.

\subsection{Ethical Considerations}
\label{sec:ethicial}
%%%%%%%%%%%%%%%%%%%%%%%%%%%%%%%%%%%%%%%%%%
Our dataset includes publicly available off-chain user profiles, social interactions and on-chain transactions. 
To address privacy concerns, we strictly follow established ethical standards~\cite{2012-dittrich-mraf}, collect only public data, and operate hubs non-intrusively at our own expense and following the guidelines issued by the Farcaster administrators~\cite{farcaster_hub_tutorial}.
Notably, wallet addresses alone offer stronger pseudonymity than social identifiers like \acp{fid} or \acp{fname}, making it harder to link transaction histories to personal identities. 
This study was reviewed and received a waiver from the authors’ institutional ethics committee.
\section{Token Economy Scale and Token Incentive Diversity.}
\label{sec:rq1_user_base_token_incentive_mechanisms}
%%%%%%%%%%%%%%%%%%%%%%%%%%%%%%%%%%%%%
We answer \textbf{RQ1} by exploring the scale and diversity of the Farcaster token economy. 
First, we assess the role tokenomics play in Farcaster's growth and user activity.
We then identify the most popular and impactful tokens.
Finally, we analyze the incentive mechanisms that use these tokens.

\subsection{Token-related Initiatives Driving Wallet Bindings.}
\label{sec:rq1_user_base_driving_events}

\begin{figure}[htbp]
    \centering
    \includegraphics[width=0.98\linewidth]{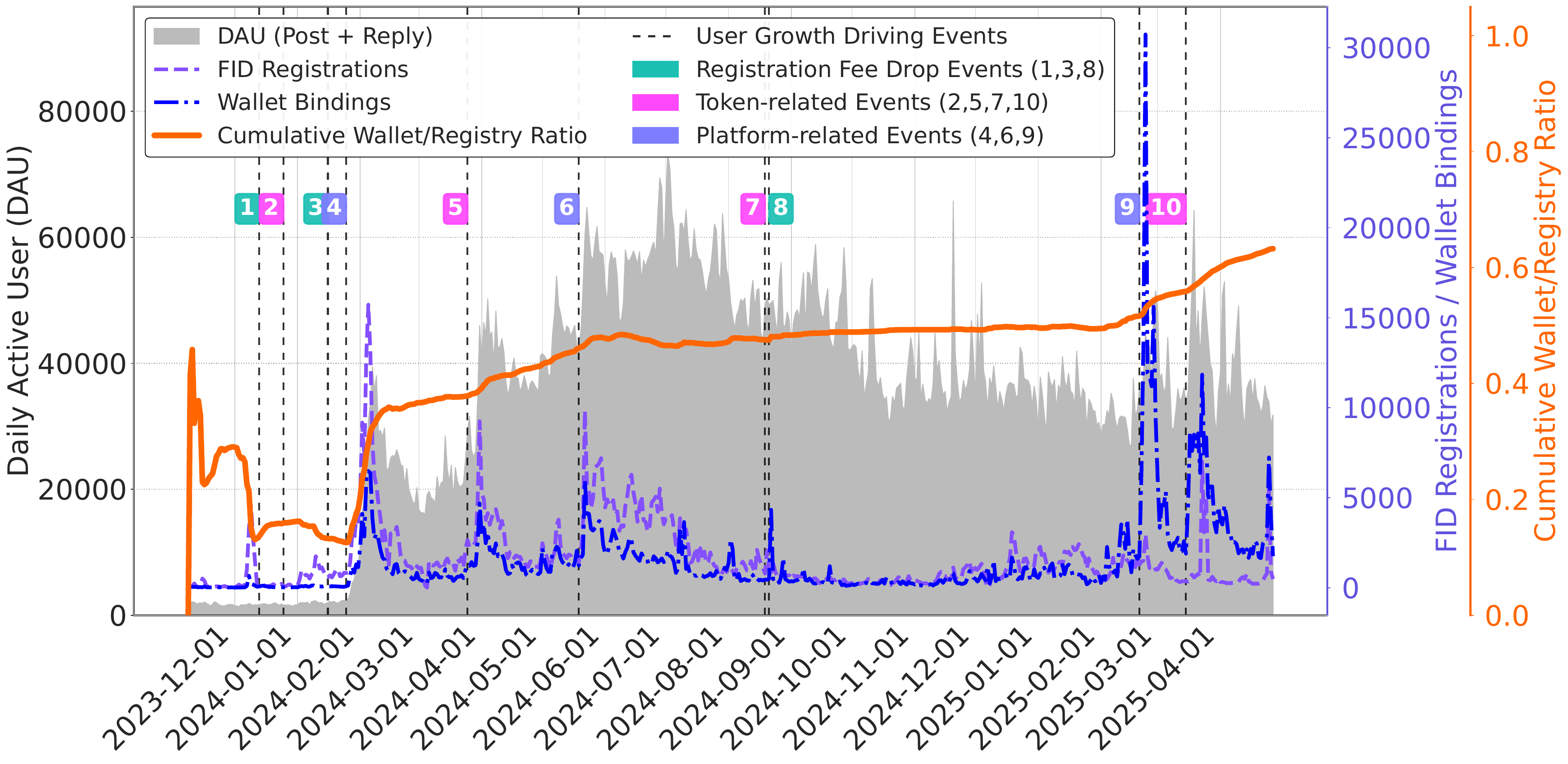}
    \caption{Daily engagement metrics and user growth on Farcaster.}
    \label{fig:DAU_walletRatio}
    \Description{Daily engagement metrics and user growth on Farcaster, marked with top 10 wallet binding surges and their related events.}
\end{figure}

\Cref{fig:DAU_walletRatio} presents daily user activity, platform growth, and user involvement in tokenomics. 
The platform experienced a steady activity growth, reaching a maximum of 73,180 daily active users (DAU) on July 2, 2024. 
Since then, the DAU stabilized at $\approx 42k$.

The new FID registrations and wallet binding show highly bursty behavior.
Registration/binding spikes occur during token-related events or new platform feature introductions. 
This includes DEGEN airdrops\footnote{
Airdrop is the free distribution of tokens to eligible wallets, often to promote token adoption or reward early users.} announcement~\cite{degen_history_top_meme,degen_official_website} (at \Circled{2} and \Circled{5}) or launch of new tokens that went viral (MOXIE~\cite{moxie_official_website} at \Circled{7} and DRB~\cite{clanker_grok_drb} at \Circled{10}).
At \Circled{4}, Farcaster launched its token-focused mini-apps~\cite{farcaster_miniapp_doc}, while at \Circled{9}, the platform introduced its official crypto wallet~\cite{farcaster_wallet_connection}.
The only token-unrelated event with a significant impact occurred at \Circled{6}, when Farcaster raised \$150M in funding~\cite{farcaster_150m_funding}, following an advertisement campaign (Farcaster Conference 2024)~\cite{farcaster_farcon_2024}.

This suggests that tokenomics is an important factor driving the Farcaster userbase. 
The platform decreased its registration fees multiple times from \$7 to \$5 in December 2023 (\Circled{1}), to \$3 in January 2024 (\Circled{3}), and to \$2 in August 2024 (\Circled{8})~\cite{storage_fee_changes}.
Surprisingly, those reductions did not significantly impact the new user registrations. 
The exact amount of the fees seems irrelevant for the new users who are mostly attracted by new features or the possibility of obtaining valuable tokens.
This is further confirmed by the high rate of users who bound a token wallet to their account. 
The ratio is steadily increasing since late January 2024, reaching 64.25\% in April 2025.
We provide an expanded correlation analysis between platform growth and real-world events in \Cref{app:rq1_social_token_related_events}.

%%%%%%%%%%%%%%%%%%%%%%%%%%%%%%%%%%%%%%%%%%%%%%%
\subsection{Prevalent Token Detection.}
\label{sec:rq1_prevalent_token_detection}
%%%%%%%%%%%%%%%%%%%%%%%%%%%%%%%%%%%%%%%%%%%%%%%
While flexible wallet binding to user accounts enhances economic autonomy and interoperability, one trade-off is that it simultaneously floods user wallets with numerous tokens unrelated to the Farcaster ecosystem. 
This introduces significant noise into our Farcaster incentive analysis. 
Therefore, we next examine the tokens circulating within the Farcaster ecosystem to discover methods for filtering this noise and identifying prevalent tokens that are genuinely relevant and impactful to the Farcaster social network. 

We identify 440{,}274 distinct tokens held in \ac{fid}-linked wallets. 
Yet, most exhibit limited activity: 99\% (435,871) tokens have fewer than 390 holders (by \acp{fid}) and fewer than 1{,}065 transactions, while the remaining 1\% (4,403) tokens account for 93.35\% of all holders and 94.58\% of all transactions (detailed in \Cref{app:rq1_overall_token_distribution}).
Furthermore, many tokens are widely distributed by just a small number of wallets, indicating a spam-like behavior without community adoption~\cite{torres2019art}.
This is common when token creators airdrop tokens to expand their popularity~\cite{allen2023airdrop,makridis2023rise,altcoinbuzz_farcaster}.
60\% (258,138) of tokens are never sent by a single \ac{fid}-bound wallet, and $>$99\% of tokens (434,094) involve fewer than 191 unique \ac{fid} senders (whether initiating a tip or executing a token swap).
These findings suggest that most tokens are passively received with limited social utility. 
We therefore strive to focus on sending activity to identify the platform's most socially engaged tokens.

For brevity, we summarize the process of selecting these tokens below, and provide more detailed description and justification in \Cref{app:rq1_prevalent_token_detection_appendix}:
\begin{enumerate*}
    \item We filter the tokens with inter-\ac{fid} transfers (\ie transacted between at least one pair of Farcaster users);
    \item we apply normalized Shannon entropy~\cite{lin2002divergence} to temporal transaction frequencies to filter out tokens with bursty, short-lived activity;
    \item we retain tokens above the 99th percentile in unique \ac{fid} senders (>$254$)\footnote{Note that this threshold of 254 unique \ac{fid} senders is derived from the 99th percentile of inter-\ac{fid} transfers and therefore differs slightly from the threshold of 191, which is the 99th percentile for overall transfers.}, filtering out those primarily distributed via airdrops rather than active social engagement.
    \item based on the transaction graph, we calculate the clustering coefficients \cite{sciencedirect_clustering} and select 0.3-0.6 as criteria ~\cite{Vasiliauskaite2020UnderstandingCV} to verify community-driven usage patterns.
\end{enumerate*}

Following this four-step process, we identify four prevalent tokens (DEGEN, MOXIE, HIGHER, and TN100X) issued by third-party developers as social rewards~\cite{degen_official_website, moxie_official_website, higher_official_doc, tn100x_tokenomics}. 
We additionally include USDC, a stablecoin incorporated into Farcaster’s official reward mechanisms ~\cite{altcoinbuzz_farcaster,farcaster_official_usdc_reward}. 
It is also used for user-to-user tipping as part of the platform’s official design~\cite{bitget_farcaster_tips}.
For our subsequent investigation, we use these five tokens as the primary subjects of study.\footnote{To avoid confusion caused by the numerous tokens sharing the same name and counterfeit tokens, the contract addresses for these five tokens are provided in \Cref{app:token_incentive_distribution}.}

%%%%%%%%%%%%%%% prevalent token table %%%%%%%%%%%%%%%%%%%%%
\sisetup{group-separator = {,}, group-minimum-digits = 4}
\begin{table}[htbp]
\centering
\caption{Prevalent tokens meeting the filtering criteria, sorted by overall transaction count (frequency).}
\label{tab:prevalent_token}
\footnotesize
\setlength{\tabcolsep}{1.6pt}
\renewcommand{\arraystretch}{0.99}
\begin{tabular*}{\textwidth}{@{\extracolsep{\fill}}lrrrrrrr}
    \toprule
    Token&
    \makecell{Holders} &
    \makecell{Total Txns} &
    \makecell{Inter-\ac{fid} Txns} &
    \makecell{\ac{fid} Sender} &
    \makecell{Clustering\\Coeff.} &
    \makecell{Token Age\\(wks)} &
    \makecell{Entropy\\(Norm)} \\
    \midrule
    \multicolumn{4}{r}{} & 
    \multicolumn{1}{r}{\footnotesize{\textbf{$\geq$ 254 (99\textsuperscript{th})}}} & 
    \multicolumn{1}{r}{\footnotesize{\textbf{$\in$ [0.3, 0.6]}}} & 
    \multicolumn{1}{r}{\footnotesize{\textbf{$\geq$ 26}}} & 
    \multicolumn{1}{r}{\footnotesize{\textbf{$\geq$ 0.9}}} \\
    \cmidrule(lr){5-8}
    DEGEN   & \num{152908} & \num{3337952} & \num{173772} & \textbf{\num{27723}} & \textbf{0.32} & \textbf{\num{73}} & \textbf{0.93} \\
    MOXIE   & \num{43742}  & \num{1810849} & \num{138728} & \textbf{\num{9002}}  & \textbf{0.58} & \textbf{\num{44}} & \textbf{0.92} \\
    HIGHER  & \num{32692}  & \num{320749}  & \num{51596}  & \textbf{\num{1153}}  & \textbf{0.41} & \textbf{\num{65}} & \textbf{0.90} \\
    TN100X  & \num{16409}  & \num{193678}  & \num{9996}   & \textbf{\num{1838}}  & \textbf{0.36} & \textbf{\num{69}} & \textbf{0.92} \\
    \bottomrule
    \addlinespace[1.0ex]
    \multicolumn{8}{@{}l}{\textit{USDC is included, meeting all but the clustering coefficient criterion.}} \\
    \addlinespace[0.5ex]
    \midrule
    USDC  & \num{216050} & \num{8768648} & \num{473801} & \textbf{\num{29464}} & \textcolor{-red}{0.23} & \textbf{\num{86}} & \textbf{0.93} \\
    \bottomrule
\end{tabular*}
\end{table}

\Cref{tab:prevalent_token} presents the primary transaction metrics and filtering criteria assessment for these five tokens. 
Notably, USDC only fails to meet the clustering coefficient criterion, with a value of 0.23 slightly below the lower threshold of 0.3, while satisfying all other three criteria. 
This indicates that USDC exhibits a relatively looser community structure compared to the four social reward tokens, which may be attributed to its additional use case as a stablecoin in payment scenarios rather than social interactions.
Furthermore, it is worth noting that DEGEN's holder count ($\approx$ 153k) ranks second only to USDC ($\approx$ 216k), surpassing the other three social tokens by a considerable margin (by $\approx$ 3.5 to 9.3 times).
Similarly, the number of \ac{fid} senders for DEGEN approaches that of USDC (27,723 vs. 29,464). 
These metrics demonstrate that DEGEN, being the earliest launched among the four social reward tokens, along with USDC (13 weeks older than DEGEN), has achieved the strongest network effects and highest community recognition among all tokens on Farcaster.

\subsection{Categorizing Incentive Mechanisms.}
\label{sec:rq1_category_incentive_mechanisms}
%%%%%%%%%%%%%%%%%%%%%%%%%%%%%%%
Finally, we investigate the incentive mechanisms that use these five tokens.
We analyze the official documentation~\cite{paybot_website, degen_official_website,moxie_official_website}, transaction history, and the smart contracts used for token distribution (detailed in \Cref{app:token_incentive_distribution}).
We then classify the incentive mechanisms into two main categories---tipping and algorithmic rewards, with algorithmic rewards further subdivided into third-party and official-led initiatives.

\paragraph{Inter-\ac{fid} Tipping.} In this mechanism, users directly send each other tokens using direct transfers:
\begin{enumerate*}
 \item direct blockchain transfers to the wallet address displayed on a recipient's profile; or  
  \item intermediary mini-apps (\eg \texttt{@paybot} \cite{paybot_website}) that enable socially-driven interactions (similar to the donate function in YouTube).\footnote{YouTube's fan funding feature: \url{https://www.youtube.com/intl/en/creators/fanfunding/}}
\end{enumerate*}
All 5 prevalent tokens are used in this mechanism.

\begin{figure}[htbp]
    \centering
    \begin{subfigure}{0.49\linewidth}
        \centering
        \includegraphics[width=\linewidth]{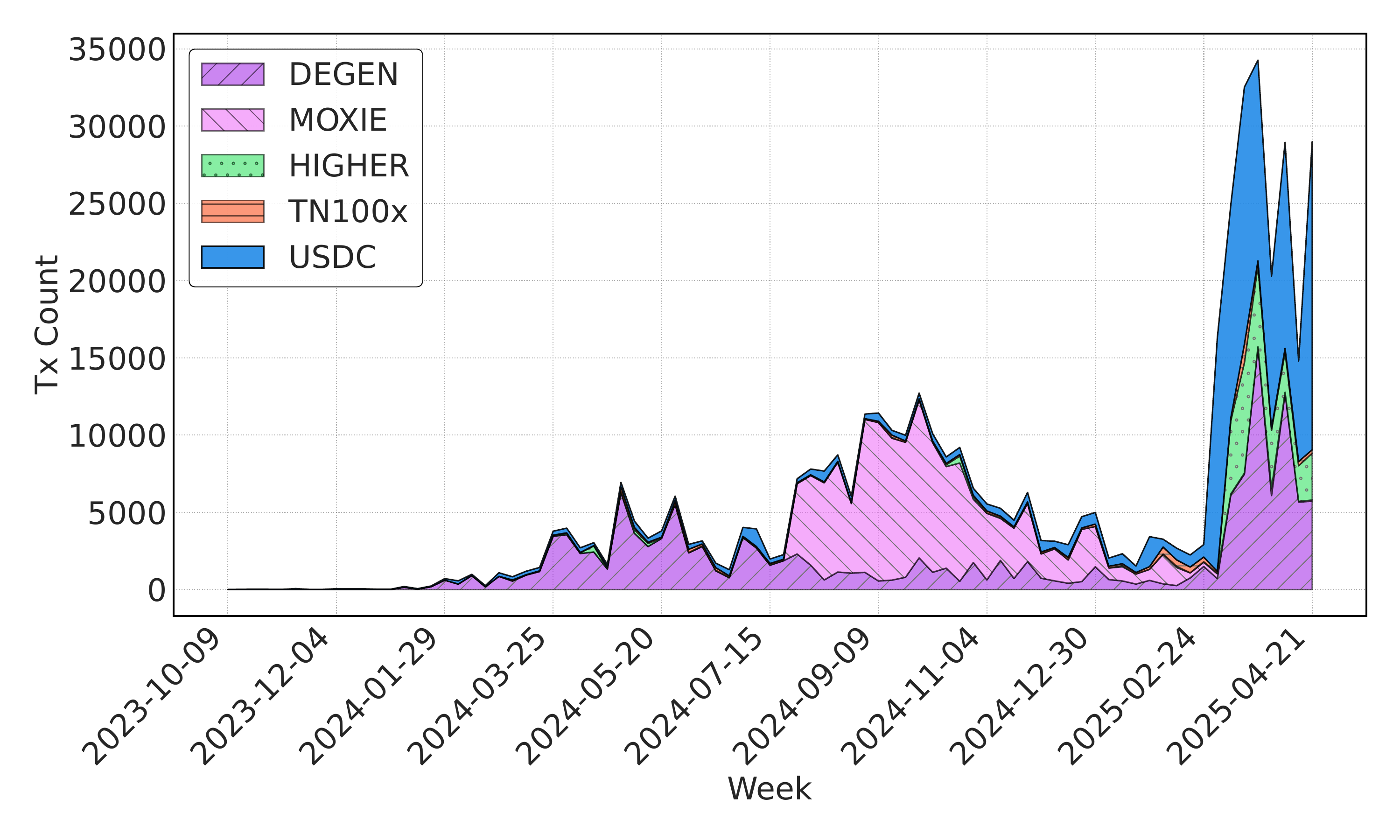}
        \caption{Tipping frequencies for five prevalent tokens.
        }
    \label{fig:mechanism_a_frequency_overtime}
    \end{subfigure}
    \hfill
    \begin{subfigure}{0.49\linewidth}
        \centering
        \includegraphics[width=\linewidth]{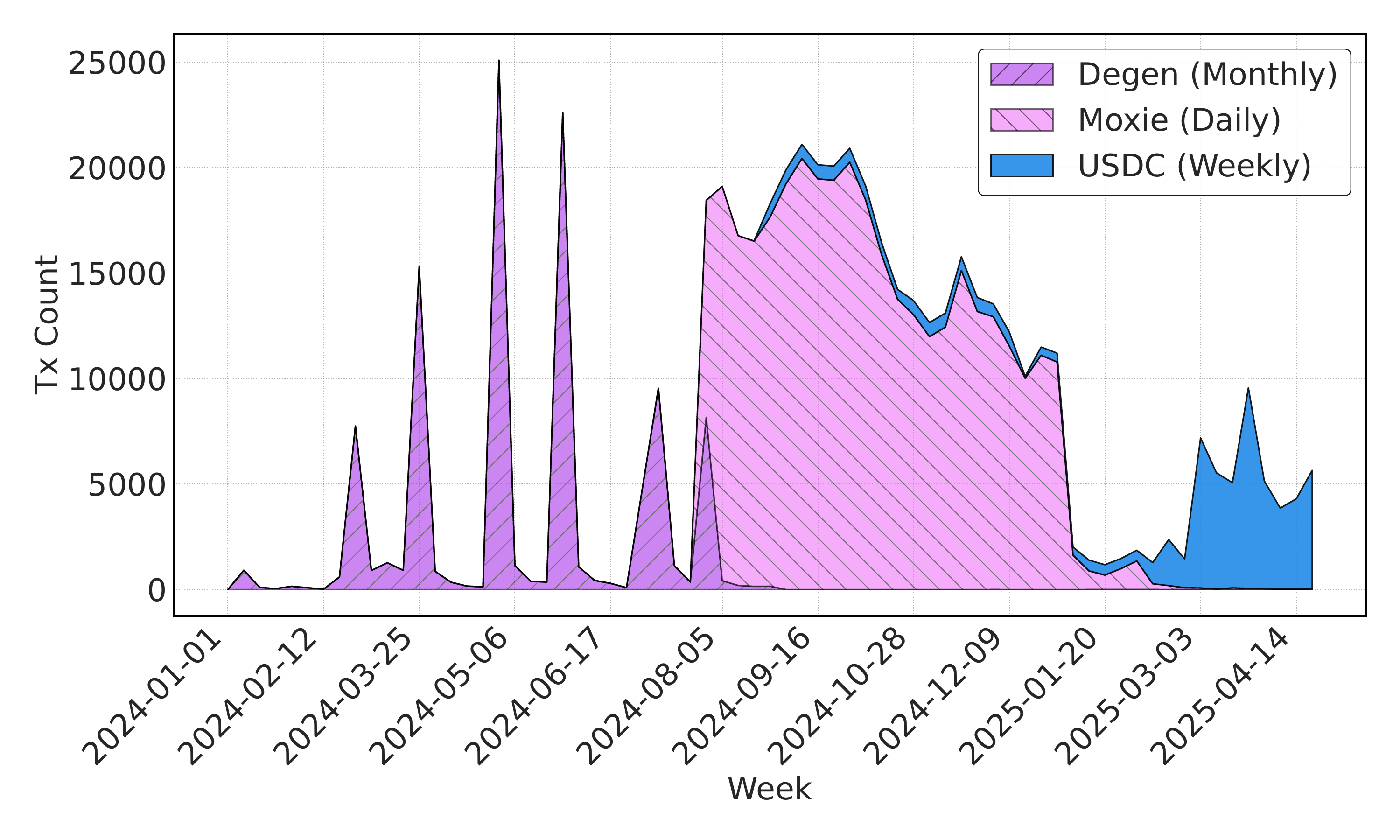}
        \caption{Rewarding frequencies for algorithmic tokens.}
    \label{fig:mechanism_b_c_frequency_overtime}
    \end{subfigure}
    \caption{Stacked area charts of weekly aggregated transaction frequencies by token across mechanisms.
    }
\label{fig:incentive_mechanism_frequency_a_b_c}
\Description{Stacked area charts of weekly aggregated transaction frequencies by token across mechanisms.}
\end{figure}

\paragraph{Third-party Algorithmic Rewards.} Farcaster enables any third party to launch tokens with bespoke distribution rules. 
These tokens are typically distributed via dedicated smart contracts designed to enhance user engagement. 
Such contracts often incorporate staking-based mechanisms\footnote{
Locking tokens in smart contracts for a set period to qualify for rewards or receive benefits like boosted scores.}
to mitigate undesired behaviors, including reward farming\footnote{A small group of users engages in circular reward-giving amongst themselves to exploit token reserves.} and sell-off pressure.\footnote{Upon receiving token rewards, users immediately exchange them for more established cryptocurrencies (\eg USDC, ETH).} 
We observe DEGEN\footnote{
DEGEN uses a nomination-based system where users reply to posts with messages like \enquote{100 \texttt{\$DEGEN}} to nominate others. These are collected monthly to determine token rewards for post creators, resulting in the spike pattern shown in \Cref{fig:mechanism_b_c_frequency_overtime}.
}~\cite{degen_official_website} and MOXIE\footnote{
MOXIE’s algorithm linearly weights posting, replying, and token staking in its reward function, making it more prone to metric gaming~\cite{moxie_official_website,moxie_scoring_system}.
}~\cite{moxie_official_website} being distributed through this mechanism.

\paragraph{Official Algorithmic Rewards.}
We distinguish the \emph{official} algorithmic reward mechanism, implemented by Farcaster's administration through the USDC stablecoin~\cite{farcaster_official_usdc_reward}.\footnote{Farcaster uses a black-box algorithm to mitigate farming and gaming behaviors, as noted by the co-founder: \url{https://farcaster.xyz/v/0x0e31071c}}
The mechanism provides weekly rewards to top-performing users based on engagement metrics.\footnote{These rewards follow a tiered structure, ranging from \$1 to \$300, allocated to qualified users across different ranking tiers.}

\paragraph{Mechanism Comparison.}
%\label{sec:rq1_mechanism_comparison}
%%%%%%%%%%%%%%%%%%%%%%%%%%%%%%%%%%%%%%
We first analyze user coverage and temporal transaction dynamics for the above three reward mechanisms.
Collectively, these mechanisms reach a total of 103,666 unique recipients, accounting for 11.59\% of all \acp{fid}.
More specifically, this figure corresponds to 17.56\% of active users, defined as individuals who have posted at least once.
This indicates a relatively high adoption rate given the diversity and scale of the user base, suggesting these incentive mechanisms play a substantial role in overall system usage.
Interestingly, \emph{Inter-\ac{fid} Tipping} and \emph{Third-party Algorithmic Reward} mechanisms reach 6.01\% and 6.43\% of all \acp{fid}, respectively, surpassing the \emph{Official Algorithmic Reward} (3.15\%). 
This indicates that community-driven incentive mechanisms achieve broader user coverage and incorporation than centralized, protocol-driven rewards.

% %%%%%%%%%stacked charts mechanism abc %%%%%%%%%

We next examine the individual tokens underpinning these mechanisms. 
\Cref{fig:mechanism_a_frequency_overtime} depicts the transaction frequencies of \emph{Inter-\ac{fid} Tipping} across five major tokens (as referenced in \Cref{sec:rq1_prevalent_token_detection}), while \Cref{fig:mechanism_b_c_frequency_overtime} illustrates the transaction frequencies for both third-party and official algorithmic reward mechanisms (note that only three out of five major tokens --- DEGEN, MOXIE and USDC --- are distributed within the algorithmic mechanism).

From the figures, we find that algorithmic rewards exhibit temporal patterns distinct from those of tipping. 
The frequencies of algorithmic rewards demonstrate pronounced episodic spikes, each corresponding to the initiation and duration of reward projects. 
By contrast, tipping frequencies display a more consistent and sustained temporal profile, closely tracking the fluctuations in daily active user (DAU) (as shown in \Cref{fig:DAU_walletRatio}). 
This contrast underscores the project-driven nature of algorithmic rewards versus the organic, user-driven dynamics of tipping.
%%%%%%%%%%%%%%%%%%%%%%%%%%%%%%%%%%%%%%%%%%%%%
\section{Socioeconomic Risks in Farcaster's Incentives} 
\label{sec:rq2_socioeconomic_risks}
%%%%%%%%%%%%%%%%%%%%%%%%%%%%%%%%%%%%%%%%%%%%%
Previous studies have shown that financial rewards, despite their potential to boost engagement, can inadvertently encourage negative behaviors (\eg entry barriers, reward concentration)~\cite{li2019incentivized,ba2022role,steemit2024richgetricher}. 
Following \textbf{RQ2}, we investigate whether Farcaster's incentive mechanism exhibits similar socioeconomic risks. 
Particularly, we focus on three example behaviors: new user participation, potential reward concentration, and echo chamber formation across incentive mechanisms.

%%%%%%% stacked area chart %%%%%%%%%%%
\begin{figure}[htbp]
    \centering
    \begin{subfigure}{0.49\linewidth}
        \centering
        \includegraphics[width=\linewidth]{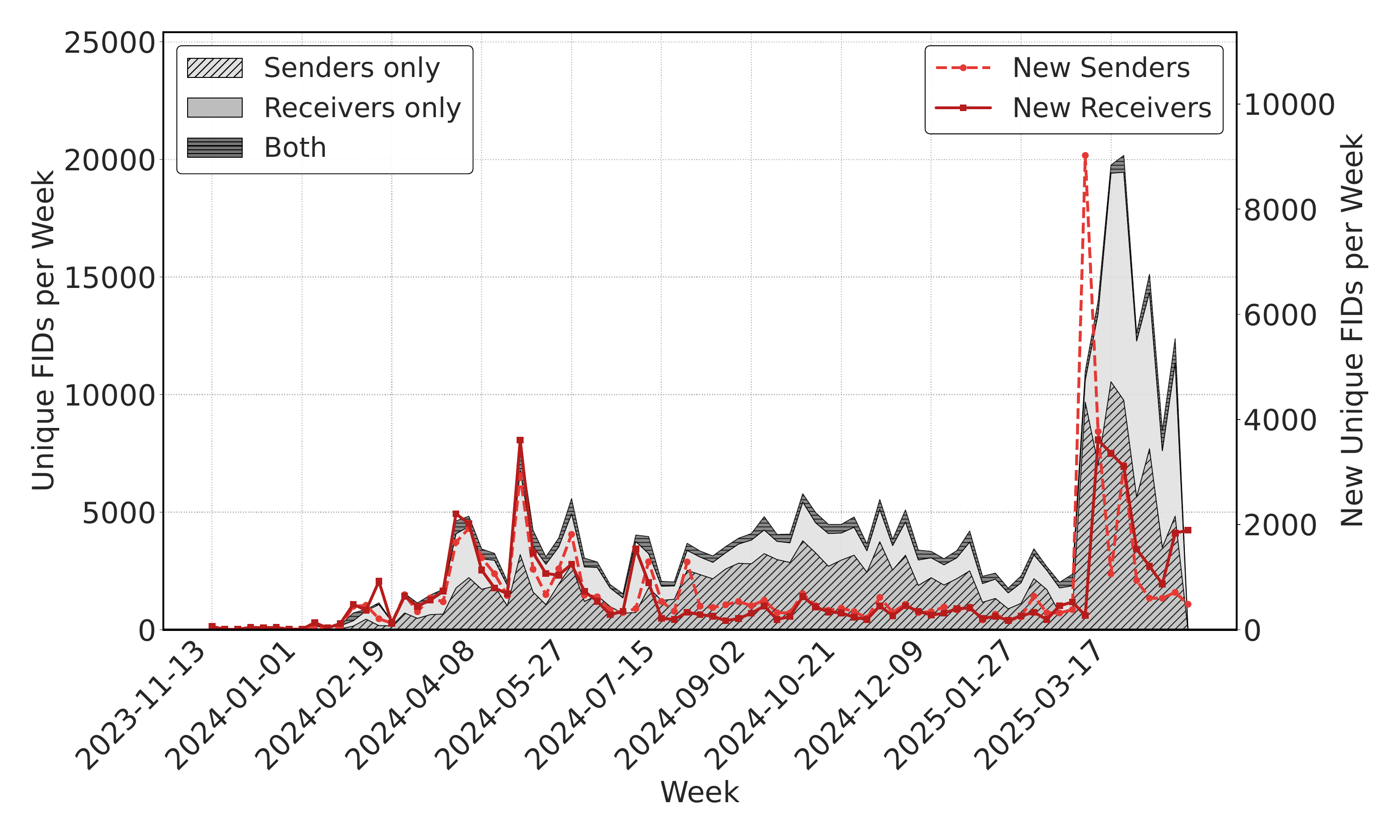}
        \caption{Inter-\ac{fid} Tipping.}
        \label{fig:mechanism_a_coverage_overtime}
    \end{subfigure}
    \hfill
    \begin{subfigure}{0.49\linewidth}
        \centering
        \includegraphics[width=\linewidth]{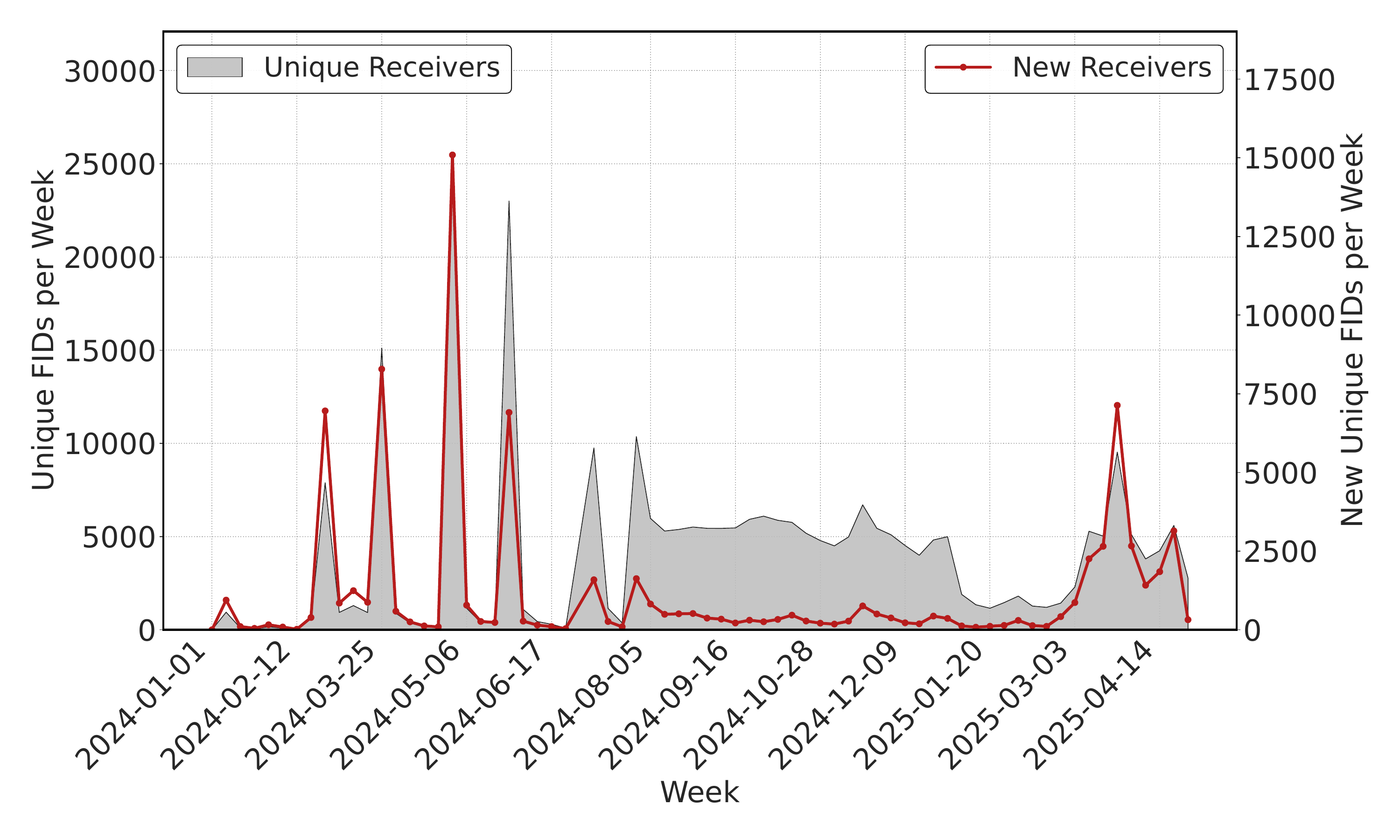}
        \caption{Algorithmic Reward.}
        \label{fig:mechanism_b_c_coverage_overtime}
    \end{subfigure}
    \caption{Temporal dynamics of new user participation in (a) tipping and (b) algorithmic rewarding mechanisms.}
    \label{fig:incentive_mechanism_coverage_a_b_c}
    \Description{Temporal dynamics of new user participation in (a) tipping and (b) algorithmic rewarding mechanisms. 
    The shaded areas represent the weekly number of unique participants---either senders (diagonal pattern), receivers (solid fill), or both (horizontal pattern). 
    The solid line indicates the number of new reward receivers appearing each week. 
    This visualization captures the dynamics of new participant inflow over time.}
\end{figure}

%%%%%%%%%%%%%%%%%%%%%%%%%%%%%%%%%%%%%%%%%%%%%
\subsection{New User Participation Rates}
\label{sec:rq2_incentive_inclusion}
%%%%%%%%%%%%%%%%%%%%%%%%%%%%%%%%%%%%%%%%%%%%%
To evaluate whether incentive mechanisms encourage broader participation (inclusivity) for newcomers or create barriers to entry (thereby offering more reward opportunities to incumbent recipients), we analyze the temporal patterns of user inclusion. 
This inclusion is measured by the rate of weekly new reward receivers to weekly total receivers.

\paragraph{Data and Methodology.} 
We perform a temporal analysis by calculating the weekly counts of unique senders and receivers (by \acp{fid}) for each type of reward, including eight token-mechanism pairs defined in \Cref{sec:rq1_category_incentive_mechanisms}. 
We also identify users who act as both senders and receivers within the same week. 
Additionally, we track the weekly influx of \emph{new} receivers and senders, defined as those receiving or sending the specific reward for the first time that week. 
This longitudinal analysis reveals new user participation patterns across different incentive mechanisms.

%%%%%%%% tipping_unique_senders_receivers%%%%%% 
\newcommand*\rot{\rotatebox{90}}
\begin{table}
    \centering
    \caption{Weekly user tipping statistics surrounding the Farcaster wallet launch.}
    \label{tab:tipping_inclusion_stats}
    \footnotesize
    \setlength{\tabcolsep}{2pt}
    \renewcommand{\arraystretch}{0.99}
    \begin{tabular}{lrrrr}
        \toprule
        \textbf{Metric}        & \textbf{Pre-launch Mean} & \textbf{Pre-launch Median} & \textbf{Post-launch Mean} & \textbf{Post-launch Median} \\
        \midrule
        Unique Senders        & 1,823 & 1,779 & 7,960 & 8,014 \\
        Unique Receivers      & 1,220 & 1,117 & 6,849 & 7,198 \\
        New Senders           &   528 &   407 & 2,480 & 1,001 \\
        New Receivers         &   513 &   334 & 1,975 & 1,687 \\
        \bottomrule
    \end{tabular}
\end{table}

\paragraph{Results.}
\Cref{fig:incentive_mechanism_coverage_a_b_c} shows the weekly count of \acp{fid} by their types.
The stacked area charts display the number of unique senders (diagonal), receivers (solid), and users acting as both (horizontal), with overlaid lines representing weekly new receivers (solid red) and new senders (dashed red).

For \emph{Inter-\ac{fid} Tipping} (see \Cref{fig:mechanism_a_coverage_overtime} and \Cref{tab:tipping_inclusion_stats}), the total sender and receiver counts fluctuate synchronously and the tipping user base is significantly enlarged due to the Farcaster wallet launch (in late Feb 2025). 
An anomaly occurred during the significant tipping surge following the Farcaster wallet launch: the weekly new sender count spiked sharply (reaching 9,023 for the week of March 3 2025), far exceeding new receivers (which remained low at 270, similar to pre-launch levels). 
The following week, new senders dropped to 3,772 while new receivers rose to 3,611, and both metrics quickly resynchronized. 
Further breakdown (see \Cref{fig:stacked_and_line_each_token_abc} in \Cref{app:rq2_socioeconomic_risks_appendix}) reveals this spike was mainly driven by USDC tipping.
We conjecture this is due to official campaigns encouraging users to send USDC to activate wallet features or qualify for airdrops~\cite{altcoinbuzz_farcaster}.

It is also worth noting that each week, only a small fraction ($Mean = 9.76\%, SD = 5.83\%, median = 8.84\%$) of users engage in both sending and receiving.
This indicates that most tipping flows are unidirectional instead of reciprocal.

For \emph{Algorithmic Rewards} (see \Cref{fig:mechanism_b_c_coverage_overtime}), 
recall that the three tokens (DEGEN, MOXIE, and USDC) were distributed in distinct time windows (as shown in \Cref{fig:mechanism_b_c_frequency_overtime}), allowing the aggregated data to still reveal clear trends. 
From Jan to August 2024, DEGEN’s algorithmic rewards followed a monthly claim pattern (as mentioned in \Cref{sec:rq1_category_incentive_mechanisms}).
For DEGEN, both weekly total receivers and weekly new receivers increased in the first four months (Jan-April 2024). 
The peak of new user participation rate (59.24\%) was reached in late April (total 25,475; new 15,091), after which both weekly total and new receivers declined --- with new receivers dropping more rapidly. 
By late May, new receivers accounted for only 30\% of total receivers (23,005 vs. 6,908), and by June and July, this dropped to $\approx$ 16\%, after which DEGEN algorithmic rewards ceased.

During the subsequent MOXIE reward period (about 30 weeks from Aug 2024 to Jan 2025), the weekly gap between total receivers (mean 4,293) and new receivers (mean 328) was much wider: new receivers account for only 7.6\%, indicating far less inclusion in MOXIE's algorithmic reward project, with more reward opportunities offered to incumbent receivers. 
This stands in sharp contrast to DEGEN, suggesting the user-driven nomination-based reward design of DEGEN was more inclusive than MOXIE's behavioral scoring approach (detailed in \Cref{sec:rq1_category_incentive_mechanisms}).
Finally, the official USDC algorithmic reward (from Feb 2025 to present) also exhibited acceptable inclusivity: new receivers accounted for 48.45\% of total receivers on average (2,477 vs. 5,112), with both metrics (weekly total and new receivers) moving in parallel.

This suggests that each token is used in quite distinct manners, with different degrees of inclusivity for attracting new users. 
This arguably highlights the benefits of the pluralistic approach taken by Farcaster: by providing a variety of entry points, it reduces the entry barrier thus mitigating the risk of low user engagement inherent in a single-tokenomic system.

%%%%%%%%%%%%%%%%%%%%%%%%%%%%%%%%%%%%%%%%%%%%%
\subsection{Income Inequality and Wealth Concentration}
\label{sec:rq2_rich_get_richer}
%%%%%%%%%%%%%%%%%%%%%%%%%%%%%%%%%%%%%%%%%%%%%
Previous research has shown that single-token incentive mechanisms can lead to income inequality and wealth concentration (Gini coefficient > 0.9) across decentralized networks~\cite{li2019incentivized,memo_set_in_stone,michienzi2024wealth}.
This motivates us to assess whether Farcaster’s pluralistic token incentive ecosystem (where user-to-user tipping and developer-led algorithmic rewards coexist) also faces the same challenges, resulting in wealth concentration.

%%%%% gini table %%%%%%
\begin{table}[htbp]
\centering
\caption{Income distribution and inequality metrics across three incentive mechanisms in Farcaster.}
\label{tab:token_mech_gini}
\footnotesize
\setlength{\tabcolsep}{2pt}
\renewcommand{\arraystretch}{0.99}
\begin{tabular}{lrrrrr rr r}
\toprule
& \multicolumn{5}{c}{\textbf{Inter-\ac{fid} Tipping Reward}} 
& \multicolumn{2}{c}{\textbf{3rd-Party Algo. Reward}} 
& \multicolumn{1}{c}{\textbf{Off. Algo. Reward}}
\\
\cmidrule(lr){2-6} \cmidrule(lr){7-8} \cmidrule(lr){9-9}
\textbf{Metric}
& \textbf{Degen}
& \textbf{Higher}
& \textbf{USDC}
& \textbf{Moxie}
& \textbf{Tn100x}
& \multicolumn{1}{c}{\qquad \textbf{Degen}}
& \multicolumn{1}{c}{\qquad \textbf{Moxie}}
& \textbf{USDC} \\
\midrule
Gini Coeff.
& 0.8304 & 0.9382 & 0.8631 & 0.7246 & 0.8277
& \multicolumn{1}{r}{\qquad 0.8433} & \multicolumn{1}{r}{\qquad 0.9248} & 0.8598 \\
Total
& 99,141.39 & 4,132.93 & 94,788.73 & 86,136.68 & 5,543.84
& \multicolumn{1}{r}{\qquad 49,612,724.35} & \multicolumn{1}{r}{\qquad 1,657,380.21} & 517,831.34 \\
Max
& 2,061.85 & 616.44 & 999.99 & 2,040.63 & 457.78
& \multicolumn{1}{r}{\qquad 492,133.29} & \multicolumn{1}{r}{\qquad 15,542.65} & 1,772.83 \\
Min
& 0.00 & 0.00 & 0.00 & 0.00 & 0.00
& \multicolumn{1}{r}{\qquad 0.00} & \multicolumn{1}{r}{\qquad 0.00} & 0.01 \\
Mean
& 3.13 & 0.29 & 2.73 & 9.97 & 3.11
& \multicolumn{1}{r}{\qquad 1,050.81} & \multicolumn{1}{r}{\qquad 77.32} & 15.49 \\
Median
& 0.19 & 0.02 & 0.11 & 3.89 & 0.07
& \multicolumn{1}{r}{\qquad 111.03} & \multicolumn{1}{r}{\qquad 1.95} & 1.00 \\
\bottomrule
\end{tabular}
\end{table}

%%%%%%% ECDF %%%%%%%%%%%
\begin{figure}[htbp]
    \centering
    \begin{subfigure}{0.49\linewidth}
        \centering
        \includegraphics[width=\linewidth]{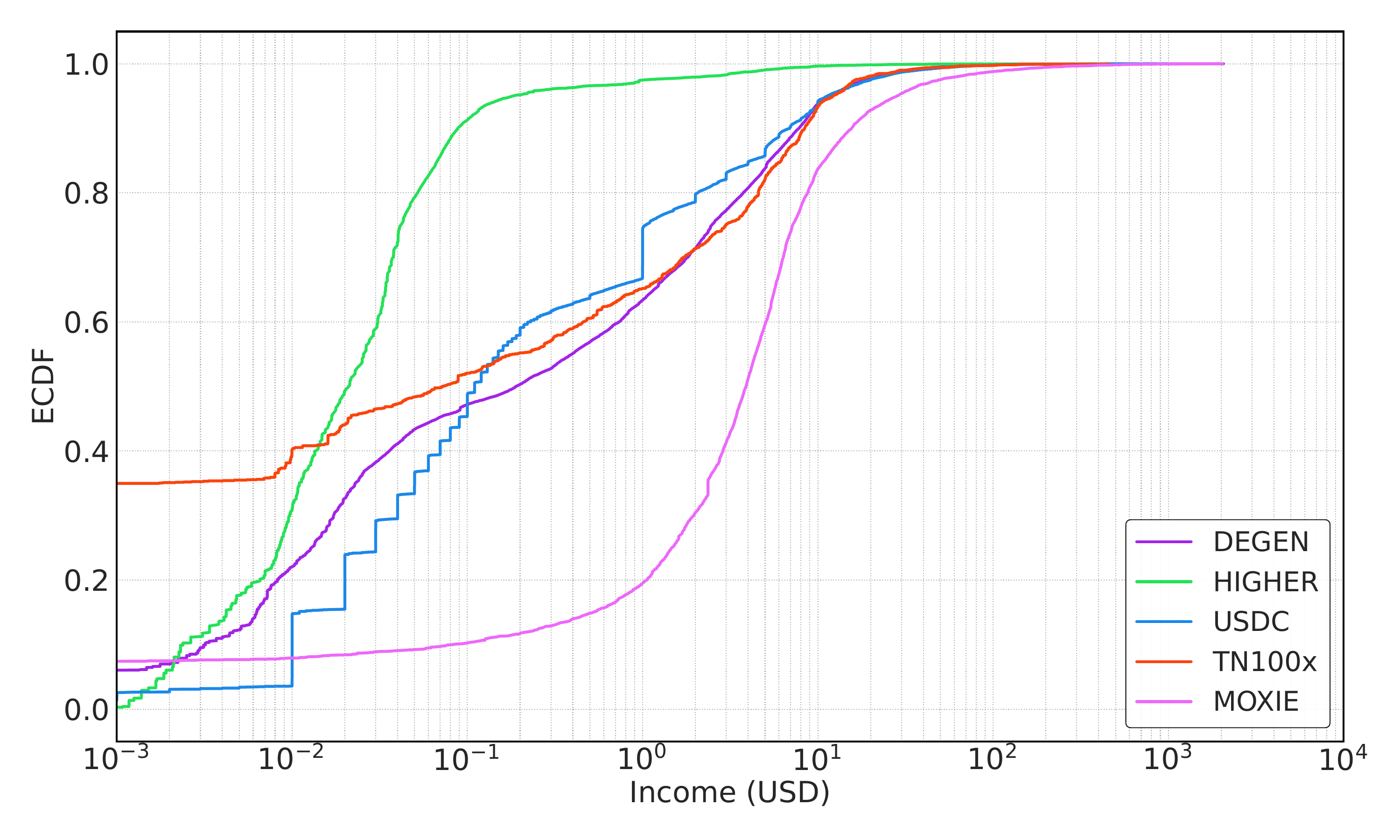}
        \caption{Inter-\ac{fid} tipping.}
        \label{fig:mechanism_a_income_ecdf}
    \end{subfigure}
    \hfill
    \begin{subfigure}{0.49\linewidth}
        \centering
        \includegraphics[width=\linewidth]{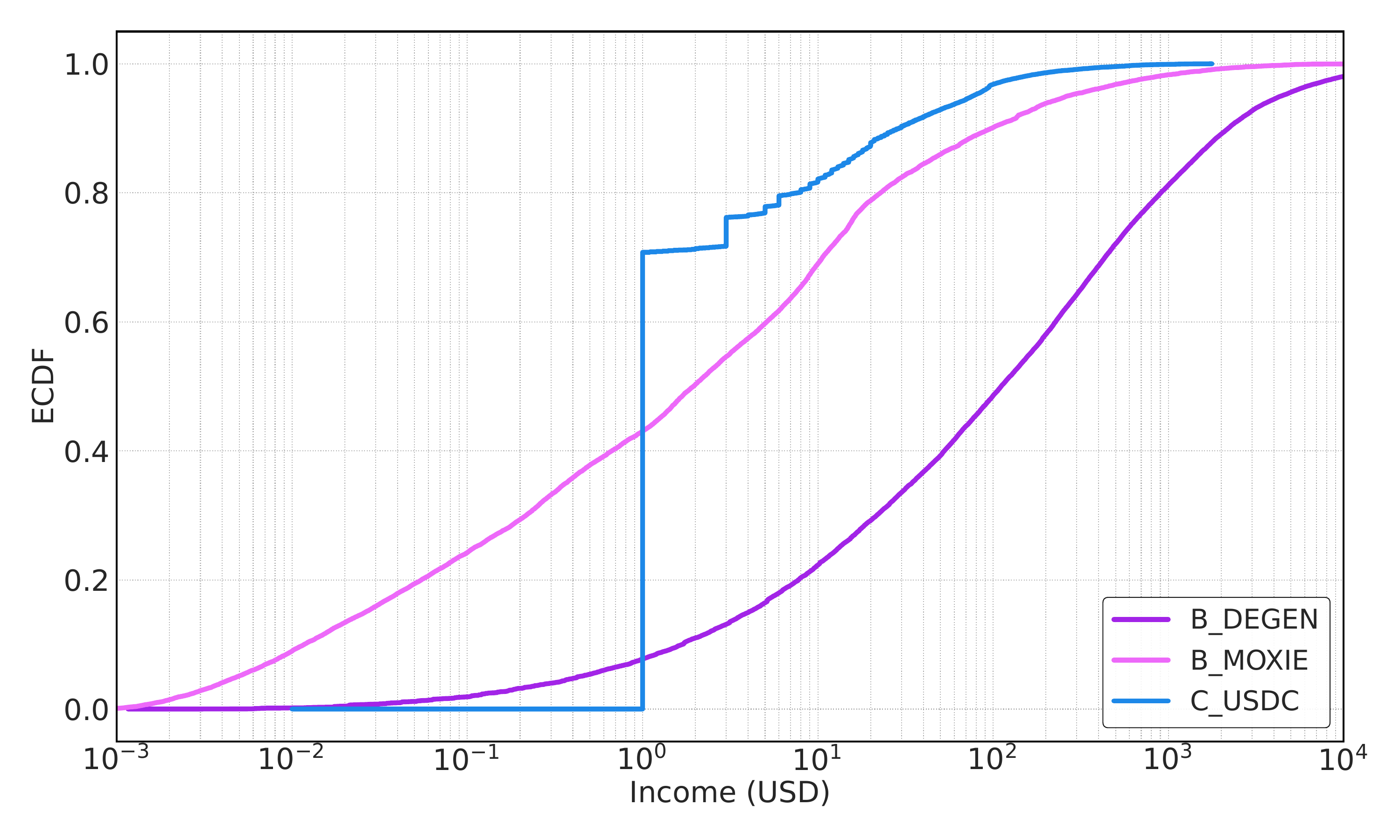}
        \caption{Algorithmic rewarding.}
        \label{fig:mechanism_b_c_income_ecdf}
    \end{subfigure}
    \caption{Income distribution (USD) across incentive mechanisms.}
    \label{fig:incentive_mechanism_income_ecdf}
    \Description{Two ECDF plots visualizing the income distribution (USD) across incentive mechanisms.}
\end{figure}

\paragraph{Data and Methodology.}
To measure wealth concentration, for tokens other than USDC (a stablecoin), we first collect daily average price data for DEGEN, MOXIE, HIGHER and TN100X. 
We estimate each user's income by multiplying the received token amount by the average daily USD price on the day of receipt. 
This provides a practical approximation, as users may exchange their tokens at any time.
Finally, we measure the concentration of each token-mechanism pair.

\paragraph{Results.}
\Cref{tab:token_mech_gini} summarizes key statistics for all major incentive mechanisms and \Cref{fig:incentive_mechanism_income_ecdf} shows the \ac{ecdf} of user income (in USD) for each token-mechanism pair.  
It reveals that despite Farcaster’s pluralistic approach (designed to potentially mitigate income inequality by offering more reward-receiving opportunities to a broader user base), significant wealth concentration persists across almost all token-mechanism pairs. 
The consistently high Gini coefficients (mostly 0.82-0.94 in \Cref{tab:token_mech_gini}, with one outlier at 0.72), although lower than the extreme concentration (Gini $\approx$ 0.99) observed in Steemit's single-token economy~\cite{steemit2024richgetricher}, suggest that while tipping and algorithmic rewards coexist in the ecosystem, their independent operation may still reintroduce the centralization issues that plagued earlier token-based social platforms.

For \emph{Inter-\ac{fid} Tipping},  it is noteworthy that income from the HIGHER token exhibits a right-skewed distribution (more low-income users), with the highest Gini coefficient of 0.94. 
This high degree of inequality is further underscored by the fact that 98\% of users received less than \$1, and 80\% received less than \$0.05. 
This imbalance is consistent with HIGHER possessing the lowest number of unique senders and the most skewed sender-to-receiver ratio (1:12.5) among all tipping tokens examined (detailed in \Cref{tab:token_mech_metrics}). 
Critically, we notice that 92.4\% of HIGHER tipping transactions originate from only two bot accounts. 
These bots reward trivial amounts of HIGHER to users during specific interactions, such as content replies or lottery drawings. 
Consequently, the distribution of HIGHER is heavily concentrated among low-value recipients and is primarily driven by automated bot activity rather than organic peer-to-peer engagement.

%%%%%%%%%%%%% stats for mechanisms %%%%%%%%%%%
\begin{table}[htbp]
\centering
\caption{Token activity metrics across three incentive mechanisms on Farcaster.}
\label{tab:token_mech_metrics}
\footnotesize
\setlength{\tabcolsep}{1.8pt}
\renewcommand{\arraystretch}{0.99}
\begin{tabular}{lrrrrr rr r}
\toprule
& \multicolumn{5}{c}{\textbf{Inter-\ac{fid} Tipping}} 
& \multicolumn{2}{c}{\textbf{3rd-Party Algo. Reward}} 
& \multicolumn{1}{c}{\textbf{Off. Algo. Reward}}
\\
\cmidrule(lr){2-6} \cmidrule(lr){7-8} \cmidrule(lr){9-9}
\textbf{Metric}
& \textbf{DEGEN}
& \textbf{HIGHER}
& \textbf{USDC}
& \textbf{MOXIE}
& \textbf{TN100X}
& \multicolumn{1}{c}{\qquad \textbf{DEGEN}}
& \multicolumn{1}{c}{\qquad \textbf{MOXIE}}
& \textbf{USDC} \\
\midrule
\# Unique Sender
& 27,519 & 1,267 & 30,004 & 9,212 & 1,859
& \multicolumn{1}{r}{\qquad --} & \multicolumn{1}{r}{\qquad --} & -- \\
\# Unique Receiver
& 40,836 & 15,849 & 15,872 & 5,252 & 3,459
& \multicolumn{1}{r}{\qquad 47,748} & \multicolumn{1}{r}{\qquad 21,505} & 28,181 \\
\# Total Transaction
& 196,555 & 58,449 & 596,213 & 139,516 & 11,026
& \multicolumn{1}{r}{\qquad 101,232} & \multicolumn{1}{r}{\qquad 354,952} & 81,812 \\
\bottomrule
\end{tabular}
\end{table}

For \emph{Algorithmic Rewards}, the distribution patterns reflect a more structured approach than tipping.
USDC, for example, uses a tiered reward scheme based on weekly behavioral rankings, distributing between \$1 to \$300 (see \Cref{sec:rq1_category_incentive_mechanisms}). 
Notably, 75\% of users receive the minimum reward of \$1.
While both USDC and MOXIE rely on similar behavioral scoring algorithms (detailed in \Cref{sec:rq1_category_incentive_mechanisms}), their metric selection and openness differ significantly. MOXIE's algorithm explicitly weights posting, replying, and token staking, making it more susceptible to metric gaming and the rich-get-richer phenomenon~\cite{moxie_official_website,moxie_scoring_system}. In contrast, USDC's scoring algorithm remains opaque but predominantly includes social behavior signals, without any wealth status metrics, making it more resistant to gaming while ensuring more opportunity to baseline rewards for a broader user base~\cite{farcaster_official_usdc_reward}.
This aligns with research showing that modest, guaranteed incentives can outperform larger, uncertain rewards in driving participation~\cite{ichimiya2023evaluation}. 

Consequently, MOXIE's algorithmic rewards show more pronounced income inequality (Gini: 0.92) contrasted with USDC (Gini: 0.86).
This is likely due to MOXIE's transparent scoring system that allows strategic users to optimize their behavior for maximum rewards, as well as its token-stake boosting scores, resulting in the rich-get-richer effect.
This extreme concentration in MOXIE's algorithmic rewards echoes its poor inclusivity metrics observed in \Cref{sec:rq2_incentive_inclusion}, as new users face barriers to participation while rich incumbents monopolize the reward opportunities.

That said, MOXIE also presents an interesting case where its redistribution mechanism effectively mitigates initial wealth concentration. 
While its algorithmic rewards show high inequality (Gini: 0.92) in initial distribution, its unique follower-followee redistribution mechanism~\cite{moxie_official_website} --- where a portion of rewards (designated by the followee, \eg 20\%) received by followees flows to followers in the form of \emph{Inter-\ac{fid} Tipping}---contributes to more balanced secondary distribution for MOXIE tipping (Gini: 0.72). 
This suggests that carefully designed redistribution rules \emph{can} help address wealth concentration issues even when primary reward allocation is highly skewed.

%%%%%%%%%%%%%%%%%%%%%%%%%%%%%%%%%%%%%%%%%%%%%
\subsection{Echo Chamber Effect in Tipping}
\label{sec:rq2_echo_chamber}
%%%%%%%%%%%%%%%%%%%%%%%%%%%%%%%%%%%%%%%%%%%%%
While \emph{Inter-\ac{fid} Tipping} facilitates value exchange and content monetization, it may inadvertently amplify echo chamber effects within social networks. 
This might drive users to create smaller social communities, driven by trends in reward-giving. 
Thus, in the context of tipping behavior, we define an \emph{echo chamber} as a closed loop of economic value circulation where tipping flows predominantly remain within tight-knit communities rather than across diverse user groups~\cite{dimaggio2012network}. 
Such economic echo chambers could potentially lead to the concentration of tipping flows among a small subset of users, reducing exposure to diverse social content and limiting the platform's ability to sustain a broad and inclusive incentive model.

\paragraph{Data and Methodology.}
To investigate potential echo chamber effects, we begin by examining the temporal dynamics between following and tipping relationships (\Cref{tab:echo_tip_follow_relation}). 
Specifically, we compare the timestamp of the first tip between pairs of users with the timestamp of their follow relationship (if any). 
We classify tipping interactions into three categories based on the timing of follow relationships: 
\begin{enumerate*}
    \item Followed before first tip,  
    \item Followed after first tip, 
    and
    \item Never followed (\ie tipping between users who never established a follow relationship), accounting for 55.61\%, 11.97\%, and 32.42\% of all tips, respectively.
\end{enumerate*} 
This distribution motivates a further analysis of whether tipping interactions, especially those without underlying social relationships (\ie \emph{Never Followed}), tend to occur within existing echo chambers or bridge across them.

To explore this, we construct the Farcaster social graph based on follow relationships, resulting in a directed network with 883,712 nodes and 159 million edges --- we refer to this as \enquote{Follow network} (see \Cref{tab:echo_network_comparison} in \Cref{app:rq2_socioeconomic_risks_appendix} for more details). 
We then incorporate the tipping relationships between pairs of users onto this network as additional edges to form the combined \enquote{Follow + Tip} network.  
The tipping relationships correspond to 55,847 edges.\footnote{To ensure robust analysis, we exclude the lottery tipping bot (FID: 987581, Fname: Warpslot) to focus on organic content-driven tipping interactions.}
To assess whether tipping rewards circulate within or across echo chambers, we identify communities within the follow network --- \ie groups of users with dense follow relationships each serving as a potential echo chamber.

We use two community detection approaches: \emph{NetworKit}'s Louvain modularity optimization~\cite{networkit_louvain} and \emph{Infomap}'s information flow-based partitioning~\cite{infomap}.
Due to the inherent randomness in NetworKit's implementation, metrics subject to variation are reported as either means or ranges from three independent runs (in \Cref{tab:echo_tip_follow_relation}).
Finally, we map tipping relationships (tip edges) onto the community structure to evaluate whether economic rewards tend to remain within follower communities or flow across them.
Since tipping edges are overlaid on top of the follow network, we also assess whether they substantially alter the underlying community structure. 
To quantify the extent to which community structures persist across different network configurations (\ie Follow vs. Follow+Tip), we use two standard metrics: Normalized Mutual Information (NMI) and Adjusted Mutual Information (AMI).\footnote{NMI measures the similarity between two clusterings but may overstate agreement by not accounting for chance overlap. 
AMI corrects for this by adjusting for the expected similarity under random labelings to yield a more conservative measure of structural alignment.}

%%%%%%%%%%% echo chamber tables %%%%%%%%%%%%%%%
\begin{table}[t]
\centering
\begin{minipage}{0.4\textwidth}
  \centering
  \caption{Tipping and following relationship.}
  \label{tab:echo_tip_follow_relation}
  \footnotesize
  \setlength{\tabcolsep}{1.5pt}
  \renewcommand{\arraystretch}{0.95}
  \begin{tabular}{lllll}
    
    \multicolumn{4}{l}{\textbf{Louvain}} \\
    \midrule
    Following Status & \# & \% & Inter-comm. Ratio\\
    \midrule
    Never Followed & 55,847 & 32.42\%  & [45.06\%, 51.48\%] \\
    Followed Before First Tip & 95,790 & 55.61\%  & [22.24\%, 26.13\%] \\
    Followed After First Tip & 20,621 & 11.97\%  & [22.47\%, 25.26\%] \\
    \midrule
    \multicolumn{4}{l}{\textbf{Infomap}} \\
    \midrule
    Never Followed &   -- & -- & 74.68\% \\
    Followed Before First Tip &   -- & --  & 56.10\% \\
    Followed After First Tip &   -- & -- & 59.66\% \\
    \bottomrule
  \end{tabular}
\end{minipage}
\hfill
\begin{minipage}{0.4\textwidth}
  \centering
  \caption{Network overlap metrics.}
  \label{tab:echo_network_overlap}
  % \normalsize
  \footnotesize
  \setlength{\tabcolsep}{1.5pt}
  \renewcommand{\arraystretch}{0.95}
  \begin{tabular}{llll}
    \multicolumn{4}{l}{\textbf{Louvain}} \\
    \midrule
    Network Pair & NMI & AMI & Max Overlap\\
    \midrule
    Follow vs. Combined &  0.73 &  0.73 &  0.925 \\
    Follow vs. Tip &  0.33 &  0.26 &  0.380 \\
    Tip vs. Combined &  0.31 &  0.24 &  0.694 \\
    \midrule
    \multicolumn{4}{l}{\textbf{Infomap}} \\
    \midrule
    Follow vs. Combined &  0.91 &  0.91 &  0.928 \\
    Follow vs. Tip &  0.19 &  0.13 &  0.519 \\
    Tip vs. Combined &  0.21 &  0.15 &  0.227 \\
    \bottomrule
  \end{tabular}
\end{minipage}
\end{table}

%%%%%%% echo stats %%%%%%%
% Table 5 & 6

\iffalse  %%%%%%%%%%%%%%%%%%%%%%%%%%
\begin{table}[t]
\centering
\begin{minipage}{0.48\textwidth}
  \centering
  \caption{Tipping and following relationship}
  \label{tab:echo_tip_follow_relation}
  \normalsize
  \setlength{\tabcolsep}{1.5pt}
  \renewcommand{\arraystretch}{0.95}
  \begin{tabular}{lllll}
    \toprule
    Following Status & Count & Percentage & Inter-community Ratio\\
    \midrule
    Never Followed & 55,847 & 32.42\%  & [45.06\%, 51.48\%] \\
    Followed Before First Tip & 95,790 & 55.61\%  & [22.24\%, 26.13\%] \\
    Followed After First Tip & 20,621 & 11.97\%  & [22.47\%, 25.26\%] \\
    \bottomrule
  \end{tabular}
\end{minipage}
\hfill
\begin{minipage}{0.48\textwidth}
  \centering
  \caption{Network overlap metrics}
  \label{tab:echo_network_overlap}
  \normalsize
  \setlength{\tabcolsep}{1.5pt}
  \renewcommand{\arraystretch}{0.95}
  \begin{tabular}{llll}
    \toprule
    Network Pair & NMI & AMI & Max Overlap Rate\\
    \midrule
    Follow vs. Combined &  0.73 &  0.73 &  0.925 \\
    Follow vs. Tip &  0.33 &  0.26 &  0.380 \\
    Tip vs. Combined &  0.31 &  0.24 &  0.694 \\
    \bottomrule
  \end{tabular}
\end{minipage}
\end{table}
\fi %%%%%%%%%%%%%%%%%%%%%%%%%%

% \vspace{1.5em}

\paragraph{Results.} 
Using Louvain (Infomap) detection, we find that 33\% (62.5\%) of all tips between \acp{fid} cross community boundaries while the remaining 67\% (37.5\%) stay within communities,  indicating that the inference about an overall echo-chamber effect is method-dependent and thus not supported by consistent evidence.
We show full results for the relationship between tipping behavior and communities in \Cref{tab:echo_community_detection} in \Cref{app:rq2_socioeconomic_risks_appendix}.

The stronger inter-community tipping observed under Infomap reflects its finer-grained community resolution. 
Infomap produces hierarchical clusters at multiple levels, \ie Levels 0, 1, 2, and 3. 
We focus on Level~1 for our analysis, because it strikes a balance between overly coarse groupings (\eg a single dominant community at Level~0) and overly fragmented structures at finer levels. 
At this level, the largest community detected by Infomap contains 330,867 users, compared to approximately 462,230 in Louvain.

We also observe a clear difference in tipping behavior based on the underlying follow relationship between users. 
As shown in~\Cref{tab:echo_tip_follow_relation}, tipping between users who never followed each other is substantially more likely to cross community boundaries: 45–51\% under Louvain and 74.68\% under Infomap, with a ratio 1.3--2 times higher than following pairs. 
In contrast, tips between users with an existing follow relationship are more likely to remain within the same community --- only 22--26\% (vs. 45--51\%) cross-community under Louvain and 56.1\% (vs. 74.68\%) under Infomap. 

These differences in community-level tipping behavior are consistent with structural differences in how communities are formed under each network. 
As shown in~\Cref{tab:echo_network_overlap}, Louvain and Infomap produce highly similar communities when applied to the follow and combined graphs (NMI = 0.73 and 0.91, respectively), suggesting that tipping edges have limited impact on the overall community structure. 
This is also shown by the similar network metrics between Follow-only and Follow+Tip networks (~\Cref{tab:echo_network_comparison} in the \Cref{app:rq2_socioeconomic_risks_appendix}.). 
However, both algorithms yield substantially lower overlap between follow and tip networks (\eg AMI = 0.26 for Louvain, 0.13 for Infomap), indicating that tipping relationships form a distinct layer of interaction. 
Thus, while follow links define stable community boundaries, tipping behaviors can cross these boundaries, particularly under finer-grained community partitions.

%%%%%%%%%%%%%%%%%%%%%%%%%%%%%%%%%%%%%%%%%%%%%
\section{Effectiveness of Token Incentives on Social Activities}
\label{sec:rq3_token_social_causality}
%%%%%%%%%%%%%%%%%%%%%%%%%%%%%%%%%%%%%%%%%%%%%
Building upon our findings from \Cref{sec:rq1_user_base_token_incentive_mechanisms} (RQ1) regarding the prevalence and diversity of token adoption, and \Cref{sec:rq2_socioeconomic_risks} (RQ2) concerning socioeconomic risks, we finally investigate whether token incentives effectively encourage subsequent social engagement \textbf{(RQ3)}, as this is the ultimate goal of the incentive design.

Given the criticism faced by previous platforms (\eg Steemit) for coordinated low-quality content farming~\cite{li2019incentivized,ba2022role}, we specifically focus on whether Farcaster's pluralistic token incentive ecosystem fosters greater user engagement.
To answer this, we investigate the causal impact of Farcaster's token incentives on social behavior through two complementary approaches~\cite{anderson2023social,balduf2025bootstrapping}: binary treatment analysis and continuous treatment analysis.

%%%%%%%%%%%%%%%%%%%%%%%%%%%%%%%%%%%%%%%%%%%%%%%%%%%%%%
\subsection{Binary Treatment: Recipients vs. Non-Recipients.}
\label{sec:binary_psm_did}
%%%%%%%%%%%%%%%%%%%%%%%%%%%%%%%%%%%%%%%%%%%%%%%%%%%%%%

\paragraph{Overview}
We begin by using the binary treatment (\eg receipt of a token reward) to compare reward recipients versus non-recipients, to measure the social impact of token rewards.\footnote{
We include wallet binding as a baseline binary treatment to assess how participation in the token economy affects user behavior (see \Cref{tab:causal-effect-summary}).}
Our analysis spans November 7, 2023, to April 27, 2025, examining five tokens (DEGEN, MOXIE, HIGHER, TN100X, USDC) across the three incentive mechanisms: \emph{Inter-\ac{fid} Tipping}, \emph{Third-party Algorithmic Rewards}, and Farcaster's \emph{Official Algorithmic Rewards}. 
This generates eight token-mechanism pairs (as shown in \Cref{sec:rq1_category_incentive_mechanisms}), whose effects we analyze on nine social activities: posting, and bidirectional interactions in replying, liking, re-posting, and following.

Our methodology aggregates data on a weekly basis, 
We implement \ac{psm} and \ac{did} analysis using a temporal alignment approach, where the week of each user's wallet binding or initial reward reception is designated as T+0,  with a four-week observation window before and after ---  dividing the timeline into pre-treatment (T-4 to T-1) and post-treatment (T+1 to T+4) windows.  
We can then compare activity levels before \vs after. 
This window size aligns with established practices in previous causal inference studies and provides sufficient time to observe short-term behavior changes while minimizing confounding temporal effects~\cite{balduf2025bootstrapping}. 
We next explain how we implement \ac{psm} and \ac{did}.

\paragraph{Propensity Score Matching (PSM)}
To compare the impact of receiving rewards, we employ \ac{psm} to construct comparable treatment and control groups by matching users with similar pre-treatment characteristics. 
We validate matching quality by examining standardized mean differences (SMD) of covariates (\ie observed pre-treatment characteristics that may influence treatment or outcome) between matched groups. 
SMD is calculated as the difference in means between treatment and control groups divided by the pooled standard deviation, with values below 0.1 indicating successful matching in relevant studies~\cite{austin2009balance}.\footnote{In the literature, higher SMD thresholds (0.25) are also proposed~\cite{austin2011introduction}.}
Our matching incorporates comprehensive covariates, covering social activity metrics (account age when receiving the token reward, weekly aggregated posting frequency, bidirectional following, replying, liking, and re-posting frequencies) and token reward features (weekly aggregated reception frequencies across all token-mechanism pairs).

Our primary specification includes all available covariates in the \ac{psm} to ensure optimal matching between control and treatment groups. 
Diagnostic assessments demonstrate successful matching outcomes, with most covariates achieving $SMD < 0.1$ (see \Cref{fig:psm_balance_tables_approach_1_appendix} in \Cref{app:causality_effect}) and matched pair sizes representing approximately 50\% of their corresponding populations across different token-mechanism pairings (see \Cref{tab:causal-effect-summary}).
Our \ac{att} and \ac{did} regression models below incorporate time fixed effects but exclude user fixed effects, as \ac{psm} already ensures group comparability. 

\paragraph{Difference-in-Differences (DID)}
Beyond the \ac{psm}, to strengthen causal identification and account for time-varying confounders, we implement a \ac{did} analysis with parallel trend validation. 
The parallel trends assumption, which is fundamental to \ac{did}, requires that treatment and control groups exhibit similar outcome trajectories during the pre-treatment period~\cite{lechner2011estimation}. 

Our validation approach divides the event timeline into pre-treatment (T-4 to T-1) and post-treatment (T+1 to T+4) windows, where the number following T denotes the number of weeks relative to the treatment week (T+0). 
For each pre-treatment window, we estimate differential coefficients between treatment and control groups. 
These coefficients measure the additional differences in outcome variables (such as posting frequencies) between treatment and control groups at each time window T. 

In a valid parallel trend test, pre-treatment coefficients should be statistically insignificant (p-value > 0.05)~\cite{Roth2022WhatsTI}.
We additionally adopt a 25\% tolerance criterion: the parallel trends assumption is considered to hold if statistically significant pre-treatment differences appear in no more than one quarter of the pre-intervention windows. 
This allowance accounts for behavioral adjustments in anticipation of reward eligibility --- such as increased engagement aimed at maximizing reward probability --- while preserving the integrity of the identification strategy. 
This approach aligns with context-aware thresholds discussed in prior methodological work~\cite{Roth2022WhatsTI}.

\paragraph{Covariate Adjustment}
Due to high inter-correlations among social behaviors~\cite{trunfio2021conceptualising,Milli2023ChoosingTR}, unadjusted analyses risk inflating treatment effects (\ac{att}) by confounding concurrent activities (\eg an increase in posting may naturally correlate with a rise in likes and replies). 
Our initial result exhibits this, showing broad positive impacts of token incentives on most social activities. 

To mitigate this, we further employ a covariate-adjusted method that accounts for both pre- and post-treatment social and token reward features as potential confounders. 
For example, when analyzing the impact of DEGEN tipping on posting behavior, we control for all other token rewards and social activities (both pre- and post-treatment) as confounders. 
This comprehensive approach reveals that the estimated effects of token rewards (\ie net effects \ac{att}) often become smaller---and sometimes reverse direction. 
These findings suggest a substantial correlation among social behaviors. 
Therefore, with the covariate-adjusted model accounting for additional social activity as confounding factors, net effects \ac{att} more accurately reflect the independent impact of token rewards on specific behaviors, rather than capturing spillover effects through correlated activities. 
This net effect approach provides deeper mechanistic insights, enabling us to identify which token rewards drive low-quality content farming versus high-quality engagement.

\subsection{Continuous Treatment: Reward Reception Frequency.}
\label{sec:continuous_ols}
%%%%%%%%%%%%%%%%%%%%%%%%%%%%%%%%%%%%%%%%%%%%%%%%%%%%%%
To quantify the intensity effect of each additional reward on social behaviors beyond binary treatment, we analyze how reward frequency affects behavioral changes through \ac{ols} regression:
\begin{equation}
\Delta Y_i = \alpha + \beta \cdot \log(RF_i) + \gamma \cdot C_i + \epsilon_i
\end{equation}
where $i$ indexes users, $\Delta Y_i$ represents the change in social behavior metrics (calculated as post-treatment minus pre-treatment social activity frequencies), $\log(RF_i)$ is the log-transformed frequency of a certain type of token reward received, and $C_i$ includes all available pre-treatment covariates (\eg posts, replies). 
For instance, when analyzing DEGEN tipping's impact, $\Delta Y_i$ measures the change in weekly posting frequency, while $RF_i$ counts the frequency of DEGEN tips received.\footnote{Due to the complexity of comparing the amount of USD value across different tokens, we only measure and compare the token reward frequencies. See more details in \Cref{sec:trading_metric}.} 
Using this methodology, we analyze users who have received at least one instance of the relevant token reward, focusing on 4-week windows before and after alignment points.

%%%%%%%%%%%%%%%%%%%%%%%%%%%%%%%%%%%%%%%%%%%%%%%
\subsection{Results}
\label{sec:rq3_results_findings}
%%%%%%%%%%%%%%%%%%%%%%%%%%%%%%%%%%%%%%%%%%%%%%%

\Cref{tab:causal-effect-summary} presents the results of the binary treatment causal analysis (detailed in \Cref{sec:binary_psm_did}). 
We use colored symbols to denote significant effects that pass the parallel trends test (including tolerance cases, detailed in \Cref{sec:binary_psm_did}): green $\textcolor{+green}{+}$ for positive effects and red $\textcolor{-red}{-}$ for negative effects, with the number of symbols indicating significance levels (\ie $\mathbf{+}: p<0.05, \mathbf{++}: p<0.01, \mathbf{+++}: p<0.001$). 
Non-significant effects are marked with \enquote{N}. 
In~\Cref{tab:causal-effect-summary}, among the 81 treatment-outcome pairs (9 social activities × 9 treatments), we denote 6 cases (7.41\%) passing with tolerance as \enquote{C}, 12 failing cases (14.81\%) as \enquote{F}, and leave complete passes unmarked (63 cases, 77.78\%).

%%%%%%%%%%% binary treatment att tables %%%%%%%%%%%%
\begin{table}[htbp]
\centering
\caption{Causal effect summary of binary treatments on social features.}
\label{tab:causal-effect-summary}
\footnotesize
\begin{tabular*}{\textwidth}{l@{\extracolsep{\fill}}*{9}{c}}
    \toprule
    & \textbf{Wallet} 
    & \multicolumn{5}{c}{\textbf{Inter-\ac{fid} Tipping}}
    & \multicolumn{3}{c}{\textbf{Algo. Rewards}} \\
    \cmidrule(lr){2-2} \cmidrule(lr){3-7} \cmidrule(lr){8-10}
    \textbf{Action} 
    & \textbf{Binding} 
    & \textbf{DEGEN} 
    & \textbf{HIGHER} 
    & \textbf{MOXIE} 
    & \textbf{TN100X} 
    & \textbf{USDC} 
    & \textbf{DEGEN} 
    & \textbf{MOXIE} 
    & \textbf{USDC} \\
    \midrule
    post                                & $F,+++$                       & $\textcolor{+green}{+}$      & $C,\textcolor{+green}{+++}$    & $\textcolor{+green}{+}$     & $N$                         & $N$                       & $C,\textcolor{+green}{+++}$   & $F,+++$                       & $N$   \\
    reply\_out                          & $N$                           & $\textcolor{+green}{+}$      & $\textcolor{-red}{-}$          & $\textcolor{+green}{+}$     & $N$                         & $N$                       & $N$                           & $N$                           & $N$  \\
    reply\_in                           & $N$                           & $N$                          & $N$                            & $N$                         & $\textcolor{+green}{+++}$   & $N$                       & $\textcolor{+green}{+++}$     & $\textcolor{+green}{+++}$     & $N$ \\
    like\_out                           & $F,N$                         & $N$                          & $N$                            & $\textcolor{-red}{-}$       & $N$                         & $\textcolor{-red}{--}$    & $\textcolor{+green}{+++}$     & $\textcolor{-red}{---}$       & $\textcolor{+green}{+++}$  \\
    like\_in                            & $N$                           & $\textcolor{+green}{+++}$    & $N$                            & $N$                         & $C,\textcolor{-red}{--}$    & $N$                       & $N$                           & $F,N$                         & $\textcolor{+green}{+}$ \\
    repost\_out                        & $C,\textcolor{+green}{+++}$   & $\textcolor{-red}{-}$        & $N$                            & $N$                         & $N$                         & $\textcolor{-red}{--}$    & $N$                           & $N$                           & $\textcolor{-red}{--}$  \\
    repost\_in                         & $N$                           & $N$                          & $N$                            & $N$                         & $N$                         & $N$                       & $\textcolor{-red}{--}$        & $\textcolor{-red}{---}$       & $\textcolor{-red}{---}$ \\
    follow\_out       & $F,+++$                       & $F,---$                      & $F,---$                        & $C,\textcolor{-red}{-}$     & $F,---$                     & $\textcolor{-red}{---}$   & $F,+++$                       & $F,+++$                       & $F,---$   \\
    follow\_in                          & $\textcolor{+green}{+++}$     & $C,\textcolor{+green}{+++}$  & $N$                            & $\textcolor{-red}{--}$      & $N$                         & $\textcolor{+green}{+++}$ & $\textcolor{+green}{+++}$     & $F,N$                         & $\textcolor{+green}{+++}$ \\
    \midrule
    populaion\_size & 574829 & 40836 & 15849 & 5252 & 3459 & 15872 & 47748 & 21505 & 28181 \\
    matched\_pairs & 48799 & 16817 & 7795 & 2643 & 2257 & 7119 & 15100 & 12260 & 7217 \\
    \bottomrule
\end{tabular*}

\vspace{2mm}
\begin{flushleft}
\small
\textit{
\textbf{Symbols:} \\
$\mathbf{+}$ \& $\mathbf{-}$: Positive \& negative causal effects (measured by ATT significance levels ($p<0.05, p<0.01, p<0.001$)); \\
$\textcolor{+green}{+}$ \& $\textcolor{-red}{-}$: Positive \& negative causal effects with parallel trend pre-test passed (including deviation tolerance);\\
$\mathbf{N}$: Average Treatment Effect on the Treated (ATT) not significant in post-treatment period;\\
$\mathbf{C}$: Deviation exists in parallel trend pre-test (only 1 week deviation);\\
$\mathbf{F}$: Parallel pre-test fails (more than 1 week deviation).
}
\end{flushleft}
\end{table}
%%%%%%%%%%%%%%%%%%%%%%%%%%%%%%%%%%%%%%%%%%

\Cref{tab:intensity_effect_olsReg_summary_simplified} summarize the results of the continuous treatment analysis (detailed in \Cref{sec:continuous_ols}), where the significant regression coefficients are highlighted in bold. 
Note, both the binary and continuous effects must be in the same direction (either both positive or both negative). 
Therefore, only the regression coefficients satisfying these criteria are further highlighted with color (green for positive effects, red for negative effects).
A more detailed result table, including $R^2$ and standard errors (SE), is provided in \Cref{tab:intensity_effect_olsReg_summary_complete} in \Cref{app:causality_effect}.

%%%%%%%%%%%%% regression reward intensity %%%%%%%%%%%%
\begin{table}[htbp]
\centering
\caption{
Regression summary of continuous treatment intensity with social activities.
}
\label{tab:intensity_effect_olsReg_summary_simplified}
\footnotesize
\setlength{\tabcolsep}{1.6pt}
\renewcommand{\arraystretch}{0.99}
\begin{tabular}{l*{8}{c}}
    \toprule
    & \multicolumn{5}{c}{\textbf{Inter-\ac{fid} Tipping}} 
    & \multicolumn{3}{c}{\textbf{Algo. Rewards}} \\
    \cmidrule(lr){2-6} \cmidrule(lr){7-9}
    \textbf{Action} 
    & \textbf{DEGEN} 
    & \textbf{TN100X} 
    & \textbf{HIGHER} 
    & \textbf{MOXIE} 
    & \textbf{USDC} 
    & \textbf{DEGEN} 
    & \textbf{MOXIE} 
    & \textbf{USDC} \\
    \midrule
post 
    & \textcolor{+green}{\textbf{1.5703***}} & 5.0001 & \textcolor{+green}{\textbf{1.9701***}} & -0.0517 & -0.0805 & \textcolor{+green}{\textbf{12.7407***}} & \textbf{15.2254***} & \textbf{8.2309***}\\
reply\_out   
    & -4.4867 & 8.6610 & -3.8922 & 10.6198 & -0.3703 & 9.4023 & \textbf{59.2029***} & \textbf{-19.1090***} \\
reply\_in     
    & 1.7207 & -35.1615 & -2.4438 & 10.4600 & \textbf{-10.2936**} & -4.8940 & \textcolor{+green}{\textbf{57.1112***}} & 5.1841 \\
like\_out      
    & -1.4999 & 6.4651 & 1.4243 & -4.1602 & \textbf{7.5337**} & -9.7807 & \textbf{22.8327***} & \textcolor{+green}{\textbf{32.0449***}} \\
like\_in       
    & -1.2148 & 25.2958 & \textbf{17.9290***} & \textbf{-27.5476***} & 0.5102 & \textbf{-77.0576***} & \textbf{7.7791**} & \textcolor{+green}{\textbf{73.9531***}} \\
repost\_out     
    & 0.4519 & 7.1819 & \textbf{2.4716*} & -0.9295 & -0.0732 & -5.6276 & -0.1039 & \textcolor{-red}{\textbf{-9.1389***}} \\
repost\_in      
    & 1.7003 & -5.8862 & \textbf{-5.2017***} & \textbf{5.6276**} & \textbf{8.7852*} & 5.0918 & \textcolor{-red}{\textbf{-6.2899***}} & \textcolor{-red}{\textbf{-30.2670***}} \\
follow\_out     
    & 0.2749 & -0.9579 & 0.0211 & \textbf{14.5753***} & 0.6082 & 0.9927 & \textbf{20.3066***} & \textbf{-54.1936***} \\
follow\_in    
    & -12.4111 & -57.8221 & \textbf{-19.9274***} & \textbf{9.9655***} & 15.6006 & \textcolor{+green}{\textbf{196.8301***}} & \textbf{6.3755**} & \textcolor{+green}{\textbf{242.8866***}} \\
    \midrule
    population\_size & 40836 & 3459 & 15849 & 5252  & 15872 & 47748 & 21505 & 28181 \\
    sample\_size & 38532 & 3438 & 13854 & 5041 & 13879 & 47748 & 21482 & 27484 \\
    \bottomrule
\end{tabular}
\par\vspace{2mm}
\noindent\small
\textit{\textbf{Symbols:} $^{***}p<0.001$, $^{**}p<0.05$, $^{*}p<0.01$. Significant coefficients with corresponding significant causal effects are highlighted both in bold and color.}
\end{table}

Our causal analyses reveal several key patterns in how different token incentive mechanisms shape user behavior on Farcaster. 
These findings span three main dimensions: 
\begin{enumerate*}
    \item the quantity-quality trade-off in content engagement,
    \item the dynamics of social network growth, and
    \item the intensity effects of repeated token rewards.
\end{enumerate*}
Through these analyses, we uncover both intended and inadvertent consequences of token-based incentive mechanisms.

\paragraph{Trade-off between Engagement Quantity and Quality.}
Our binary treatment analysis employing \ac{psm} and \ac{did} reveals that the initial reception of token incentives generally increases content engagement quantity (posts and replies) while showing insufficient effectiveness in improving quality (likes and re-posts).

%%%%%%%%%%%%%%%%%% DID visualization %%%%%%%%%%%%%%%%%%%%%%%
\begin{figure}[htbp]
    \centering
    \begin{subfigure}{0.45\linewidth}
        \centering
        \includegraphics[width=\linewidth]{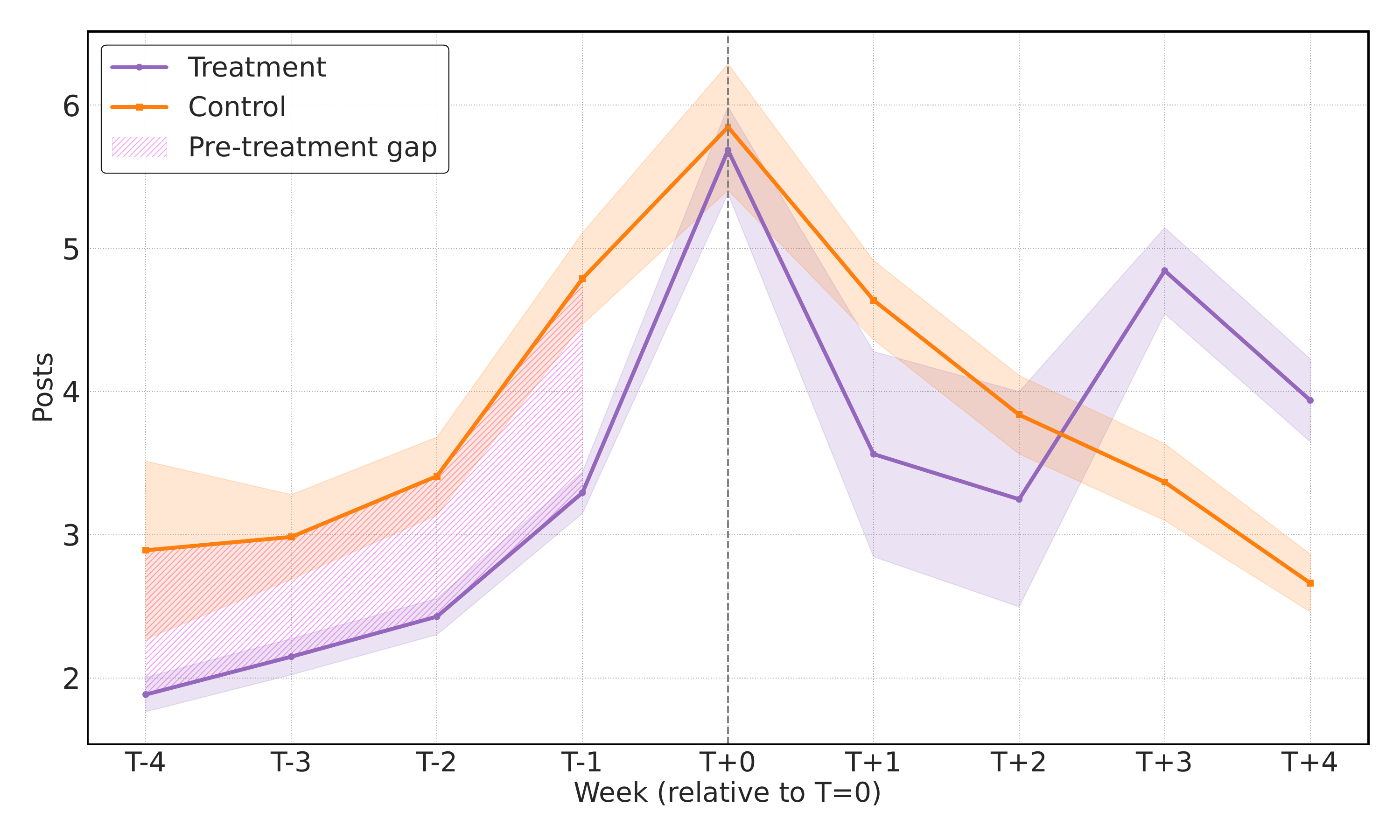}
        \caption{DEGEN (tipping)'s positive impact ($\textcolor{+green}{+}$) on weekly post frequencies.}
        \label{fig:did_parallel_trend_degen_A_1_post}
    \end{subfigure}
    \hfill
    \begin{subfigure}{0.45\linewidth}
        \centering
        \includegraphics[width=\linewidth]{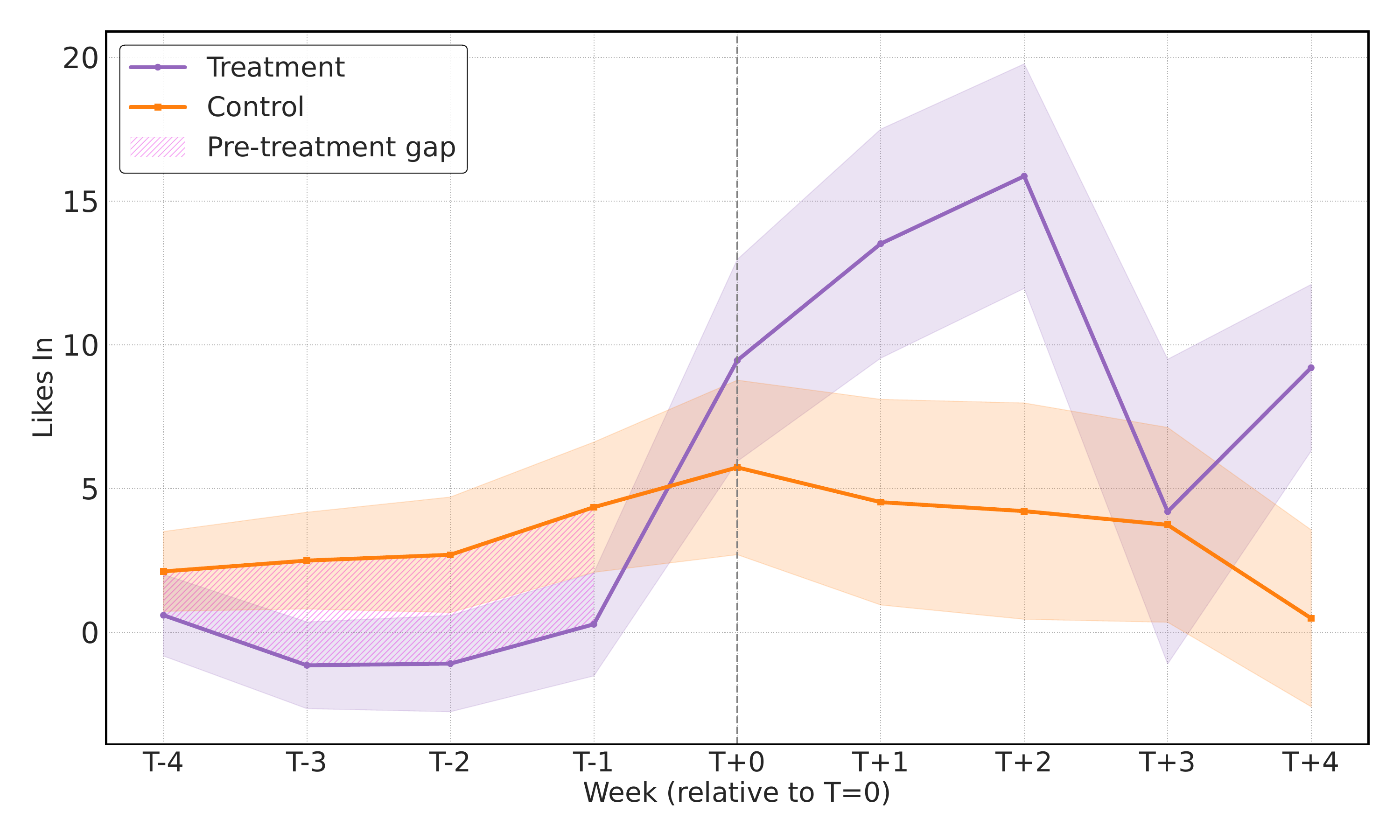}
        \caption{DEGEN (tipping)'s positive impact ($\textcolor{+green}{+++}$) on weekly inbound like frequencies.}
        \label{fig:did_parallel_trend_degen_A_1_like_in}
    \end{subfigure}
    \hfill
    \begin{subfigure}{0.45\linewidth}
        \centering
        \includegraphics[width=\linewidth]{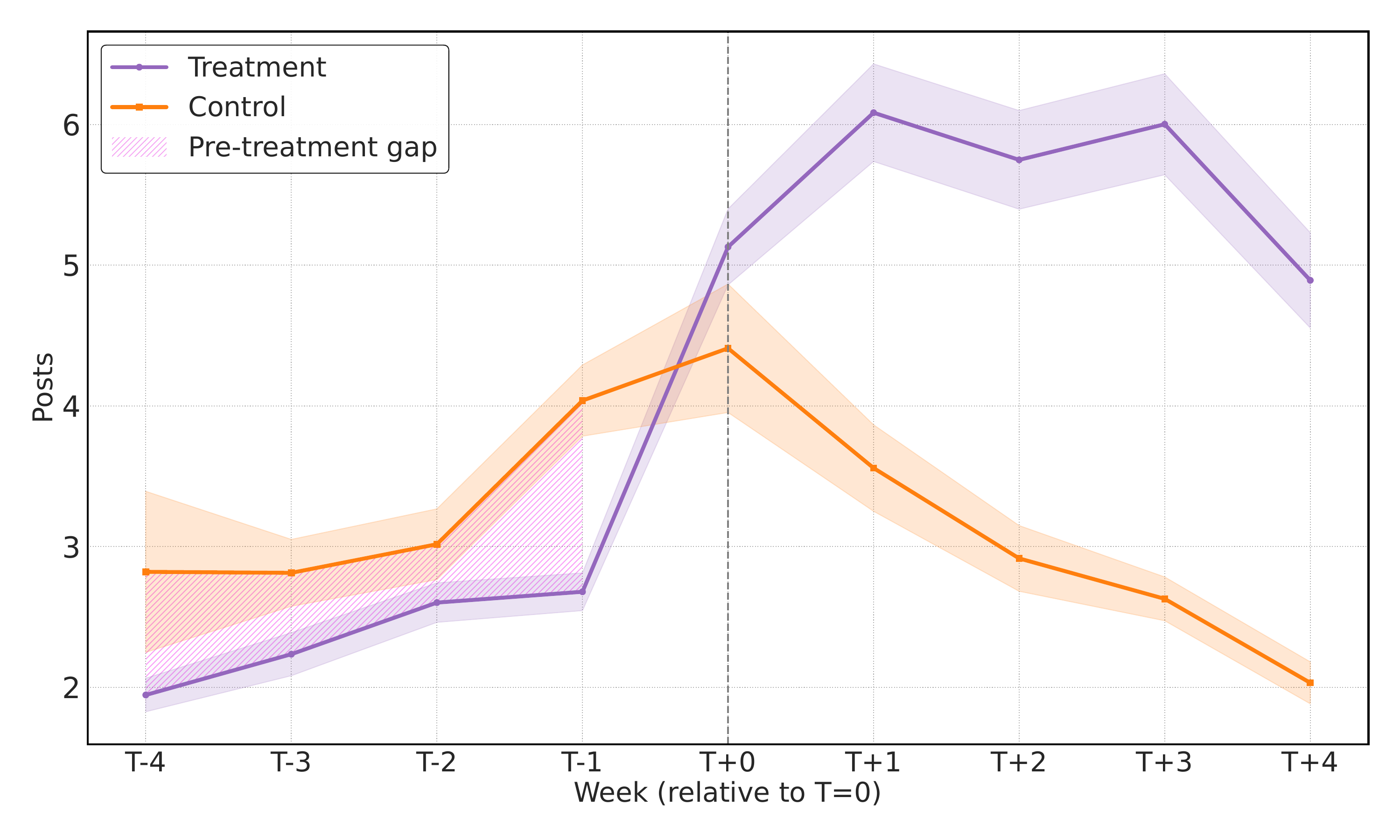}
        \caption{DEGEN (algorithmic reward)'s positive impact ($\textcolor{+green}{+++}$) on weekly inbound post frequencies.}
        \label{fig:did_parallel_trend_degen_B_1_post}
    \end{subfigure}
    \hfill
    \begin{subfigure}{0.45\linewidth}
        \centering
        \includegraphics[width=\linewidth]{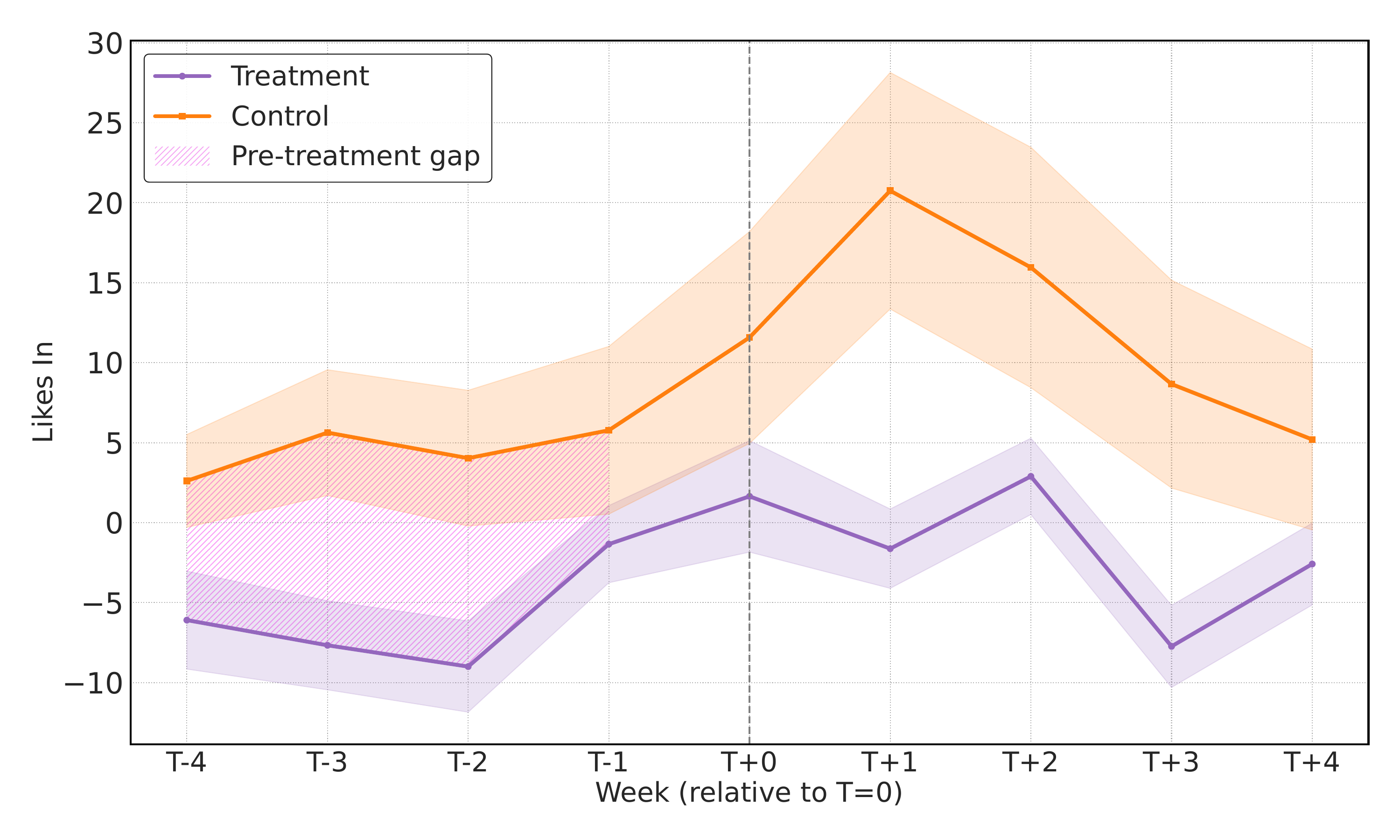}
        \caption{USDC (algorithmic reward)'s positive impact ($\textcolor{+green}{+}$) on weekly inbound like frequencies.}
        \label{fig:did_parallel_trend_usdc_C_1_like_in}
    \end{subfigure}
    \caption{Difference-in-Differences (DID) visualizations.}
    \label{fig:did_parallel_trends_plots_DegenAB_12_post}
    \Description{Difference-in-Differences (DID) visualizations (DEGEN tipping, along with DEGEN and USDC algorithmic rewarding as illustrative examples) on users' weekly social behaviors. 
    Parallel trend assumptions are validated for all panels. 
    All plots are covariate-adjusted, incorporating covariate adjustments for all available social and token features while excluding the outcome variable and treatment indicators.}
\end{figure}

To illustrate these findings, we present \ac{did} visualizations (\Cref{fig:did_parallel_trends_plots_DegenAB_12_post}) for cases that both pass the parallel trends test and show significant positive \ac{att} results (detailed in \Cref{sec:binary_psm_did}). 
We focus on three representative token-mechanism pairs as binary treatments: DEGEN tipping, DEGEN algorithmic reward, and USDC algorithmic reward. 
This selection enables us to compare both the effectiveness of different mechanisms (tipping vs. algorithmic) for the same token (DEGEN) and different token types (volatile DEGEN vs. stablecoin USDC) under the same mechanism (algorithmic). 
Complete results are available in \Cref{tab:causal-effect-summary}. 

\Cref{fig:did_parallel_trend_degen_A_1_post} illustrates that DEGEN tipping positive, albeit delayed, effects (\textcolor{+green}{+}, $p<0.05$) on weekly post frequencies after the tipping reception (T+0), with the increase beginning around T+2. The parallel trend assumption is satisfied (see the shaded pre-treatment gap between the control group line and the treatment group line). 
Similar positive effects are observed for other tipping tokens, with HIGHER and MOXIE showing significant positive impacts (\textcolor{+green}{+}, $p<0.05$), while TN100x and USDC show no significant effects. 
\Cref{fig:did_parallel_trend_degen_B_1_post} demonstrates that DEGEN as an algorithmic reward shows even stronger positive effects on weekly post frequencies (\textcolor{+green}{+++}, $p<0.001$). 
The visualizations (\Cref{fig:did_parallel_trend_degen_A_1_post,fig:did_parallel_trend_degen_B_1_post}) clearly show a more pronounced treatment effect for DEGEN algorithmic rewards compared to DEGEN tipping. 
While Moxie's algorithmic reward also shows positive \ac{att} (\textcolor{+green}{+++}, $p<0.001$), it fails the parallel trends assumption test (see \Cref{tab:causal-effect-summary}). 
In contrast, Farcaster's official USDC algorithmic rewards show no significant effects on posts and replies.
According to these observations, \emph{Third-party Algorithmic Rewards} using volatile tokens as the medium (DEGEN and MOXIE) actually demonstrate stronger positive effects on content engagement quantity compared to both \emph{Inter-\ac{fid} Tipping} mechanisms (all five tokens) and the \emph{Official Algorithmic Rewards} distributing stablecoins (USDC).

Compared to quantity metrics, token incentives show very limited improvement in content quality, measured by received likes and re-posts. 
Only DEGEN tipping (\Cref{fig:did_parallel_trend_degen_A_1_like_in} (\textcolor{+green}{+++}, $p<0.001$)) and USDC algorithmic rewards (\Cref{fig:did_parallel_trend_usdc_C_1_like_in} (\textcolor{+green}{+}, $p<0.05$)) significantly increase recipients' received likes.
This may be attributed to DEGEN and USDC's network effects as the most-traded reward token on Farcaster (discussed in \Cref{sec:rq1_prevalent_token_detection}). 
However, these effects do not generalize to other tipping or third-party algorithmic tokens. 
Moreover, no tipping tokens show significant effects on re-post gains, while algorithmic reward tokens even show negative effects (see \Cref{tab:causal-effect-summary}). 
This suggests token incentives not only fail to promote high-quality and share-worthy content, but may even have counter-productive effects, potentially echoing previous literature's findings on financial rewards' crowd-out effects on quality content due to strategic farming behaviors prioritizing quantity over quality~\cite{garnefeld2012explicit,liu2015can,qiao2017incentive,zhang2025time}.\footnote{These malicious behaviors have been reported by a Farcaster co-founder (\url{https://farcaster.xyz/v/0x0e31071c}) and both DEGEN and MOXIE developers (\url{https://x.com/degentokenbase/status/1802985205021466790}, \url{https://farcaster.xyz/dwr.eth/0x8bfde087})}

\paragraph{Effects on Follower Growth}
Beyond content interactions, we next examine follower growth. 
Wallet binding, the pre-requisite token economy participation behavior, serving as a baseline binary treatment, shows significant positive effects on follower growth (\textcolor{+green}{+++}, $p<0.001$), indicating users participating in Farcaster's token economy gain more followers than non-participants. 
This effect extends to both DEGEN and USDC across tipping and algorithmic mechanisms, reinforcing their unique network effect and social recognition status. 
However, all token rewards, as well as wallet binding, show neutral or negative effects on outbound following (follow-out) behavior, suggesting token economy participants are more likely to focus on self-promotion than expanding social connections, potentially exacerbating echo chamber effects.

\paragraph{Token Incentive Intensity Effects}
Our continuous treatment analysis employing \ac{ols} regression provides insights into the cumulative effects of token rewards.
Comparing causal analysis results (\Cref{tab:causal-effect-summary}) with significant and directionally consistent \ac{ols} regression coefficients, we highlight significant intensity effects in \Cref{tab:intensity_effect_olsReg_summary_simplified} --- examining whether higher reward frequency correlates with stronger social behavior impacts. 
The tipping mechanisms show minimal intensity effects, with only DEGEN and HIGHER showing slight positive effects on posting (coefficients: DEGEN 1.57, HIGHER 1.97, indicating less than 2 additional weekly posts per reward).

In contrast, algorithmic rewards demonstrate substantial intensity effects across most social behaviors. 
DEGEN and USDC algorithmic rewards show particularly strong follower growth effects ($\approx$ 197 and 243 additional weekly followers per reward respectively, echoing the findings in the previous paragraph), while MOXIE shows no such effect. 
DEGEN and MOXIE algorithmic rewards show significant positive intensity effects on content quantity ($\approx$ 13 additional weekly posts per DEGEN reward, $\approx$ 57 additional replies per MOXIE reward) but no effects on quality metrics.

USDC algorithmic rewards demonstrate a more nuanced impact pattern: while showing strong positive intensity effects on like-based interactions ($\approx +32$ likes given and $\approx +74$ received per reward), they simultaneously exhibit significant negative effects on content sharing ($\approx -9$ re-posts given and $\approx -30$ received). 
Combined with the neutral effects on posting frequency, this pattern suggests that Farcaster's official USDC algorithmic rewards may shift user behavior toward producing content that attracts quick, surface-level engagement (likes) rather than content worthy of redistribution (re-posts). 

This behavioral shift aligns with previous research on monetary incentives in social platforms~\cite{qiao2017incentive,yan2017financial,khern2018extrinsic,huang2023effect,zhang2025time}, where extrinsic rewards can potentially alter content creation motivations from intrinsic quality pursuit to reward optimization. 
The divergence between like-based and repost-based engagement particularly highlights how token incentives might inadvertently promote content optimized for immediate reaction rather than lasting value that users want to preserve and share with their networks.
%%%%%%%%%%%%%%%%%%%%%%%%%%%%%%%%%%%%%%%%
\section{Related Work}
\label{sec:related_work}

%%%%%%%%%%%%%%%%%%%%%%%%%%%%%%%%%%%%%%%%
The study of how incentives influence user behaviors and network dynamics on social platforms is a well-established area, situated at the intersection of behavioral economics and online social networks~\cite{micheli2021effect,singh2011mechanism,lindstrom2021computational}.
Prior research frameworks have examined how social and financial incentives shape user participation and network evolution~\cite{acuff2018applying,jackson2011overview}, utilizing controlled field experiments~\cite{sun2018designing}, laboratory simulations~\cite{kapoor2023does,globig2023changing}, observational data analyses~\cite{lindstrom2021computational,ba2022role}, and quasi-experimental approaches~\cite{yu2022paying,burtch2015social}. 
Measurement metrics include engagement indicators (likes, re-posts, replies)~\cite{Furini2024XAA,Donkers2025UnderstandingOP,trunfio2021conceptualising}, content quality (accuracy, complexity, informativeness)~\cite{ceylan2023sharing,paridar2023more}, and network-level effects such as clustering and propagation~\cite{centola2010spread,rand2011dynamic,zheng2024implications}.
Within this context, our study leverages observational data from Farcaster, employing \ac{psm} and \ac{did} as quasi-experimental approaches to examine the influence of token incentives on social engagement indicators, using likes and re-posts as proxies for content quality.

Studies of traditional centralized social platforms demonstrate that
monetary incentives reliably increase the quantity of social engagement behaviors, particularly under performance-contingent schemes~\cite{yu2018impact,yu2022paying,wang2023recognition,kapoor2023does,globig2023changing,ceylan2023sharing}, though effects on content quality and novelty are mixed~\cite{qiao2017incentive,yan2017financial,khern2018extrinsic,huang2023effect,zhang2025time}.
Moderating factors such as demographics, user characteristics, social status, and platform context critically shape incentive effectiveness~\cite{anderson2023social,gruzd2024lure}, while combined monetary and social incentives often yield superior outcomes~\cite{scekic2013incentives,micheli2021effect, paridar2023more}. 
Temporal analyses reveal strong short-term engagement boosts but potential long-term habituation effects (\ie frequent users develop reduced sensitivity to social rewards over time, while occasional users remain highly responsive~\cite{anderson2023social}), crowding-out effects (\ie monetary incentives reduce
intrinsic motivation, negatively impacting content quality)~\cite{garnefeld2012explicit,liu2015can,qiao2017incentive,zhang2025time}, and inequality amplification~\cite{dimaggio2012network}. 
These findings underscore the complexity of incentive design and user heterogeneity in digital environments.

In blockchain-based decentralized social platforms, Steemit~\cite{steemit_doc} remains the most studied~\cite{ba2022role,li2019incentivized,delkhosh2023impact,steemit2024richgetricher,michienzi2024wealth}.
Steemit's proprietary blockchain and platform-mandated token mechanism enabled early advances in decentralized incentive design, eliminating transaction fees and facilitating high-throughput reward distribution.
However, these design choices unintentionally introduced critical vulnerabilities, including susceptibility to farming and collusion~\cite{ba2022role}, bot misuses~\cite{delkhosh2023impact}, centralization of rewards, and exacerbation of economic stratification~\cite{li2019incentivized,steemit2024richgetricher,michienzi2024wealth}.
Research by Li et al.~\cite{li2019incentivized} and Ba et al.~\cite{ba2022role} indicates that successful users adapt their content strategies to maximize rewards, often focusing on content promotion rather than creation. 
This finding raises questions about whether financial incentives optimize for platform goals or user gaming.
Ba et al.~\cite{ba2022role} further reveal strong correlations between cryptocurrency prices and user activity levels on Steemit: when token values increase, posting activity and user engagement spike correspondingly.

These prior works have focused on examining single mechanisms. However, studies suggest that the most effective incentive systems should combine multiple types of rewards rather than relying on single mechanisms~\cite{scekic2013incentives,micheli2021effect, paridar2023more}. 
Thus, our work differs from the above in that we move beyond single-incentive vulnerabilities. Instead, our research offers the first empirical analysis of Farcaster’s \emph{pluralistic} incentive ecosystem---integrating multiple tokens and diverse reward mechanisms through modular wallet binding and third-party reward projects~\cite{farcaster_official_doc,farcaster_wallet_connection,farcaster_miniapp_doc}.
Notably, we find that despite individual mechanisms retaining some prior identified shortcomings, their coexistence and complementarity show the potential to mitigate platform-wide risks.
%%%%%%%%%%%%%%%%%%%%%%%%%%%%%%%%%%%%%%%%
\section{Discussion and Conclusion}
\label{sec:discussion_conclusion}
%%%%%%%%%%%%%%%%%%%%%%%%%%%%%%%%%%%%%%%%%%%

%%%%%%%%%%%%%%%%%
\subsection{Summary of Main Findings}
\label{sec:findings}
%%%%%%%%%%%%%%%%%
We have presented the first large-scale empirical analysis of Farcaster's pluralistic token incentive ecosystem, examining how diverse reward mechanisms shape user behavior and social network structure.
Through the analysis of 574,829 wallet-linked users (64.25\% of the user base), we have revealed several critical insights about token-based incentives in decentralized social networks.
Our analysis demonstrates that while token incentives effectively drive platform growth and user participation, their differences in eligibility criteria, reward distribution structure and token types significantly impact socioeconomic outcomes.

While \emph{user-to-user tipping} represents the most flexible and common incentive mechanism, it is predominantly unidirectional (with less than 10\% of users acting as both tip receivers and senders) (see~\Cref{sec:rq2_incentive_inclusion}). 
Nevertheless, inter-community tipping occurs more frequently between non-following pairs, at a rate 1.3--2 times higher than following pairs (see~\Cref{sec:rq2_echo_chamber}). 
This pattern indicates that, in contrast to reinforcing echo chambers, tipping serves as a significant mechanism for cross-community value exchange beyond existing communities and following users.

Examining \emph{algorithmic reward} mechanisms, we observe notable differences in inclusivity (\Cref{sec:rq2_incentive_inclusion}).
DEGEN, which relies on user-driven nominations, reaches up to 70\% new participant rates.
In contrast, MOXIE relies on an open-source behavioral scoring algorithm and includes only 7.6\% of new participants.
This contrast suggests that transparent scoring systems are more susceptible to exploitation, reducing entry opportunities for new users.

However, wealth concentration persists across most mechanisms (Gini coefficients: 0.82-0.94) (see \Cref{sec:rq2_rich_get_richer}). 
Compared to user-to-user tipping, algorithmic rewards demonstrate greater inequality, primarily due to: 
\begin{enumerate*}
    \item the token staking model (\eg MOXIE and DEGEN), which amplifies incumbent advantages; and 
    \item increased vulnerability to strategic farming and gaming (\Cref{sec:rq1_category_incentive_mechanisms}).
\end{enumerate*}
Notably, MOXIE's innovative redistribution mechanism alleviates initial concentration effects (Gini: 0.72), suggesting that well-designed secondary distributions can help address wealth inequality.

Moreover, our causal analyses (see \Cref{sec:rq3_token_social_causality}) uncover fundamental trade-offs in promoting social activity via token incentives: while most rewards effectively boost content creation quantity (posts and replies), they often fail to enhance---and sometimes undermine---content quality measured by likes and re-posts.
Furthermore, we identify potentially problematic patterns in repeated reward effects. 
Users receiving multiple algorithmic rewards show asymmetric network growth, seemingly prioritizing follower acquisition over outbound following. 
They also demonstrate strategic optimization behaviors, favoring immediate reactions over share-worthy content creation. 
These findings suggest that while token incentives can drive engagement, their current implementations inadvertently encourage superficial participation over meaningful social interaction.

%%%%%%%%%%%%%%%%%
\subsection{Design Implications and Mitigation Strategies}
\label{sec:implication_strategy}
%%%%%%%%%%%%%%%%%
Our results yield three implications for sustainable incentive design. 

First, coordination across pluralistic rewards is limited. 
Token-related programs on Farcaster are launched independently, target overlapping social metrics (\eg posts, replies, and likes), and operate without shared objectives or interoperable scoring. 
This leads to inefficient spending and conflicting reputation signals. 
This suggests that there could be value in introducing a composable reputation layer, which could provide a common substrate by aggregating on-chain behavior, off-chain interactions, and temporal consistency, ideally in a privacy-preserving fashion. 
By standardizing interfaces, programs could align on value metrics (\eg novelty, informativeness, cross-community engagement), avoid redundant payouts, and specialize on complementary goals.

Second, our results indicate that Farcaster's existing incentives inadequately foster network exploration and content quality. 
This is evidenced by a tension in asymmetric network growth (see~\Cref{sec:rq3_results_findings}): all token rewards, as well as wallet binding, show neutral or negative effects on outbound following (follow-out) behavior, suggesting token economy participants are more likely to focus on self-promotion than expanding social connections, potentially exacerbating echo chamber effects. 
Simultaneously, a quantity-quality trade-off exists (see~\Cref{sec:rq3_results_findings}): while most tokens boost content creation (measured by posts and replies), they often fail to enhance content quality (measured by likes and re-posts). 
We argue that dedicating budget to exploration-aware payouts (\eg cross-community edges, first-time ties) and to quality-weighted rewards (\eg delayed distributions conditioned on downstream popularity or more advanced quality metrics) is crucial to shift behavior toward new content discovery and quality social engagement.

Third, it is clear that robustness to gaming and manipulation is essential. 
The current static, transparent, count-based scoring invites farming, collusion and wealth concentration.
Potential mitigations include periodically rotated and adversarially tested scoring functions; a mix of limited-opacity components with auditable elements; anti-collusion safeguards (staked attestations, adaptive rate limits); and redistribution mechanisms (\eg MOXIE-style follower sharing) to counter excessive wealth accumulation. 
Together, these measures could harden the system while preserving incentives for genuine, socially valuable activity.

%%%%%%%%%%%%%%%%%
\subsection{Limitations}
\label{sec:limitations_generalizability}
%%%%%%%%%%%%%%%%%
This work carries several limitations that shape the scope of inference, and inform how our results may generalize beyond the studied setting. 

First, our measurement of content quality relies on engagement-based proxies such as likes and re-posts. 
Whilst widely used in prior work~\cite{Furini2024XAA,Donkers2025UnderstandingOP,trunfio2021conceptualising}, these signals imperfectly capture intrinsic quality. 
They are affected by users’ network position, temporal visibility, and short-lived trends. 
As a result, our estimates may understate improvements in deeper dimensions --- such as informativeness, novelty, or persuasiveness --- that are less immediately reflected in reaction metadata (\eg likes and re-posts). 

Second, our \ac{did} design uses a four-week pre/post window around alignment events. Content with longer discovery cycles or diffusion paths may accrue recognition beyond this horizon. 
Consequently, our identification is more sensitive to short-term responses and may bias against detecting lagged, higher-fidelity effects. 
Methodological extensions with longer windows and complementary designs could better separate immediate reactivity from durable changes.

Third, Farcaster remains smaller and more crypto-centric than mainstream platforms (\eg Twitter/X, Facebook) and some larger \acp{dosn} (\eg Mastodon, Bluesky). 
This composition likely raises activation thresholds and reduces sensitivity to marginal token incentives, which can attenuate absolute effect sizes. 
Nonetheless, the directional insights we uncover --- wealth concentration risks, cross-community value flows, and the tension between content quantity and quality --- are expected to carry over to other platforms that deploy analogous pluralistic incentive schemes.

Regarding external validity, our token selection follows a multi-criteria, robust filtering pipeline (see~\Cref{sec:rq1_prevalent_token_detection}) that emphasizes sustained, community-relevant activity. 
This approach improves the representativeness of mechanisms most likely to shape ecosystem-level behavior, compared to analyzing arbitrary or transient tokens. 
Generalization to substantially less crypto-native populations or to ecosystems with different incentive mechanisms should be interpreted with the caveat that absolute magnitudes may differ, even if the underlying trade-offs tend to persist.

%%%%%%%%%%%%%%%%%
\subsection{Future Directions}
\label{sec:future_direction}
%%%%%%%%%%%%%%%%%

Building on our empirical findings, several avenues emerge for advancing research on token-based incentives in decentralized social networks. 

First, future studies could develop and evaluate composable reputation systems that integrate multi-token signals, enabling coordinated reward allocation across mechanisms while preserving user privacy. 
Such systems would address the observed limitations in incentive alignment, potentially through machine learning models that optimize for long-term platform health metrics like sustained engagement and content diversity.
Second, research could investigate enhanced mechanism designs that explicitly incentivize network exploration and high-quality content creation. 
This might involve experimental deployments of rewards weighted by cross-community interactions or delayed propagation signals, testing their efficacy in mitigating asymmetric growth and superficial participation patterns identified in our analysis.
Third, advancing robustness against gaming requires systematic evaluation of adaptive scoring functions, anti-collusion protocols, and redistribution schemes. 
Longitudinal studies could assess these interventions' impact on wealth concentration and inclusivity, particularly in platforms with varying scales and user demographics.
Finally, comparative analyses across diverse \acp{dosn} would elucidate how contextual factors---such as user sovereignty levels, user characteristics, and token volatility---influence incentive outcomes, informing generalizable design principles for sustainable token economies. 

%%%%%%%%%%%%%%
\subsection{Concluding Remarks}
%%%%%%%%%%%%%%
To conclude, our analysis reveals that, despite the persistent limitations of individual tokens or mechanisms, their combined presence and mutual reinforcement can effectively mitigate platform-wide vulnerabilities, such as high barriers to entry, excessive wealth concentration, the formation of echo chambers, and strategic farming behaviors that arise from a single tokenomic model. 
However, we also find that the mere coexistence of pluralistic incentives, without organic coordination and integration, fails to achieve the desired outcomes in terms of sustainable network growth and content quality.
Our findings advance understanding of token-based incentive design and provide future research directions and practical guidance for implementing reward mechanisms in social platforms.
More broadly, our results suggest that while token incentives offer promising tools for decentralized platforms, their effectiveness depends on careful mechanism design that considers both immediate behavioral impacts and longer-term social dynamics.

\begin{acks}
This work was supported in part by the Guangzhou Science and Technology Bureau (2024A03J0684), Guangdong provincial project 2023QN10X048, the Guangzhou Municipal Key Laboratory on Future Networked Systems (2024A03J0623), the Guangdong Provincial Key Lab of Integrated Communication, Sensing and Computation for Ubiquitous Internet of Things (No.2023B1212010007), the Guangzhou Municipal Science and Technology Project (2023A03J0011), Guangdong provincial project (2023ZT10X009), and the 111 Center (No. D25008).
\end{acks}


\clearpage
% ========= references =========
% \bibliographystyle{ACM-Reference-Format}
% \bibliography{reference}

% \bibliographystyle{acm}
% \bibliography{reference}

\bibliographystyle{ACM-Reference-Format}
\bibliography{reference} % name of .bib file

@String{Computing = "Computing" }

@String{Computer = "{IEEE} Computer" }

@String{Springer = "Springer-Verlag" }

@inproceedings{mastodon_challenges,
author = {Raman, Aravindh and Joglekar, Sagar and Cristofaro, Emiliano De and Sastry, Nishanth and Tyson, Gareth},
title = {Challenges in the Decentralised Web: The Mastodon Case},
year = {2019},
isbn = {9781450369480},
publisher = {Association for Computing Machinery},
address = {New York, NY, USA},
url = {https://doi.org/10.1145/3355369.3355572},
doi = {10.1145/3355369.3355572},
booktitle = {Proceedings of the Internet Measurement Conference},
pages = {217–229},
numpages = {13},
location = {Amsterdam, Netherlands},
series = {IMC '19}
}

@inproceedings{memo_set_in_stone,
author = {Zuo, Wenrui and Raman, Aravindh and Mondrag\'{o}n, Raul J and Tyson, Gareth},
title = {Set in Stone: Analysis of an Immutable Web3 Social Media Platform},
year = {2023},
isbn = {9781450394161},
publisher = {Association for Computing Machinery},
address = {New York, NY, USA},
url = {https://doi.org/10.1145/3543507.3583510},
doi = {10.1145/3543507.3583510},
booktitle = {Proceedings of the ACM Web Conference 2023},
pages = {1865–1874},
numpages = {10},
location = {Austin, TX, USA},
series = {WWW '23}
}

@article{GossipSub,
  title={GossipSub: Attack-resilient message propagation in the Filecoin and Ethereum networks},
  author={Heath, Andrew and others},
  journal={arXiv preprint arXiv:2007.02754},
  year={2020}
}

@misc{memo_blog,
  title        = {memo.cash Blog},
  year         = {2018},
  month        = apr,
  note         = {Accessed: 2025-07-29},
  author={Memo.cash},
  howpublished = {\url{https://memo.cash/blog/introducing-memo}}
}

@article{Wei2024ExploringTN,
  title={Exploring the Nostr Ecosystem: A Study of Decentralization and Resilience},
  author={Yiluo Wei and Gareth Tyson},
  journal={ArXiv},
  year={2024},
  volume={abs/2402.05709},
  url={https://api.semanticscholar.org/CorpusID:267548110}
}

@article{survey_on_perf,
author = {Anjum, Nasreen and Karamshuk, Dmytro and Shikh-Bahaei, Mohammad and Sastry, Nishanth},
title = {Survey on peer-assisted content delivery networks},
year = {2017},
issue_date = {April 2017},
publisher = {Elsevier North-Holland, Inc.},
address = {USA},
volume = {116},
number = {C},
issn = {1389-1286},
url = {https://doi.org/10.1016/j.comnet.2017.02.008},
doi = {10.1016/j.comnet.2017.02.008},
journal = {Comput. Netw.},
month = apr,
pages = {79–95},
numpages = {17}
}

@misc{farcaster_official_github,
  author       = {{Farcaster}},
  title        = {Farcaster Official Documentation - Hub Monorepo},
  year         = {2023},
  note         = {Accessed: 2025-07-29},
  howpublished = {\url{https://github.com/farcasterxyz/hub-monorepo}},
}

@misc{farcaster_official_doc,
  author = {Farcaster},
  title = {Farcaster Docs},
  note         = {Accessed: 2025-07-29},
  year         = {2025},
  url = {https://docs.farcaster.xyz/}
}

@misc{snapchain_doc,
  author       = {Farcaster},
  title        = {Snapchain Official Document},
  note         = {Accessed: 2025-07-29},
  year         = {2025},
  howpublished = {\url{https://docs.farcaster.xyz/hubble/migrating}},
}

@misc{farcaster_150m_funding,
  author       = {MK Manoylov},
  title        = {Paradigm leads \$150 million raise for web3 social media platform Farcaster},
  year         = {2024},
  month        = may,
  howpublished = {\url{https://www.theblock.co/post/295700/paradigm-150-million-raise-web3-social-platform-farcaster}},
}

@misc{farcaster_farcon_2024,
  author       = {Farcaster},
  title        = {Farcaster Conference 2024},
  note         = {Accessed: 2025-07-29},
  year         = {2024},
  howpublished = {\url{https://farcon.xyz/2024.html}},
}

@misc{farcaster_official_usdc_reward,
  author       = {Farcaster},
  title        = {Farcaster Offical USDC Reward},
  note         = {Accessed: 2025-07-29},
  year         = {2025},
  howpublished = {\url{https://warpcast.notion.site/Warpcast-Rewards-15f6a6c0c10180339fcdee6c9a18c338}},
}

@misc{farcaster_official_doc_hub_install,
  author       = {Farcaster},
  title        = {Hub Installation Configuration},
  note         = {Accessed: 2025-07-29},
  year         = {2025},
  howpublished = {\url{https://docs.farcaster.xyz/hubble/install}},
}

@misc{farcaster_hub_tutorial,
  author       = {Farcaster},
  title        = {Farcaster Hub Setup Tuturial},
  note         = {Accessed: 2025-07-29},
  year         = {2025},
  howpublished = {\url{https://docs.farcaster.xyz/hubble/tutorials}},
}

@misc{farcaster_token_links,
  author       = {Farcaster},
  title        = {Token Links},
  note         = {Accessed: 2025-07-29},
  year         = {2025},
  howpublished = {\url{https://warpcast.notion.site/Token-Links-17e6a6c0c10180a68c4dce055b64d12b}},
}

@misc{farcaster_token_index,
  author       = {Farcaster},
  title        = {Token Index},
  note         = {Accessed: 2025-07-29},
  year         = {2025},
  howpublished = {\url{https://www.farcaster.in/}},
}

@misc{farcaster_token_coingecko,
  author       = {Coingecko},
  title        = {Farcaster Token List},
  note         = {Accessed: 2025-07-29},
  year         = {2025},
  howpublished = {\url{https://www.coingecko.com/en/categories/farcaster-ecosystem}},
}

@misc{alchemy_transfer_api,
  author       = {Alchemy},
  title        = {Alchemy's API for Token Transfer History},
  note         = {Accessed: 2025-07-29},
  year         = {2025},
  howpublished = {\url{https://www.alchemy.com/docs/reference/transfers-api-quickstart}},
}

@misc{uniswap_dex,
  author       = {Uniswap},
  title        = {The Decentralized Exchange Application},
  note         = {Accessed: 2025-07-29},
  year         = {2025},
  howpublished = {\url{https://app.uniswap.org/}},
}

@misc{clanker_website,
  author       = {Clanker},
  title        = {The Official Website of Clanker},
  note         = {Accessed: 2025-07-29},
  year         = {2025},
  howpublished = {\url{https://www.clanker.world/}},
}

@misc{clanker_doc,
  author       = {Clanker},
  title        = {Clanker Official Document},
  note         = {Accessed: 2025-07-29},
  year         = {2025},
  howpublished = {\url{https://clanker.gitbook.io/clanker-documentation/users/deploying-a-token}},
}

@misc{clanker_dune_compare,
  author       = {Clanker},
  title        = {Data Dashboard: Comparing Clanker with Other Token Launchpads},
  note         = {Accessed: 2025-07-29},
  year         = {2025},
  howpublished = {\url{https://dune.com/clanker_protection_team/clanker-vs-others}},
}

@misc{clanker_lum,
  author       = {Itsmechaseb},
  title        = {The Dawn of AI Autonomy: How Two AI Agents Created a \$70M Cryptocurrency},
  year         = {2024},
  month        = nov,
  howpublished = {\url{https://x.com/itsmechaseb/status/1857511608810901941?s=46&t=ywALktyM5ZCyt-zigN5Big}},
}

@misc{clanker_anon_vitalik,
  author       = {TechFlow},
  title        = {Vitalik and Jesse are buying one after another, what is the background of the ANON in the Base ecosystem?},
  year         = {2024},
  month        = nov,
  howpublished = {\url{https://en.theblockbeats.news/news/55854}},
}

@misc{clanker_grok_drb,
  author       = {Ournetwork},
  title        = {Coverage on Clanker, Farcaster, Lens, Kaito, World, Layer3 and Basenames},
  year         = {2025},
  month        = mar,
  howpublished = {\url{https://www.ournetwork.xyz/p/on-320-onchain-culture?utm_campaign=ournetwork}},
}

@misc{degen_official_website,
  author       = {Degen},
  title        = {Degen Tokenomics Document},
  note         = {Accessed: 2025-07-29},
  year         = {2025},
  howpublished = {\url{https://www.degen.tips/}},
}

@misc{degen_history_top_meme,
  author       = {Vladimir},
  title        = {What is DEGEN? The History of The Top Memecoin on Base},
  year         = {2024},
  month        = apr,
  howpublished = {\url{https://zerion.io/blog/the-true-history-of-degen/}},
}

@misc{degen_archive,
  author       = {Folklore},
  title        = {The DEGEN Archives},
  year         = {2024},
  month        = mar,
  howpublished = {\url{https://folklore.mirror.xyz/q9fjoEXpgDH8YGm_EaN3y8jhJfc_kRz7kBW-4IEya7A}},
}

@misc{moxie_official_website,
  author       = {MOXIE},
  title        = {MOXIE Official Website},
  note         = {Accessed: 2025-07-29},
  year         = {2025},
  howpublished = {\url{https://www.moxie.xyz/}},
}

@misc{moxie_scoring_system,
  author       = {MOXIE},
  title        = {FarScores, FarBoost, and Cast Scores},
  note         = {Accessed: 2025-07-29},
  year         = {2025},
  howpublished = {\url{https://github.com/Airstack-xyz/docs/blob/main/social-capital-value-and-social-capital-scores.md}},
}

@misc{tn100x_tokenomics,
  author       = {Ham.fun},
  title        = {TN100X Tokenomics},
  note         = {Accessed: 2025-07-29},
  year         = {2025},
  howpublished = {\url{https://docs.ham.fun/docs/tokenomics/}},
}

@misc{higher_official_doc,
  author       = {Higher},
  title        = {HIGHER Official Greenpaper},
  note         = {Accessed: 2025-07-29},
  year         = {2025},
  howpublished = {\url{https://www.aimhigher.net/greenpaper}},
}

@misc{farther_official_doc,
  author       = {Father},
  title        = {Farther Official Website},
  note         = {Accessed: 2025-07-29},
  year         = {2025},
  howpublished = {\url{https://farther.social/}},
}

@misc{openrank_score,
  author       = {OpenRank},
  title        = {Farcaster Scoring System},
  note         = {Accessed: 2025-07-29},
  year         = {2025},
  howpublished = {\url{https://docs.openrank.com/integrations/farcaster}},
}

@misc{farcaster_contract_addresses,
  author       = {Farcaster},
  title        = {Farcaster Contract Addresses for Identity Registration and Management},
  note         = {Accessed: 2025-07-29},
  year         = {2025},
  howpublished = {\url{https://github.com/farcasterxyz/contracts}},
}

@misc{storage_registry_contract_address,
  author       = {Farcaster},
  title        = {Farcaster Storage Registry Contract},
  note         = {Accessed: 2025-07-29},
  year         = {2025},
  howpublished = {\url{https://optimistic.etherscan.io/address/0x00000000fcce7f938e7ae6d3c335bd6a1a7c593d}},
}

@misc{storage_fee_changes,
  author       = {Dune},
  title        = {Farcaster Storage Fee Changes},
  note         = {Accessed: 2025-07-29},
  year         = {2025},
  howpublished = {\url{https://dune.com/data/farcaster_optimism.storageregistry_evt_setprice}},
}

@misc{hackmd_farcaster_hub_costs,
  author       = {Farcaster},
  title        = {Decentralization of Hubs},
  year         = {2022},
  month        = dec,
  howpublished = {\url{https://hackmd.io/@farcasterxyz/ry0QL4M4o\#Hub-Costs}},
}

@misc{base_chain_90_percentage,
  author       = {Pixelhack},
  title        = {Dune Dashboard of Farcaster users transactions by chain},
  note         = {Accessed: 2025-07-29},
  year         = {2025},
  howpublished = {\url{https://dune.com/queries/3027146/5036846}},
}

@misc{base_farconomy,
  author       = {Luc},
  title        = {Dune Dashboard of Farcaster Token},
  note         = {Accessed: 2025-07-29},
  year         = {2025},
  howpublished = {\url{https://www.dune.com/0xluc/farconomy}},
}

@misc{memocash_official_doc,
  author       = {Jason C},
  title        = {Memo.Cash Official Documentation},
  note         = {Accessed: 2025-07-29},
  year         = {2025},
  howpublished = {\url{https://jasonc.me/blog/what-is-memo}},
}

@misc{neynar_official_dash,
  author       = {Neynar},
  title        = {Farcaster Official Dashboard},
  note         = {Accessed: 2025-07-29},
  year         = {2025},
  howpublished = {\url{https://dune.com/neynarxyz/farcaster}},
}

@misc{deso_official_doc,
  author       = {DeSo},
  title        = {DeSo Official Documentation},
  note         = {Accessed: 2025-07-29},
  year         = {2025},
  howpublished = {\url{https://docs.deso.org/}},
}

@misc{lens_official_doc,
  author       = {Lens},
  title        = {Lens Official Documentation},
  note         = {Accessed: 2025-07-29},
  year         = {2025},
  howpublished = {\url{https://lens.xyz/docs/chain/overview}},
}

@misc{lens_vs_farcaster,
  author       = {Filarm},
  title        = {Comparison of Lens and Farcaster},
  note         = {Accessed: 2025-07-29},
  year         = {2025},
  howpublished = {\url{https://dune.com/filarm/lens-vs-farcaster}},
}

@misc{steemit_doc,
  author       = {Steemit},
  title        = {Steemit Official Documentation},
  note         = {Accessed: 2025-07-29},
  year         = {2025},
  howpublished = {\url{https://steemit.com/guide/@steemitblog/steemit-a-guide-for-newcomers}},
}

@misc{zora_doc,
  author       = {Zora},
  title        = {Zora Official Documentation},
  year         = {2024},
  month        = apr,
  howpublished = {\url{https://zerion.io/blog/guide-to-the-zora-ecosystem/}},
}

@misc{zora_dune_dashboard,
  author       = {Jhackworth},
  title        = {Zora Data Dashboard},
  note         = {Accessed: 2025-07-29},
  year         = {2025},
  howpublished = {\url{https://dune.com/jhackworth/zora-network}},
}

@misc{farcaster_miniapp_doc,
  author       = {Farcaster},
  title        = {Farcaster Mini-app Documentation},
  note         = {Accessed: 2025-07-29},
  year         = {2025},
  howpublished = {\url{https://miniapps.farcaster.xyz/}},
}

@misc{paybot_website,
  author       = {Paycaster},
  title        = {Paybot Documentation},
  note         = {Accessed: 2025-07-29},
  year         = {2025},
  howpublished = {\url{https://paycaster.co/}},
}

@misc{farcaster_wallet_connection,
  author       = {Farcaster},
  title        = {External Wallet Connection},
  note         = {Accessed: 2025-07-29},
  year         = {2025},
  howpublished = {\url{https://docs.farcaster.xyz/developers/guides/writing/verify-address}},
}

@misc{base_doc,
  author       = {Base},
  title        = {Base Chain Documentation},
  note         = {Accessed: 2025-07-29},
  year         = {2025},
  howpublished = {\url{https://docs.base.org/}},
}

@misc{eth_whitepaper,
  author       = {Ethereum},
  title        = {The Ethereum address format and why it matters when using MetaMask},
  note         = {Accessed: 2025-07-29},
  year         = {2025},
  howpublished = {\url{https://ethereum.org/en/whitepaper/}},
}

@misc{evm_addresses,
  author       = {MetaMask},
  title        = {Ethereum Whitepaper},
  note         = {Accessed: 2025-07-29},
  year         = {2025},
  howpublished = {\url{https://support.metamask.io/start/learn/the-ethereum-address-format-and-why-it-matters-when-using-metamask/}},
}

@misc{optimism_doc,
  author       = {Optimism},
  title        = {Optimism Official Documentation},
  note         = {Accessed: 2025-07-29},
  year         = {2025},
  howpublished = {\url{https://docs.optimism.io/}},
}

@misc{solana_doc,
  author       = {Solana},
  title        = {EVM vs. SVM},
  note         = {Accessed: 2025-07-29},
  year         = {2025},
  howpublished = {\url{https://solana.com/developers/evm-to-svm/smart-contracts}},
}

@misc{farcaster_solana,
  author       = {Farcaster},
  title        = {What's New: May 21, 2025 (0.0.49) Introduced Wallet Standard integration for Solana wallets},
  note         = {Accessed: 2025-07-29},
  year         = {2025},
  howpublished = {\url{https://miniapps.farcaster.xyz/docs/sdk/changelog\#may-21-2025-0049}},
}

@misc{diff_sync_project,
  author       = {Farcaster},
  title        = {Diff-Sync Description},
  year         = {2024},
  month        = apr,
  howpublished = {\url{https://github.com/farcasterxyz/protocol/discussions/163}},
}

@inproceedings{raman2019challenges,
  title={Challenges in the decentralised web: The mastodon case},
  author={Raman, Aravindh and Joglekar, Sagar and Cristofaro, Emiliano De and Sastry, Nishanth and Tyson, Gareth},
  booktitle={Proceedings of the internet measurement conference},
  pages={217--229},
  year={2019}
}

@article{jeong2025navigating,
  title={Navigating Decentralized Online Social Networks: An Overview of Technical and Societal Challenges in Architectural Choices},
  author={Jeong, Ujun and Ng, Lynnette Hui Xian and Carley, Kathleen M and Liu, Huan},
  journal={arXiv preprint arXiv:2504.00071},
  year={2025}
}

@techreport{2012-dittrich-mraf,
  author = {Dittrich, D and Kenneally, E},
  title = {{The Menlo Report: Ethical Principles Guiding Information and Communication Technology Research}},
  institution = {U.S. Department of Homeland Security},
  year = {2012},
  month = {August},
  doi = {https://catalog.caida.org/paper/2012_menlo_report_actual_formatted}
}

@article{lin2002divergence,
  title={Divergence measures based on the Shannon entropy},
  author={Lin, Jianhua},
  journal={IEEE Transactions on Information theory},
  volume={37},
  number={1},
  pages={145--151},
  year={2002},
  publisher={IEEE}
}

@article{10.3390/e24050683, 
    author = {Zhang, J. and Shi, J.}, 
    title = {Asymptotic normality for plug-in estimators of generalized shannon’s entropy}, journal = {Entropy}, 
    year = {2022}, 
    volume = {24}, 
    issue = {5}, 
    pages = {683}, 
    doi = {10.3390/e24050683} 
}

@article{10.1371/journal.pone.0202202,
  author = {Liang, J. and Li, L. and Zeng, D.},
  title = {Evolutionary dynamics of cryptocurrency transaction networks: an empirical study},
  journal = {Plos One},
  year = {2018},
  volume = {13},
  issue = {8},
  pages = {e0202202},
  doi = {10.1371/journal.pone.0202202}
}

@article{10.1177/02601079241265744,
  author = {Mardan, M. and Khosravipour, I.},
  title = {Dynamic evolution analysis of cryptocurrency market: a network science study},
  journal = {Journal of Interdisciplinary Economics},
  year = {2024},
  doi = {10.1177/02601079241265744}
}

@article{tao2021complex,
  title={Complex network analysis of the bitcoin transaction network},
  author={Tao, Bishenghui and Dai, Hong-Ning and Wu, Jiajing and Ho, Ivan Wang-Hei and Zheng, Zibin and Cheang, Chak Fong},
  journal={IEEE Transactions on Circuits and Systems II: Express Briefs},
  volume={69},
  number={3},
  pages={1009--1013},
  year={2021},
  publisher={IEEE}
}

@article{liang2022end,
  title={The end of social media? How data attraction model in the algorithmic media reshapes the attention economy},
  author={Liang, Meng},
  journal={Media, Culture \& Society},
  volume={44},
  number={6},
  pages={1110--1131},
  year={2022},
  publisher={SAGE Publications Sage UK: London, England}
}

@article{Cai2024InnovationOM,
  title={Innovation of Media Business Models in the Digital Economy Era},
  author={Shanglin Cai},
  journal={SHS Web of Conferences},
  year={2024},
  url={https://api.semanticscholar.org/CorpusID:272071211}
}

@article{Vasiliauskaite2020UnderstandingCV,
  title={Understanding complexity via network theory: a gentle introduction},
  author={Vaiva Vasiliauskaite and Fernando E. Rosas},
  journal={arXiv: Physics and Society},
  year={2020},
  url={https://api.semanticscholar.org/CorpusID:216867636}
}

@inproceedings{torres2019art,
  title={The art of the scam: Demystifying honeypots in ethereum smart contracts},
  author={Torres, Christof Ferreira and Steichen, Mathis and others},
  booktitle={28th USENIX Security Symposium (USENIX Security 19)},
  pages={1591--1607},
  year={2019}
}

@article{Roth2022WhatsTI,
  title={What’s trending in difference-in-differences? A synthesis of the recent econometrics literature},
  author={Jonathan Roth and Pedro H. C. Sant’Anna and Alyssa M. Bilinski and John Poe},
  journal={Journal of Econometrics},
  year={2022},
  url={https://api.semanticscholar.org/CorpusID:245668975}
}

@misc{networkit_louvain,
  author       = {Networkit},
  title        = {Community Detection with NetworKit},
  note         = {Accessed: 2025-07-29},
  year         = {2025},
  howpublished = {\url{https://networkit.github.io/dev-docs/notebooks/Community.html}}
}

@misc{infomap,
  author       = {Infomap Online},
  title        = {Network community detection using the Map Equation framework},
  note         = {Accessed: 2025-07-29},
  year         = {2025},
  howpublished = {\url{https://www.mapequation.org/infomap/}}
}

@article{micheli2021effect,
  title={The effect of centralized financial and social incentives on cooperative behavior and its underlying neural mechanisms},
  author={Micheli, Leticia and Stallen, Mirre and Sanfey, Alan G},
  journal={Brain sciences},
  volume={11},
  number={3},
  pages={317},
  year={2021},
  publisher={MDPI}
}

@incollection{singh2011mechanism,
  title={Mechanism design for incentivizing social media contributions},
  author={Singh, Vivek K and Jain, Ramesh and Kankanhalli, Mohan},
  booktitle={Social media modeling and computing},
  pages={121--143},
  year={2011},
  publisher={Springer}
}

@article{lindstrom2021computational,
  title={A computational reward learning account of social media engagement},
  author={Lindstr{\"o}m, Bj{\"o}rn and Bellander, Martin and Schultner, David T and Chang, Allen and Tobler, Philippe N and Amodio, David M},
  journal={Nature communications},
  volume={12},
  number={1},
  pages={1311},
  year={2021},
  publisher={Nature Publishing Group UK London}
}

@article{acuff2018applying,
  title={Applying behavioral economic theory to problematic Internet use: An initial investigation.},
  author={Acuff, Samuel F and MacKillop, James and Murphy, James G},
  journal={Psychology of Addictive Behaviors},
  volume={32},
  number={7},
  pages={846},
  year={2018},
  publisher={American Psychological Association}
}

@article{jackson2011overview,
  title={An overview of social networks and economic applications},
  author={Jackson, Matthew O},
  journal={Handbook of social economics},
  volume={1},
  pages={511--585},
  year={2011},
  publisher={Elsevier}
}

@article{sun2018designing,
  title={Designing promotional incentive to embrace social sharing: Evidence from field and online experiments},
  author={Sun, Tianshu and Viswanathan, Siva and Huang, Ni and Zheleva, Elena},
  journal={Forthcoming at MIS Quarterly},
  year={2018}
}

@article{kapoor2023does,
  title={Does incentivization promote sharing “true” content online?},
  author={Kapoor, Hansika and Rezaei, Sarah and Gurjar, Swanaya and Tagat, Anirudh and George, Denny and Budhwar, Yash and Puthillam, Arathy},
  journal={Harvard Kennedy School Misinformation Review},
  year={2023}
}

@article{globig2023changing,
  title={Changing the incentive structure of social media platforms to halt the spread of misinformation},
  author={Globig, Laura K and Holtz, Nora and Sharot, Tali},
  journal={Elife},
  volume={12},
  pages={e85767},
  year={2023},
  publisher={eLife Sciences Publications Limited}
}

@article{Milli2023ChoosingTR,
  title={Choosing the Right Weights: Balancing Value, Strategy, and Noise in Recommender Systems},
  author={Smitha Milli and Emma Pierson and Nikhil Garg},
  journal={ArXiv},
  year={2023},
  volume={abs/2305.17428},
  url={https://api.semanticscholar.org/CorpusID:258960306}
}

@article{Furini2024XAA,
  title={X as a Passive Sensor to Identify Opinion Leaders: A Novel Method for Balancing Visibility and Community Engagement},
  author={Marco Furini},
  journal={Sensors (Basel, Switzerland)},
  year={2024},
  volume={24},
  url={https://api.semanticscholar.org/CorpusID:267092962}
}

@inproceedings{Donkers2025UnderstandingOP,
  title={Understanding Online Polarization Through Human-Agent Interaction in a Synthetic LLM-Based Social Network},
  author={Tim Donkers and J{\"u}rgen Ziegler},
  booktitle={International Conference on Web and Social Media},
  year={2025},
  url={https://api.semanticscholar.org/CorpusID:279226933}
}

@article{trunfio2021conceptualising,
  title={Conceptualising and measuring social media engagement: A systematic literature review},
  author={Trunfio, Mariapina and Rossi, Simona},
  journal={Italian Journal of Marketing},
  volume={2021},
  number={3},
  pages={267--292},
  year={2021},
  publisher={Springer}
}

@article{ceylan2023sharing,
  title={Sharing of misinformation is habitual, not just lazy or biased},
  author={Ceylan, Gizem and Anderson, Ian A and Wood, Wendy},
  journal={Proceedings of the National Academy of Sciences},
  volume={120},
  number={4},
  pages={e2216614120},
  year={2023},
  publisher={National Academy of Sciences}
}

@article{paridar2023more,
  title={More, faster, and better? effects of rewards on incentivizing the creation of user-generated content},
  author={Paridar, Mahsa and Ameri, Mina and Honka, Elisabeth},
  journal={Effects of Rewards on Incentivizing the Creation of User-Generated Content (December 31, 2023)},
  year={2023}
}

@article{scekic2013incentives,
  title={Incentives and rewarding in social computing},
  author={Scekic, Ognjen and Truong, Hong-Linh and Dustdar, Schahram},
  journal={Communications of the ACM},
  volume={56},
  number={6},
  pages={72--82},
  year={2013},
  publisher={ACM New York, NY, USA}
}

@article{dimaggio2012network,
  title={Network effects and social inequality},
  author={DiMaggio, Paul and Garip, Filiz},
  journal={Annual review of sociology},
  volume={38},
  number={1},
  pages={93--118},
  year={2012},
  publisher={Annual Reviews}
}

@article{centola2010spread,
  title={The spread of behavior in an online social network experiment},
  author={Centola, Damon},
  journal={science},
  volume={329},
  number={5996},
  pages={1194--1197},
  year={2010},
  publisher={American Association for the Advancement of Science}
}

@article{rand2011dynamic,
  title={Dynamic social networks promote cooperation in experiments with humans},
  author={Rand, David G and Arbesman, Samuel and Christakis, Nicholas A},
  journal={Proceedings of the National Academy of Sciences},
  volume={108},
  number={48},
  pages={19193--19198},
  year={2011},
  publisher={National Academy of Sciences}
}

@article{zheng2024implications,
  title={Implications of social network structures on socially influenced decision-making},
  author={Zheng, Rui and Ospina-Forero, Luis and Chen, Yu-wang},
  journal={Decision},
  volume={51},
  number={1},
  pages={85--103},
  year={2024},
  publisher={Springer}
}

@article{gruzd2024lure,
  title={The lure of decentralized social media: Extending the UTAUT model for understanding users’ adoption of blockchain-based social media},
  author={Gruzd, Anatoliy and Saiphoo, Alyssa and Mai, Philip},
  journal={Plos one},
  volume={19},
  number={8},
  pages={e0308458},
  year={2024},
  publisher={Public Library of Science San Francisco, CA USA}
}

@article{anderson2023social,
  title={Social motivations’ limited influence on habitual behavior: Tests from social media engagement.},
  author={Anderson, Ian A and Wood, Wendy},
  journal={Motivation Science},
  volume={9},
  number={2},
  pages={107},
  year={2023},
  publisher={Educational Publishing Foundation}
}

@article{wang2023recognition,
  title={Recognition of opinion leaders in blockchain-based social networks by structural information and content contribution},
  author={Wang, Chuansheng and Tan, Xuecheng and Shi, Fulei},
  journal={PeerJ Computer Science},
  volume={9},
  pages={e1549},
  year={2023},
  publisher={PeerJ Inc.}
}

@inproceedings{li2019incentivized,
  title={Incentivized blockchain-based social media platforms: A case study of steemit},
  author={Li, Chao and Palanisamy, Balaji},
  booktitle={Proceedings of the 10th ACM conference on web science},
  pages={145--154},
  year={2019}
}

@article{ba2022role,
  title={The role of cryptocurrency in the dynamics of blockchain-based social networks: The case of steemit},
  author={Ba, Cheick Tidiane and Zignani, Matteo and Gaito, Sabrina},
  journal={PloS one},
  volume={17},
  number={6},
  pages={e0267612},
  year={2022},
  publisher={Public Library of Science San Francisco, CA USA}
}

@ARTICLE{steemit2024richgetricher,
  author={Guidi, Barbara and Michienzi, Andrea and Ricci, Laura},
  journal={IEEE Transactions on Computational Social Systems}, 
  title={Assessment of Wealth Distribution in Blockchain Online Social Media}, 
  year={2024},
  volume={11},
  number={1},
  pages={671-682},
  doi={10.1109/TCSS.2022.3228925}}

@inproceedings{michienzi2024wealth,
  title={A wealth-driven analysis of user engagement in Blockchain Online Social Media},
  author={Michienzi, Andrea and Guidi, Barbara and Ricci, Laura},
  booktitle={2024 IEEE International Conference on Pervasive Computing and Communications Workshops and other Affiliated Events (PerCom Workshops)},
  pages={637--642},
  year={2024},
  organization={IEEE}
}

@article{delkhosh2023impact,
  title={Impact of bot involvement in an incentivized blockchain-based online social media platform},
  author={Delkhosh, Fatemeh and Gopal, Ram D and Patterson, Raymond A and Yaraghi, Niam},
  journal={Journal of Management Information Systems},
  volume={40},
  number={3},
  pages={778--806},
  year={2023},
  publisher={Taylor \& Francis}
}

@article{ichimiya2023evaluation,
  title={Evaluation of response to incentive recruitment strategies in a social media-based survey},
  author={Ichimiya, Megumi and Muller-Tabanera, Hope and Cantrell, Jennifer and Bingenheimer, Jeffrey B and Gerard, Raquel and Hair, Elizabeth C and Donati, Dante and Rao, Nandan and Evans, W Douglas},
  journal={Digital health},
  volume={9},
  pages={20552076231178430},
  year={2023},
  publisher={SAGE Publications Sage UK: London, England}
}

@article{yu2022paying,
  title={When Paying for Reviews Pays Off: The Case of Performance-Contingent Monetary Rewards.},
  author={Yu, Yinan and Khern-am-nuai, Warut and Pinsonneault, Alain},
  journal={MIS Quarterly},
  volume={46},
  number={1},
  year={2022}
}

@article{burtch2015social,
  title={What are social incentives worth? A randomized field experiment in user content generation},
  author={Burtch, Gordon and Hong, Yili and Bapna, Ravi and Griskevicius, Vladas},
  year={2015}
}

@article{yan2017financial,
  title={Do financial incentives induce more online participatory behaviors?},
  author={Yan, Zhijun and Kuang, Lini and Huang, He and Yang, Han},
  year={2017}
}

@article{khern2018extrinsic,
  title={Extrinsic versus intrinsic rewards for contributing reviews in an online platform},
  author={Khern-am-nuai, Warut and Kannan, Karthik and Ghasemkhani, Hossein},
  journal={Information Systems Research},
  volume={29},
  number={4},
  pages={871--892},
  year={2018},
  publisher={INFORMS}
}

@article{qiao2017incentive,
  title={Incentive provision and pro-social behaviors},
  author={Qiao, Dandan and Lee, Shun-Yang and Whinston, Andrew and Wei, Qiang},
  year={2017}
}

@article{yu2018impact,
  title={The impact of performance-contingent monetary incentives on user-generated content contribution},
  author={Yu, Yinan and Khern-am-nuai, Warut and Pinsonneault, Alain},
  year={2018}
}

@article{garnefeld2012explicit,
  title={Explicit incentives in online communities: boon or bane?},
  author={Garnefeld, Ina and Iseke, Anja and Krebs, Alexander},
  journal={International Journal of Electronic Commerce},
  volume={17},
  number={1},
  pages={11--38},
  year={2012},
  publisher={Taylor \& Francis}
}

@article{zhang2025time,
  title={Time to Stop? An Empirical Investigation on the Consequences of Canceling Monetary Incentives on a Digital Platform},
  author={Zhang, Dongcheng and Jiang, Hanchen and Qiang, Maoshan and Zhang, Kunpeng and Qiu, Liangfei},
  journal={Information Systems Research},
  volume={36},
  number={2},
  pages={781--801},
  year={2025},
  publisher={INFORMS}
}

@article{liu2015can,
  title={Can Monetary Incentives Increase UGC Contribution? The Motivation and Competition Crowding Out},
  author={Liu, Yuewen and Feng, Juan},
  year={2015}
}

@article{huang2023effect,
  title={The Effect of Incentives on Facilitating User Engagement with Succulent Retailers’ Social Media Pages},
  author={Huang, Li-Chun},
  journal={Horticulturae},
  volume={9},
  number={8},
  pages={849},
  year={2023},
  publisher={MDPI}
}

@article{austin2009balance,
  title={Balance diagnostics for comparing the distribution of baseline covariates between treatment groups in propensity-score matched samples},
  author={Austin, Peter C},
  journal={Statistics in medicine},
  volume={28},
  number={25},
  pages={3083--3107},
  year={2009},
  publisher={Wiley Online Library}
}

@article{austin2011introduction,
  title={An introduction to propensity score methods for reducing the effects of confounding in observational studies},
  author={Austin, Peter C},
  journal={Multivariate behavioral research},
  volume={46},
  number={3},
  pages={399--424},
  year={2011},
  publisher={Taylor \& Francis}
}

@inproceedings{balduf2025bootstrapping,
  title={Bootstrapping Social Networks: Lessons from Bluesky Starter Packs},
  author={Balduf, Leonhard and Sokoto, Saidu and Baronchelli, Andrea and Castro, Ignacio and Kr{\'o}l, Micha{\l} and Tyson, Gareth and Pavlou, George and Scheuermann, Bj{\"o}rn and Ascigil, Onur},
  booktitle={Proceedings of the International AAAI Conference on Web and Social Media},
  volume={19},
  pages={178--192},
  year={2025}
}

@misc{sciencedirect_clustering,
  author = {{ScienceDirect}},
  title = {Clustering Coefficient},
  note         = {Accessed: 2025-07-29},
  year         = {2025},
  howpublished = {\url{https://www.sciencedirect.com/topics/computer-science/clustering-coefficient}}
}

@misc{altcoinbuzz_farcaster,
  author={{Victor}},
  title = {3 Ways to Earn on Farcaster: USDC, Frames, and Airdrops},
  year         = {2025},
  month        = mar,
  howpublished = {\url{https://www.altcoinbuzz.io/bitcoin-and-crypto-guide/3-ways-to-earn-on-farcaster-usdc-frames-and-airdrops/}}
}

@misc{bitget_farcaster_tips,
  author       = {{Bitget}},
  title        = {Farcaster Launches “TIPS” Feature, Supporting One-Click Sending of USDC Rewards},
  year         = {2025},
  month        = apr,
  howpublished = {\url{https://www.bitget.com/news/detail/12560604678174}},

}

@article{lechner2011estimation,
  title={The estimation of causal effects by difference-in-difference methods},
  author={Lechner, Michael and others},
  journal={Foundations and Trends{\textregistered} in Econometrics},
  volume={4},
  number={3},
  pages={165--224},
  year={2011},
  publisher={Now Publishers, Inc.}
}

@article{allen2023airdrop,
  title={Why airdrop cryptocurrency tokens?},
  author={Allen, Darcy WE and Berg, Chris and Lane, Aaron M},
  journal={Journal of Business Research},
  volume={163},
  pages={113945},
  year={2023},
  publisher={Elsevier}
}

@article{makridis2023rise,
  title={The rise of decentralized cryptocurrency exchanges: Evaluating the role of airdrops and governance tokens},
  author={Makridis, Christos A and Fr{\"o}wis, Michael and Sridhar, Kiran and B{\"o}hme, Rainer},
  journal={Journal of Corporate Finance},
  volume={79},
  pages={102358},
  year={2023},
  publisher={Elsevier}
}

% Articles V9pmc045-V9pmc125 use
\received{July 2025}
\received[revised]{September 2025}
\received[accepted]{October 2025}


\clearpage

% ========= appendix =========
\begin{appendices}
\crefalias{section}{appendix}
\crefalias{subsection}{appendix}
\FloatBarrier
%%%%%%%%%%%%%%%%%%%%%%%%%%%%%%%%%%%%%%%%%%%%%%%
\section{Token-related Events Driving User Wallet Binding.}
\label{app:rq1_social_token_related_events}
%%%%%%%%%%%%%%%%%%%%%%%%%%%%%%%%%%%%%%%%%%%%%%
To better understand how token activities drive token economy participation, this section plots the wallet binding dynamics along with influential token-related events in greater detail. 

\begin{figure}[h]
  \centering
  \includegraphics[width=0.9\linewidth]{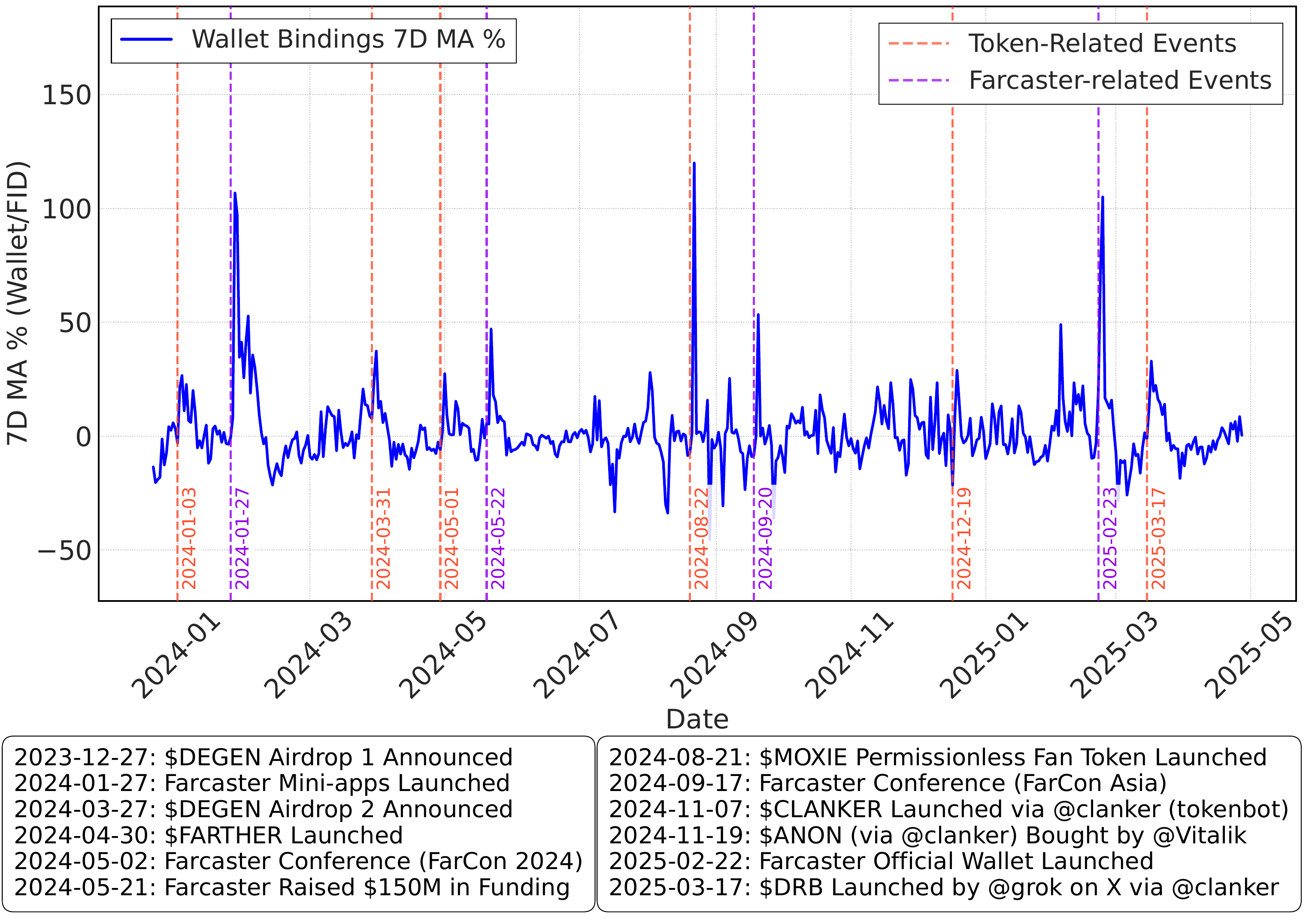}
  \caption{Top 10 surges in 7-day moving average percentage change of daily wallet bindings (highlighted by color-coded dashed lines), annotated with platform milestones and token campaigns.}
\label{fig:eth_timeline_with_surges}
  \Description{FID Registration and ETH Wallet Binding Timeline with Surges and Events}
\end{figure}

We observe a notable delay effect between the onset of growth and peak binding activities.  After experimenting with various smoothing techniques including Exponential Moving Average (EMA) and different moving average windows (3-day, 5-day, 7-day, and 10-day), we find that the 7-day moving average most effectively captures well-distributed top 10 surges that align with visual inspection of the data.

\Cref{fig:eth_timeline_with_surges} presents the 7-day moving average percentage change for wallet bindings. 
We then annotate major spikes in activity with key events identified in the Farcaster ecosystem.
Key event identification follows our mixed-methods approach: 
We do this by reviewing news from The Block Beats\footnote{The Block Beats: \url{www.theblockbeats.info}.} and posts from Farcaster's hub dataset, combined with quantitative examination of token trading frequencies among newly bound wallets around surge dates.

Through this, we identify key wallet-binding-driving events covering two categories: platform-led events and social-token-driven events (further classified into reward-token and meme-coin\footnote{Meme-coins are tokens typically created as jokes or for entertainment purposes, often inspired by internet memes or popular culture, with their value largely driven by community sentiment and social media trends.} events). 
We next introduce each category of event and discuss the corresponding surges in wallet binding activity.

\subsection{Platform-led Events.} 
Four surges in wallet bindings closely align with core Farcaster milestones: the Mini-apps launch (106.74\% surge, 2024-01-27), the \$150M funding announcement (47.06\%, 2024-05-21), the Farcaster Conference (FarCon) Asia (53.39\%, 2024-09-17), and the official Farcaster wallet launch (105.02\%, 2025-02-22). 
Notably, these platform-driven events account for the 2nd to 5th largest surges among the top 10 observed, with the Mini-apps launch ranking 2nd, the official wallet launch 3rd, and the funding announcement and FarCon Asia ranking 4th and 5th, respectively. 
This pattern demonstrates that user growth on Farcaster is driven by feature releases and platform milestones, rather than by entry barriers (\eg registration fee reductions).

\subsection{Reward-Token Events.}
In addition to platform-led milestones, token airdrop\footnote{Airdrop refers to the free distribution of cryptocurrency tokens or coins to eligible wallet addresses, often as a marketing strategy to increase protocol adoption and reward early users.} announcements---such as DEGEN (26.62\% surge, 2023-12-27; 37.34\%, 2024-03-27), \$FARTHER (27.42\%, 2024-04-30), and MOXIE (119.93\%, 2024-08-21)---represent a distinct category of events that also drive wallet binding surges. 
These initiatives incentivize user engagement through mechanisms that allocate daily token allowances based on social interactions and third-party reputation scores (\eg OpenRank scores~\cite{openrank_score}), without requiring direct financial expenditure from users. 
The distributed rewards are funded by project treasuries locked in smart contracts~\cite{degen_official_website, farther_official_doc, moxie_official_website}, encouraging both new and existing users to link wallets to their \acp{fid}.

A particularly noteworthy development occurred on August 21, 2024, when MOXIE introduced a permissionless mechanism for users to issue and auction their own profile-tokenized \emph{Fan Tokens}---a model closely aligned with Lens Protocol~\cite{lens_official_doc} and Zora~\cite{zora_doc}. 
These fan tokens can be freely traded, and holders are eligible to receive a proportion (designated by the Fan Token issuer, \eg 20\%) of the Fan Token issuer's daily MOXIE engagement rewards.
Although this announcement led to the largest observed surge in wallet bindings ($\approx$ 120\%), the underlying tokenomics and redistribution dynamics are beyond the scope of this work; in \Cref{sec:rq1_user_base_token_incentive_mechanisms}, we focus on incentive mechanisms for the initial allocation of tokens based on user engagement.

\subsection{Meme-Coin Events.} 
While reward tokens inherently contain speculative elements~\cite{degen_archive,degen_history_top_meme}, their primary design is to foster social engagement. 
In contrast, the launch of \emph{@clanker}---an AI-powered token issuance bot (\ac{fid} = 874542)---on November 9, 2024, marked a significant shift by enabling an automated pipeline for meme-coin creation~\cite{clanker_doc}. 
Users can deploy new meme-coins simply by posting with the desired token name and description while mentioning @clanker. 
The bot then deploys the meme-coin on the Base chain, establishes initial liquidity pools,\footnote{When Clanker launches a new meme-coin, it automatically creates a trading pair between ETH and the meme-coin as a liquidity pool, enabling users to trade the token.} and facilitates instant trading via the Clanker platform~\cite{clanker_website} or decentralized exchanges such as Uniswap~\cite{uniswap_dex}. 
By June 5, 2025, Clanker had enabled the creation of 280,678 meme-coins, with 28,224 (about 10\%) of these tokens observed in the on-chain transaction records of Farcaster users' wallets, attracting considerable attention for its rapid wealth effects and fee revenue model~\cite{clanker_lum, clanker_dune_compare}.

Several key meme-coins issued via Clanker token produced pronounced spikes in wallet binding: 
\begin{enumerate*}
    \item the launch of CLANKER (wallet binding increases 11.54\%, 2024-11-09, the first eponymous meme-coin by @clanker followed by Ethereum founder Vitalik Buterin's~\cite{eth_whitepaper} purchase of ANON token (a token representing anonymous internet culture) on November 19, 2024, jointly driving a 28.9\% wallet binding surge by December 19, 2024~\cite{clanker_anon_vitalik};
    
    \item the launch of \$DRB by @grok (X's AI Agent account) on March 17, 2025, resulting in a 37.34\%, which exemplifies AI-to-AI token interaction and triggered widespread discussion of the Farcaster ecosystem across X (Twitter)~\cite{clanker_grok_drb}.
\end{enumerate*}
These events, amplified by social momentum and celebrity engagement, underscore the intricate relationship between platform growth and token-based speculation.

The above demonstrates that platform innovation and token-driven incentives are critical for driving deeper user engagement. 

%%%%%%%%%%%%%%%%%%%%%%%%%%%%%%%%%%%%%%%%%%%%%%%
\FloatBarrier
\section{Skewed Token Distribution} 
\label{app:rq1_overall_token_distribution}
%%%%%%%%%%%%%%%%%%%%%%%%%%%%%%%%%%%%%%%%%%%%%%%

In ~\Cref{sec:rq1_prevalent_token_detection}, we aim to identify tokens that demonstrate sustained and widespread activity within the Farcaster ecosystem, rather than those irrelevant to Farcaster or exhibiting merely temporary bursts of activity (\eg due to spam, speculation, or airdrops).

Among all \ac{fid}-linked wallets in our dataset, we observe 440,274 distinct tokens. 
This substantial diversity stems from the interoperability between users' external wallets and the broader Ethereum ecosystem, resulting in the presence of many tokens that may have little to no direct connection to the Farcaster ecosystem. 
Therefore, in this section, we first examine the overall distribution of these tokens, which guides us in developing a systematic approach to identify prevalent tokens that maintain consistent usage patterns and meaningful relevance to social interactions on Farcaster.

\subsection{Trading Metric.}
\label{sec:trading_metric}
In traditional markets, trading activity is measured via trading volume (\eg 1,000 Tesla shares) or dollar volume (total value at $\approx$ \$400/share). 
Cryptocurrency markets similarly use token-denominated (\eg 1 ETH) and fiat-equivalent volumes (\eg \$4,000). 
However, both metrics pose challenges for cross-token analysis in blockchain systems.
Token volumes cannot be meaningfully aggregated due to vast quantity differences across cryptocurrencies (\eg 0.00001 BTC vs 10,000 DOGE, both $\approx$ \$1 equivalent). 
Fiat-equivalent aggregation is complicated by high token price volatility and limited price data availability for illiquid tokens.
We therefore focus on transaction frequency---the count of distinct transaction events per token.

\begin{figure}[h]
  \centering
  \includegraphics[width=0.9\linewidth]{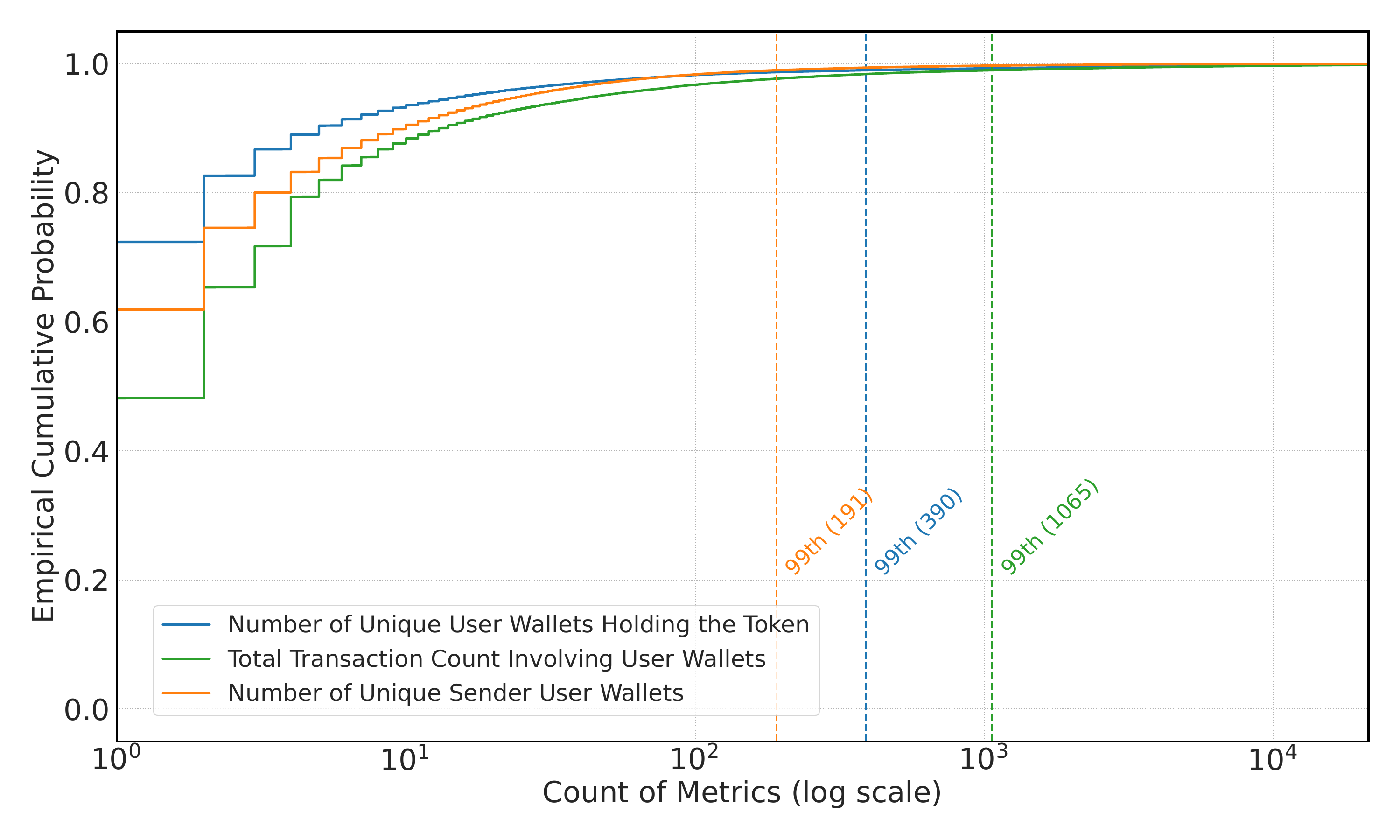}
  \caption{ECDF of token metrics on Farcaster: transaction frequency, holder count (full population: 440,274 tokens), and \ac{fid}-sender count (subset: 177,733 tokens).}
  \label{fig:token_ECDF}
  \Description{ECDF of Farcaster token metrics (transaction  frequency, holder count and FID-sender count)}
\end{figure}

We inspect three metrics: trading frequency, number of holders, and number of users as token sender (\ie the wallet actively sending out the token is linked to an \ac{fid})---for all 440,274 tokens that have appeared in Farcaster users' Ethereum wallets. 
\Cref{fig:token_ECDF} illustrates the \ac{ecdf} for these three metrics.
The overall distribution exhibits high skewness, with a small subset of tokens dominating these indicators.

\subsection{Transaction Frequency and Holder Count.}
Based on a total token population of 440,274, our analysis reveals that 99\% of tokens have no more than 390 holders ($mean \approx 63.46$, $median = 1$) and 1,065 transactions ($mean \approx 207.86$, $median = 2$)---notably low figures when compared to the potential market of 489,824 \ac{fid}-linked wallets on Farcaster. 
The concentration of activity in a small number of tokens suggests that most tokens in users' wallets are not actively used for social interactions. 
Further, the low median values (1 holder, 2 transactions) suggest that many tokens never gained meaningful adoption. 

\subsection{\ac{fid} as Token Sender.}
Furthermore, this skewness is particularly pronounced for \ac{fid} senders. 
Recall, an \ac{fid} sender means the wallet is linked to an \ac{fid}. 
Within the total population of 440,274 tokens, only 177,733 tokens (40.37\%) exhibit at least one \ac{fid}-sender interaction. 
This indicates that approximately 60\% of all tokens have never been actively sent by any Farcaster user, suggesting they are only passively received by users and have never been employed in any use cases such as tipping other users and interacting with exchanges or smart contracts. 
Furthermore, among these 177,733 tokens with at least one \ac{fid}-sender activity, 99\% have no more than 191 unique senders ($mean \approx 24.57$, $median = 1$), revealing a highly concentrated distribution of active engagement. 

This observation reveals critical insights into token circulation patterns:
Due to the transparent and non-rejectable nature of blockchain transactions, users frequently become passive token recipients through promotional airdrops or potential phishing attempts. 
Consequently, active token sending behavior, particularly from \ac{fid}-linked wallets, serves as a more reliable indicator of genuine user engagement and token utility.
This becomes especially significant when considering the scale: among 489,824 \ac{fid}-linked wallets, 99\% of inter-user traded tokens engage fewer than 191 active senders (merely 0.039\% of total \ac{fid}-linked wallets), highlighting a striking disparity between trivial and influential token circulation in the ecosystem.
This distribution pattern indicates that despite the presence of over 440K tokens in Farcaster users' Ethereum wallets, only a small subset demonstrates meaningful interaction initiated by Farcaster users, as evidenced by the highly skewed \ac{fid}-sender distribution. 
This observation motivates us to introduce a systematic approach to automatically identify and analyze prevalent tokens in the next section.

%%%%%%%%%%%%%%%%%%%%%%%%%%%%%%%%%%%%%%%%%%%%%%%
\FloatBarrier
\section{Prevalent Token Detection.}
\label{app:rq1_prevalent_token_detection_appendix}
%%%%%%%%%%%%%%%%%%%%%%%%%%%%%%%%%%%%%%%%%%%%%%%
In the exploratory analysis of token metrics, we observe that commonly used indicators such as rankings of trading volumes, transaction frequencies, and holder counts may provide insufficient or potentially misleading signals regarding a token's genuine influence. 
Indeed, malicious actors could artificially manipulate these metrics through strategic token distributions targeting user wallet addresses, thereby fabricating an illusion of market popularity~\cite{torres2019art}. 
This phenomenon poses significant challenges for token valuation that rely on these surface-level metrics.
Therefore, we propose a lightweight systematic method to differentiate between genuinely prevalent tokens and those potentially manipulated with artificially inflated indicators. 

We formalize our detection approach as a four-step algorithm, structured in the following subsections.

\subsection{Step 1: Inter-\ac{fid} Transactions (5,878 tokens remained after screening).}
%%%%%%%%%%%%%%%%%%%%%%%%%%%%%%%%%%%%%%%%%%%%%%%%%
We begin with user-to-user (inter-\ac{fid}) transactions---by extracting all transactions where both sender and receiver wallets are explicitly linked to registered \ac{fid} accounts on Farcaster. 
It is important to note that Farcaster allows users to link their existing Ethereum-compatible wallets to their accounts.
Consequently, these wallets contain transaction records that extend beyond the Farcaster ecosystem, with inter-user transactions representing only a subset of total wallet activity. 
This initial filtering identifies tokens with at least one transaction between Farcaster users, excluding tokens solely traded with external smart contracts, exchanges, or wallets lacking Farcaster social context. 

After restricting transactions where both recipient and sender wallets are linked to an \acp{fid}, we identified a subset of 5,878 tokens, constituting 1.34\% of the total 440,274 tokens, accounting for 3,354,378 transfers, representing 3.63\% of the complete dataset containing 92,287,905 token transactions.
This reveals a power-law distribution of token ecosystems, where a small fraction of tokens (1.34\%) achieve meaningful social circulation, while the vast majority of tokens (98.66\%) lack user-to-user activity and primarily operate in non-social contexts such as smart contract interactions and exchange swaps.

\subsection{Step 2: Shannon Entropy (104 tokens remained after screening).} 
%%%%%%%%%%%%%%%%%%%%%%%%%%%%%%%%%%%%%%%%%%%%%%%%%
To address the ephemeral nature of most tokens, which typically show activity only in their initial weeks, we employ \emph{Shannon entropy} to analyze weekly transaction distributions~\cite{lin2002divergence, 10.3390/e24050683}.

\paragraph{Shannon Entropy}
\label{sec:shannon_entropy}
We compute Shannon entropy over the weekly transaction frequency distribution for each token. Specifically:

\begin{itemize}
    \item \textbf{Input Data}: For each token, we construct a probability vector $\mathbf{p} = (p_1, \ldots, p_T)$ where:
    \begin{equation}
        p_t = \frac{n_t}{\sum_{i=1}^T n_i}
    \end{equation}
    with $n_t$ being the transaction count in week $t$, and $T$ the token's lifespan in weeks.
    
    \item \textbf{Entropy Calculation}:
    \begin{equation}
        H(\mathbf{p}) = -\sum_{t=1}^T p_t \log_2 p_t \quad \text{(bits)}
    \end{equation}
    
    \item \textbf{Normalization}:
    \begin{equation}
        H_{\text{norm}} = \frac{H(\mathbf{p})}{H_{\text{max}}}, \quad \text{where} \quad H_{\text{max}} = \log_2 T
    \end{equation}
\end{itemize}

Key properties:
\begin{itemize}
    \item $H_{\text{norm}} \in [0,1]$ with:
    \begin{itemize}
        \item 1: Perfectly uniform distribution
        \item 0: Single-week concentration
    \end{itemize}
    \item Threshold $H_{\text{norm}} \geq 0.9$ selects tokens with:
    \begin{equation}
        \frac{H(\mathbf{p})}{\log_2 T} \geq 0.9
    \end{equation}
\end{itemize}

We compute normalized Shannon entropy over each token's weekly transaction frequency, retaining tokens with  $H_{\text{norm}} = H(\mathbf{p})/\log_2 T \geq 0.9$.
This threshold identifies tokens with temporal uniformity (evenly distributed weekly transactions) and sustained vitality (activity sustained over a lifespan $T$ long enough for stable entropy measurement), thus filtering for those that maintain consistent engagement beyond an initial launch phase. 
This yields 793 tokens exhibiting both temporal uniformity and sustained vitality.

Nevertheless, a substantial subset of 559 tokens (70.5\% of the total 793 tokens), each with a lifespan not exceeding 5 weeks, exhibits high normalized entropy values (mean $\approx$ 0.967), despite their consistently low raw entropy (all values $< 1.6$, aligning with the mean raw entropy across the entire 793-token sample).
To address this short-period bias, we add a minimum 26-week (half-year) lifespan requirement, yielding 104 tokens.

The choice of a 26-week minimum existence requirement is not arbitrary. 
We observe that among 969 tokens with raw entropy values above 3 (99.78th percentile, N=440,274), only 19 (1.96\%, n=976) have existed for less than 26 weeks (compared to a mean existence of just $\approx$ 1.67 weeks across all 440K tokens). 
This threshold thus ensures both an adequate sample size ($\approx$ 1,000 tokens) and effectively excludes sampling insufficiency issues with newer tokens, giving appropriate weight to tokens with longer trading histories. 
In real-world applications, these calculations can be performed weekly to dynamically include tokens previously excluded by the 26-week requirement.

\subsection{Step 3: \acp{fid} as Token Senders (9 tokens remained after screening).} 
%%%%%%%%%%%%%%%%%%%%%%%%%%%%%%%%%%%%%%%%%%%%%%%%%
The 104 tokens identified in the previous steps exhibit a right-skewed distribution in their number of unique \ac{fid}-linked senders.
Here, an \ac{fid}-linked sender is defined as a wallet address explicitly associated with a registered \ac{fid}, ranging from 1 to 29,464 (mean $\approx$ 897). 
The number of \ac{fid}-linked wallet senders serves as a crucial metric for evaluating token prevalence, as it more substantially reflects genuine social interactions rather than passive reception. 
This metric's significance stems from its ability to distinguish between tokens with meaningful user engagement and those with merely superficial circulation.
Consequently, we employ the number of \ac{fid}-linked wallet senders within each token's inter-\ac{fid}-transactions as our final filtering criterion.
Using the 99th percentile threshold (254 \ac{fid}-senders), we finally identify nine tokens---four reward tokens and five blockchain network tokens---detailed in \Cref{tab:prevalent_token}.

This process yields nine tokens.
Notably, DEGEN and MOXIE correspond to significant user growth events (recall that we have discussed the top 10 events in \Cref{fig:eth_timeline_with_surges}). 
TN100X and HIGHER, two other reward tokens launched in February and March 2024 respectively, did not trigger top 10 wallet binding surges. 
However, these tokens were identified through our screening process for long-term token popularity. 
Conversely, \$FARTHER, which appeared in the Top 10 events, was not selected by our screening criteria. 
Through analysis of community content and documentation~\cite{farther_official_doc}, we discover that the \$FARTHER reward program was terminated by developers in August 2024 due to excessive user farming\footnote{User farming refers to the behavior where small groups of users or bots engage in circular reward-giving among themselves to exploit the project's token reserves} and sell-offs.\footnote{Sell-offs occur when users, upon receiving token rewards, immediately exchange them in the open market for more established cryptocurrencies (\eg USDC, ETH, or BTC) instead of utilizing them within the ecosystem for services or tipping.} 
This finding suggests that while reward tokens may generate temporary enthusiasm and transaction bursts, only a select few achieve sustained user adoption and utilization. 
The remaining five are native tokens and stablecoins commonly used on blockchain main-net (Ethereum) and scaling layers (L2s and L3s): USDC, USDT, USDbC, WETH, and L3.
These tokens were also identified by our screening methodology due to their broader market acceptance and high utilization rates.

\subsection{Step 4: Clustering Coefficient (4 tokens remained after screening).} 
%%%%%%%%%%%%%%%%%%%%%%%%%%%%%%%%%%%%%%%%%%%%%%%%%
We next calculate the average clustering coefficient for the transaction graph of each token. 
We do this for each transaction graph's largest connected component.
\emph{\ac{acc}} effectively captures community network patterns and this metric has also been used in cryptocurrency analysis for identifying artificial transaction patterns~\cite{tao2021complex, 10.1371/journal.pone.0202202, 10.1177/02601079241265744}. 
We then select all tokens that have a clustering coefficient between 0.3 and 0.6. 
We choose this because previous studies present that real-world community has a clustering coefficient around 0.45 ~\cite{Vasiliauskaite2020UnderstandingCV}. 
This leaves four remaining tokens that are considered prevalent.

\paragraph{Summary.}
Our methodology, based on social relationships (inter-\ac{fid} transaction and \ac{fid} sender), network spatial distribution (clustering), and transaction temporal distribution (entropy), successfully identifies 4 prevalent tokens within Farcaster, with strong social attributes (see \Cref{tab:prevalent_token}).
This approach is useful for identifying genuinely influential and commonly used tokens
especially for permissionless ecosystems like Farcaster, where the ability to bind external wallets and support all Ethereum-compatible tokens necessitates robust filtering mechanisms to distinguish viable tokens from low-signal noise.
This selection also aligns perfectly with our ground truth observations of the influential tokens that drive user growth in \Cref{sec:rq1_user_base_driving_events}, validating our approach. 
Moreover, given the extreme sparsity of positive samples (\ie few viable tokens among all tokens present in user wallets)
and the fact that each filtering stage employs thresholds tailored to specific scenarios, our three-dimensional framework demonstrates superior operational practicality, performance, and contextual explainability compared to machine learning-based approaches. 
These advantages make it particularly suitable for real-world applications like platform built-in token linking algorithms~\cite{farcaster_token_links} or index website ranking systems~\cite{farcaster_token_index,farcaster_token_coingecko}.

\FloatBarrier
\section{Token Incentive Distributions}
\label{app:token_incentive_distribution}
%%%%%%%%%%%%%%%%%%%%%%%%%%%%%%%%%%%%%%

\subsection{Methodology for Tracing Reward Sources}
\label{sec:reward_source_tracing_method}
%%%%%%%%%%%%%%%%%%%%%%%%%%%%%%%%%%%
To identify Inter-\ac{fid} Tipping, we implement a triple-filtering process: (i) eliminating self-transfers within the same FID, (ii) transfers between different wallets bound to the same FID, and (iii) transfers between FIDs that have historically shared wallet bindings---indicating affiliated entities. 

To identify \emph{Third-party Algorithmic Rewards}, we filter transactions originating from token issuers' official wallets/dedicated smart contracts\footnote{The identification of these addresses is based on a combination of the following methods: examining token project whitepapers, leveraging third-party data dashboards, and analyzing the transaction history of a wallet used for field experimentation in relevant token reward projects.} to FID-linked wallets. 
Similarly, the official USDC algorithmic rewarding involves transfers from Farcaster's official wallet cluster/dedicated smart contracts to FID-linked wallets.
The source wallet addresses and smart contracts for distributing algorithmic token rewards, along with the five identified prevalent token contract addresses, are listed in \Cref{tab:token_rewards_source_addresses}.

%%%%% token reward source addresses %%%%%%%
\begin{table}[htbp]
\centering
\caption{Tokens's name and contract addresses, and algorithmic token reward distribution addresses}
\label{tab:token_rewards_source_addresses}
\footnotesize
\setlength{\tabcolsep}{6pt}
\renewcommand{\arraystretch}{1.05}
\begin{tabular}{l l l}
\toprule
\textbf{Token's Namer} & \textbf{Token's Contract Address} & \textbf{Algorithmic Token Reward Source Address} \\
\midrule
DEGEN & 0x4ed4e862860bed51a9570b96d89af5e1b0efefed &
0x9f07f8a82cb1af1466252e505b7b7ddee103bc91\\ 
& & 0x7134ddcd6bc3d3f29643ec8882cd4d7d0b38379d\\
& & x88d42b6dbc10d2494a0c6c189cefc7573a6dce62\\
& & 0xa2a5c549a454a1631ff226e0cf8dc4af03a61a75\\
& & 0x0bf676823f958d0ae6af9860880fc7a327a0c582\\
& & 0xf4c9fc64ed902b08c3397c48dca0342bdfd49a32\\
& & 0x81ac315d7baf7fae2e8278dc33ec91b752560166\\
& & 0xdfeddda2bc2e75524f470991f46404ee90ae2eea \\
\addlinespace[2pt]
USDC & 0x833589fcd6edb6e08f4c7c32d4f71b54bda02913 &
0x2d93c2f74b2c4697f9ea85d0450148aa45d4d5a2\\
& & 0xd15fE25eD0Dba12fE05e7029C88b10C25e8880E3 \\
& & 0xf6ea479f30a71cc8cb28dc28f9a94246e1edc492 \\
& & 0xf25b82277d9dfdbb341fd1bd9bf197b60ae09e37 \\
& & 0xed64b7fd10f6f06def48af99e719c3cd8e9c78c6 \\
\addlinespace[2pt]
MOXIE & 0x8c9037d1ef5c6d1f6816278c7aaf5491d24cd527 & 0xfe569dc5a297de1dfbcce19273fc8267f2c52c45 \\
\addlinespace[2pt]
HIGHER & 0x0578d8a44db98b23bf096a382e016e29a5ce0ffe & -- \\
TN100X & 0x5b5dee44552546ecea05edea01dcd7be7aa6144a & -- \\
\addlinespace[2pt]

\bottomrule
\end{tabular}
\end{table}

\subsection{Intersection of Reward Participants}
\label{sec:intersection_reward_participants}
\begin{figure}[htbp]
    \centering
    \begin{subfigure}{0.60\linewidth}
        \centering
        \includegraphics[width=\linewidth]{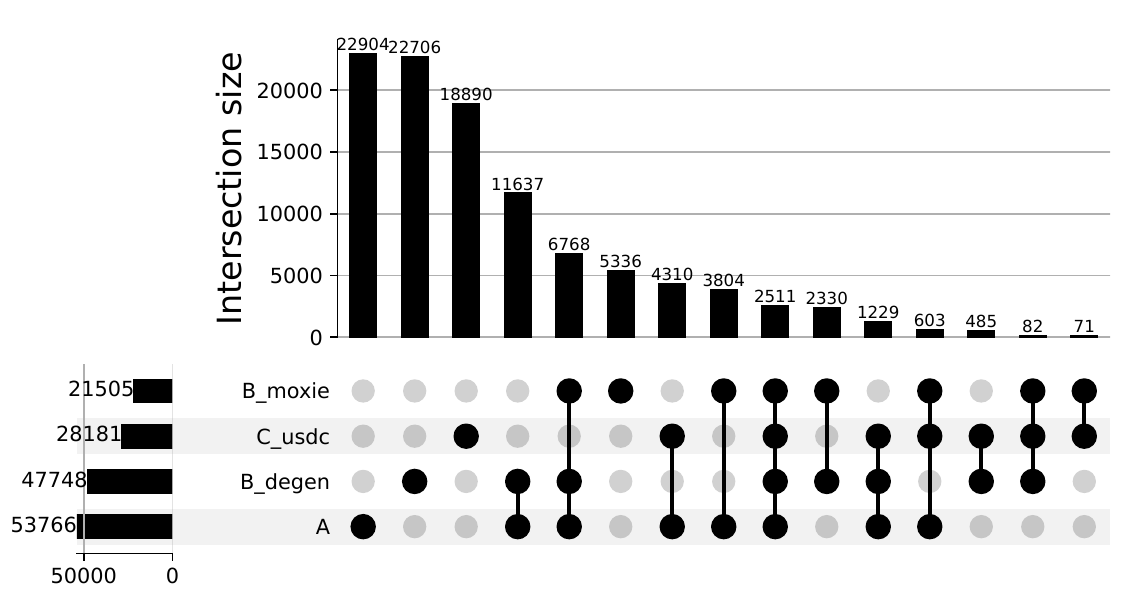}
        \caption{Receivers across Inter-\ac{fid} Tipping, Third-party Algorithmic Rewards and Official Algorithmic Rewards.}
        \label{fig:upset_plot_a}
    \end{subfigure}
    \hfill
    \begin{subfigure}{0.49\linewidth}
        \centering
        \includegraphics[width=\linewidth]{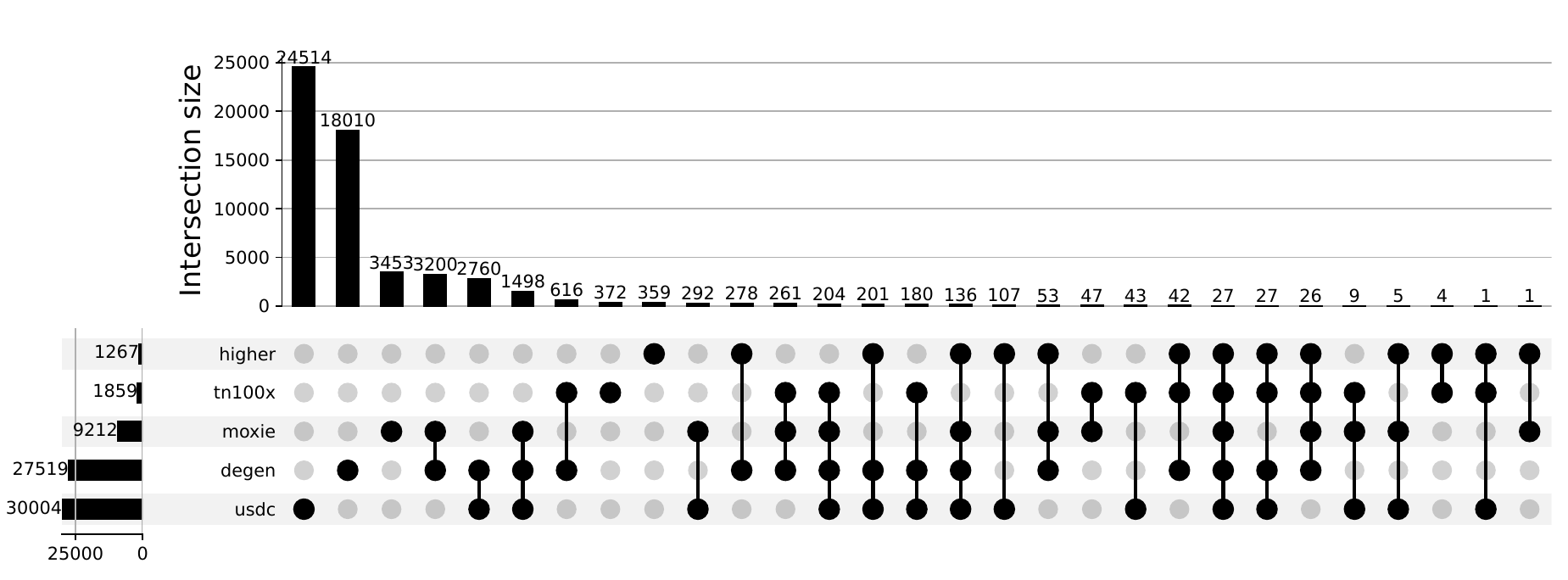}
        \caption{Senders across five tokens in Inter-\ac{fid} Tipping.}
        \label{fig:upset_plot_b}
    \end{subfigure}
    \hfill
    \begin{subfigure}{0.49\linewidth}
        \centering
        \includegraphics[width=\linewidth]{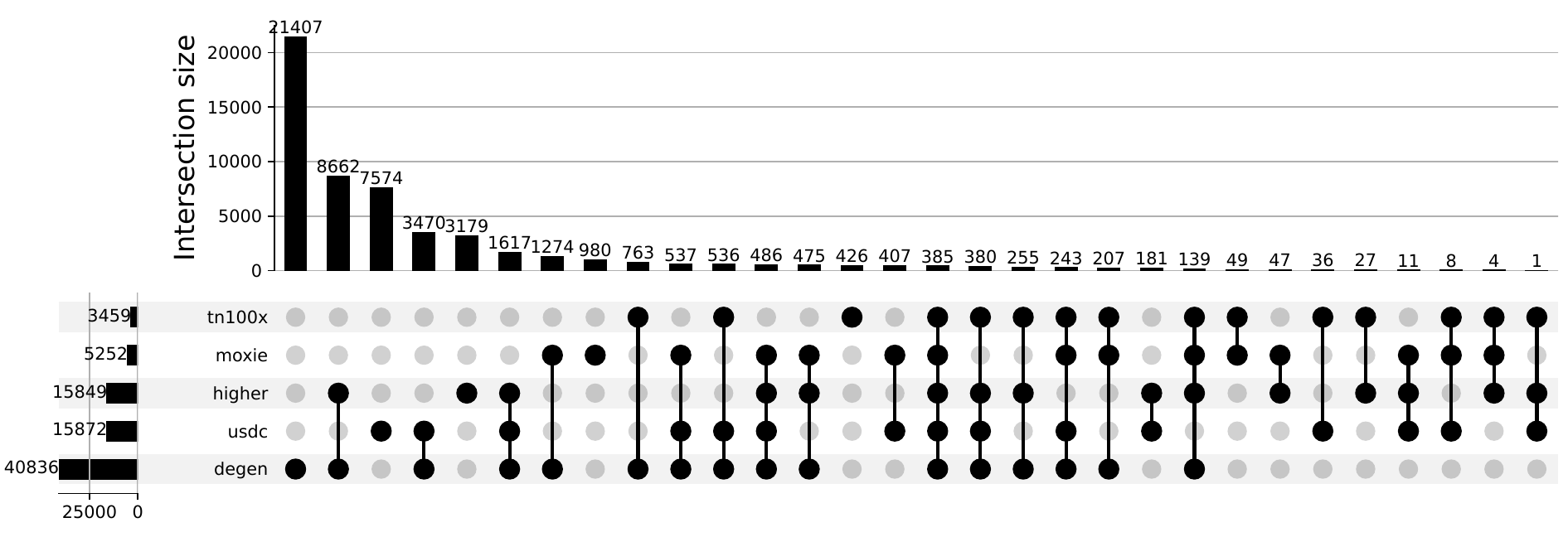}
        \caption{Receivers across five tokens in Inter-\ac{fid} Tipping.}
        \label{fig:upset_plot_c}
    \end{subfigure}
    \hfill
    \caption{Upset plots.}
    \label{fig:upset_plot}
    \Description{Upset plots show the intersections of reward participants across different incentive mechanisms and tokens. }
\end{figure}

%%%%%%%%%%%%%%%%%%%%%%%%%%%%%%%%%%%
\Cref{fig:upset_plot} illustrates the upset plots showing the intersections of reward participants across different incentive mechanisms and tokens. 
\Cref{fig:upset_plot_a} shows the overlap of receivers among three Mechanisms---\emph{Inter-\ac{fid} Tipping}, \emph{Third-party Algorithmic Rewards}, and Farcaster's \emph{Official Algorithmic Rewards}; 
\Cref{fig:upset_plot_b} depicts the overlap of senders across the five tokens within \emph{Inter-\ac{fid} Tipping}; 
\Cref{fig:upset_plot_c} displays the overlap of receivers across the five tokens within \emph{Inter-\ac{fid} Tipping}. 
These plots highlight the extent to which participants engage with multiple mechanisms or tokens.

%%%%%%%%%%%%%%%% upset plots %%%%%%%%%%%%%%

The upset plots help analyze user coverage across three reward mechanisms involving five tokens on Farcaster.
\emph{Inter-\ac{fid} Tipping} reaches 53,766 unique receivers through five tokens (6.01\% of Farcaster's 894,678 valid registered users). 
The upset plots in \Cref{fig:upset_plot} reveal minimal multi-token overlap---the largest intersection comprises 8,662 users (16.11\% of tipping receivers) who received both DEGEN and HIGHER tokens, while only 385 users (0.716\%) received all five tokens.
For algorithmic rewards, \emph{Third-party Algorithmic Rewards} (via DEGEN and MOXIE) reach 57,562 receivers (6.43\% of \acp{fid}), while the official USDC rewards engage 28,181 users (3.15\%). 
Combined, both algorithmic rewards reach 80,762 users (9.03\%), indicating moderate overlap (20.64\%) between \emph{Third-party Algorithmic Rewards} and \emph{Official Algorithmic Rewards}. 
Across all three mechanisms, the total unique recipient count is 103,666 (11.59\% of \acp{fid})---notably lower than the sum of individual mechanism counts (178,702) but higher than any single mechanism's reach. 
The primary cross-mechanism overlap consists of 11,637 users (11.23\% of all recipients) receiving both tipping and DEGEN algorithmic rewards, while only 2,511 users (2.42\%) received rewards from all three mechanisms.

These patterns demonstrate strong mechanism specialization, with users typically engaging with specific token incentives rather than participating broadly across the reward ecosystem. 
Notably, algorithmic rewards collectively reach more unique users (80,762) than tipping (53,766), suggesting that systematic incentive distribution achieves wider user coverage than peer-to-peer tipping.

% \FloatBarrier
\section{Socioeconomic Risks of Token Incentives.}
\label{app:rq2_socioeconomic_risks_appendix}

%%%%%%%%%%%%%%%%%%%%%%%%%%%%%%%%%%%%%%

\begin{figure}[htbp]
    \centering
    \begin{subfigure}{0.47\linewidth}
        \centering
        \includegraphics[width=\linewidth]{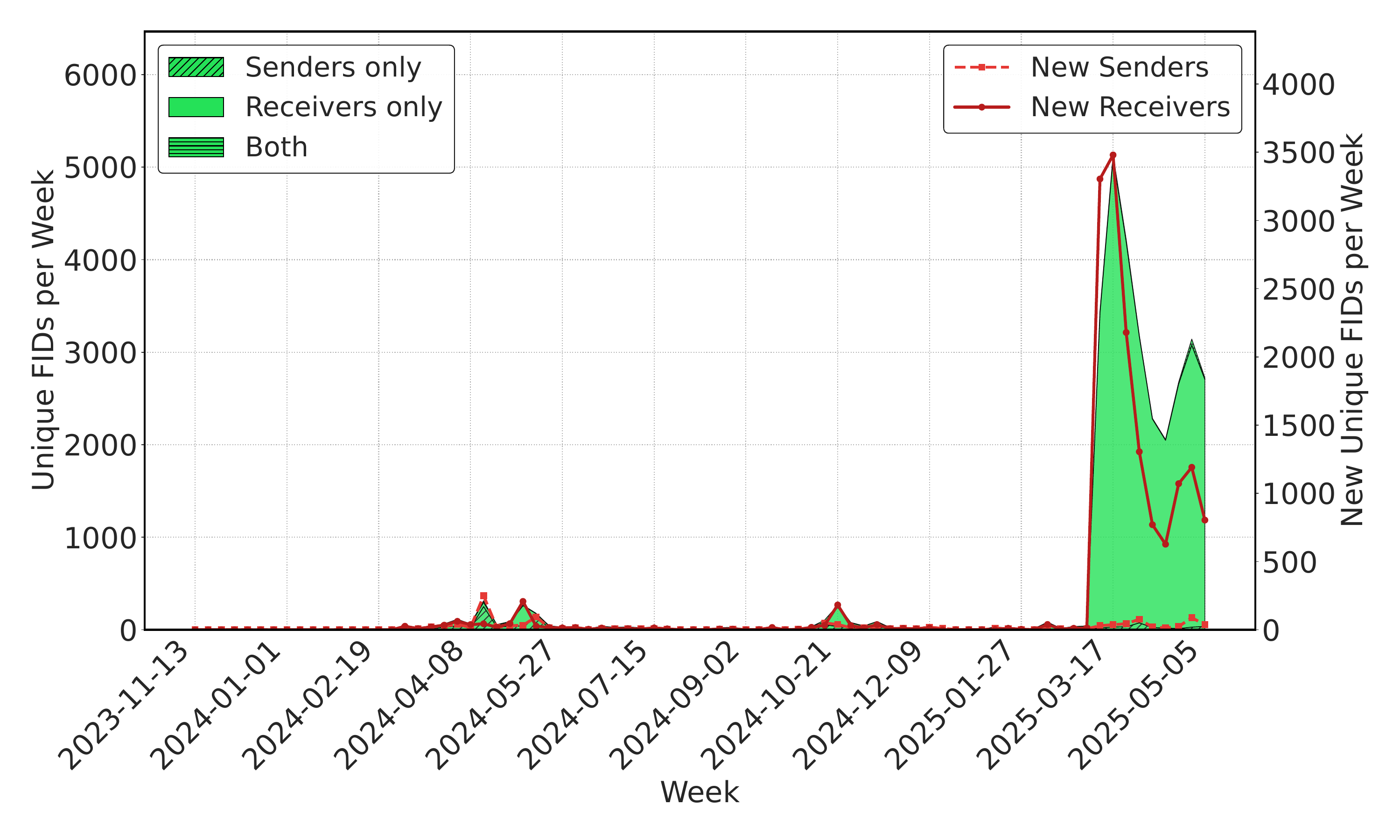}
        \caption{HIGHER (Inter-\ac{fid} Tipping).}
        \label{fig:}
    \end{subfigure}
    \hfill
    \begin{subfigure}{0.47\linewidth}
        \centering
        \includegraphics[width=\linewidth]{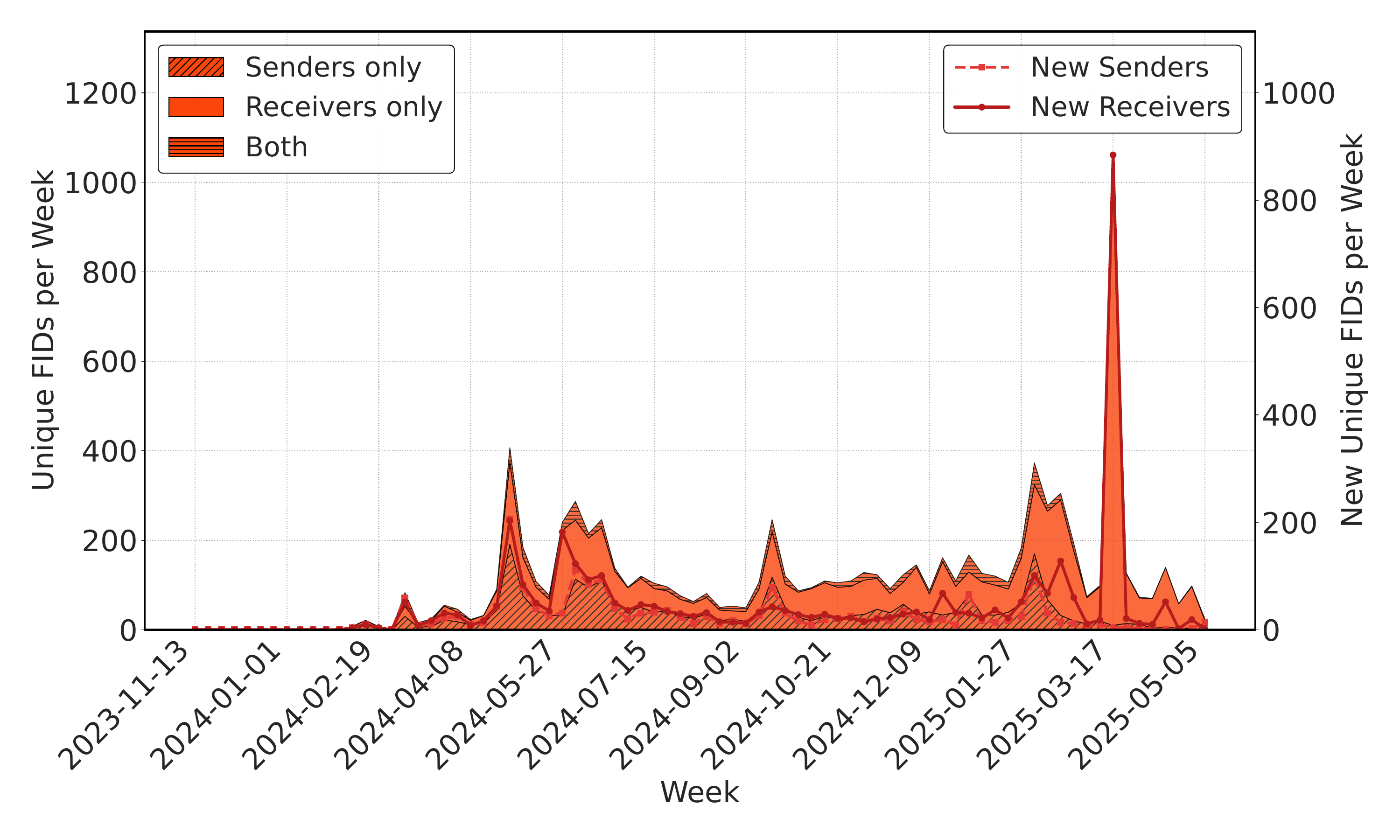}
        \caption{TN100X (Inter-\ac{fid} Tipping).}
        \label{fig:}
    \end{subfigure}
    \hfill
    \begin{subfigure}{0.47\linewidth}
        \centering
        \includegraphics[width=\linewidth]{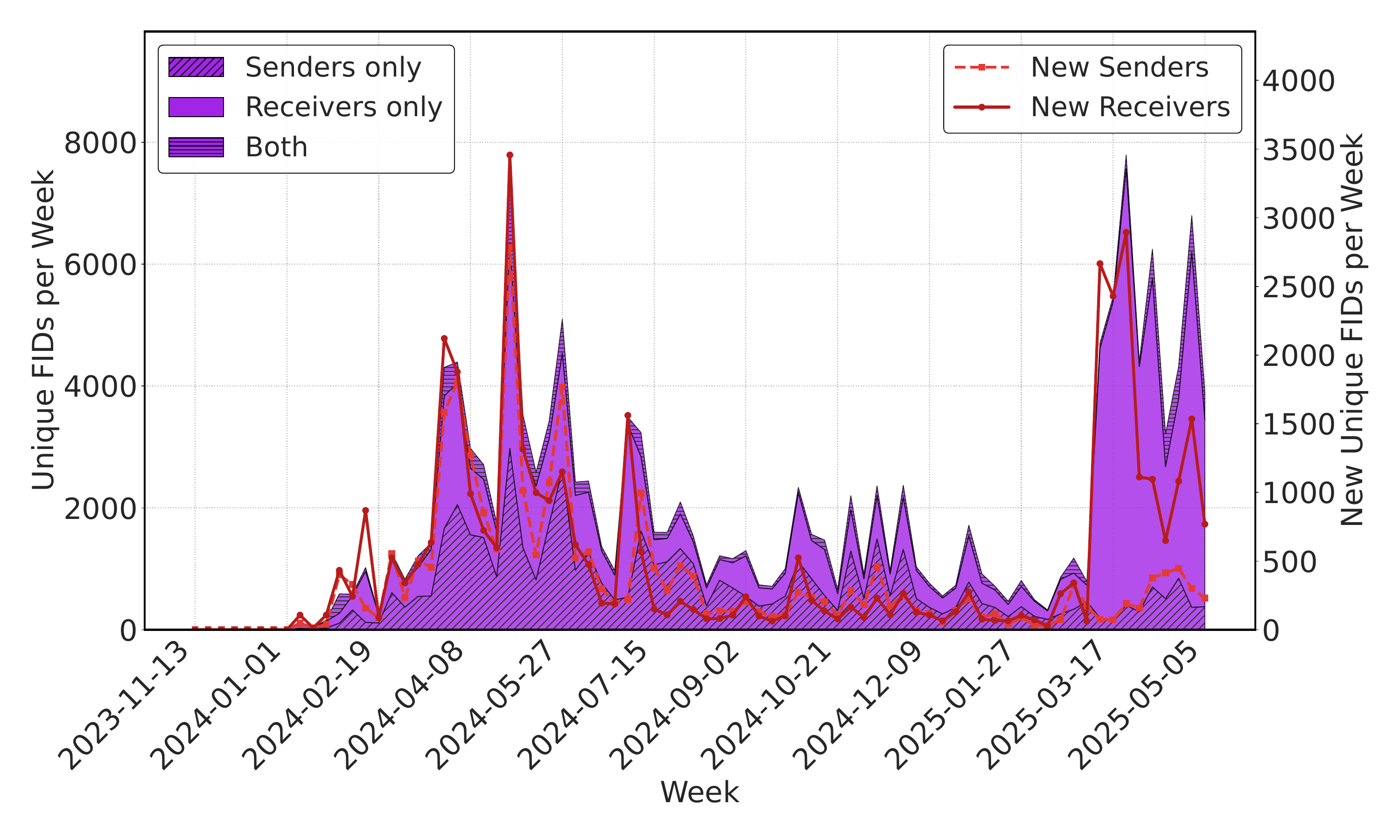}
        \caption{DEGEN (Inter-\ac{fid} Tipping).}
        \label{fig:}
    \end{subfigure}
    \hfill
    \begin{subfigure}{0.47\linewidth}
        \centering
        \includegraphics[width=\linewidth]{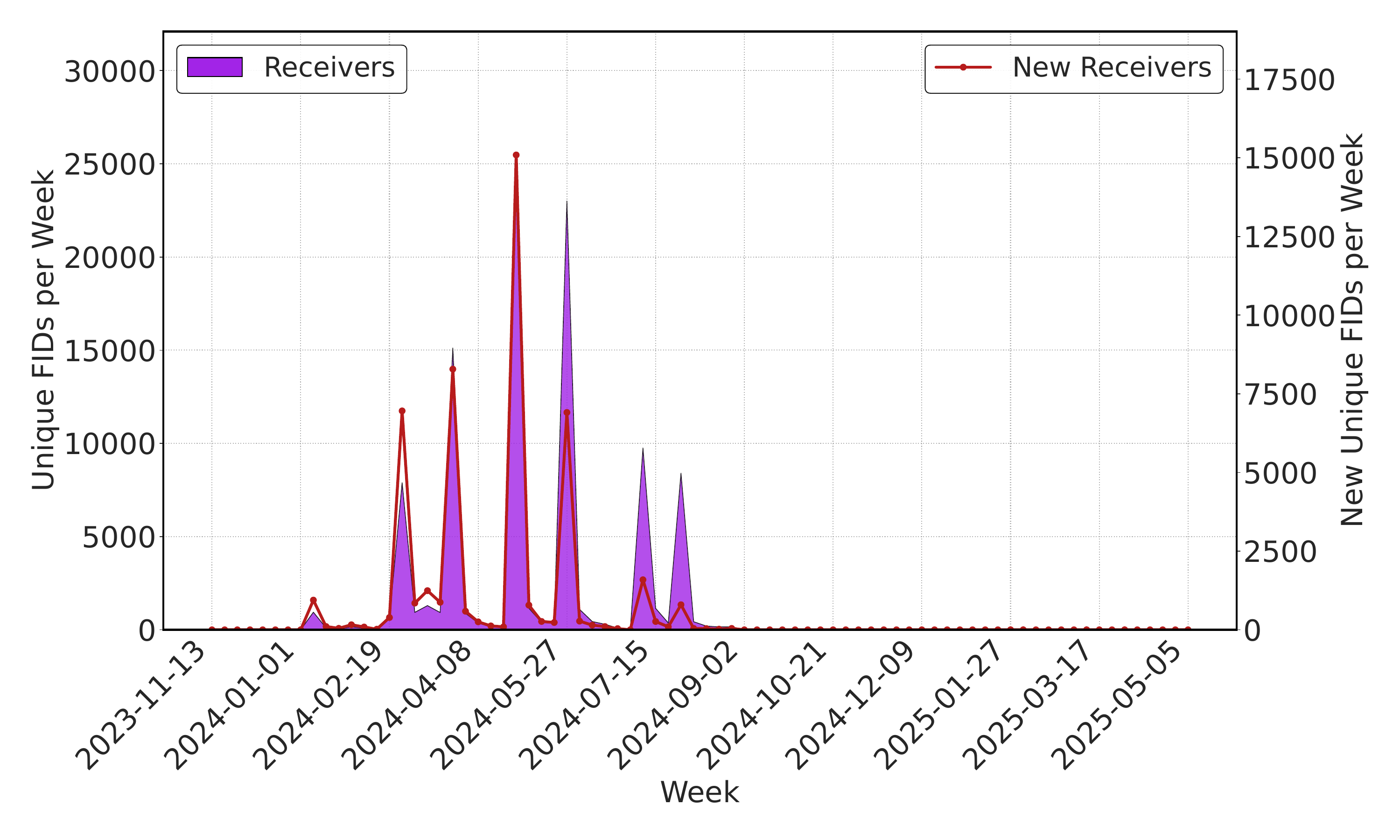}
        \caption{DEGEN (Third-party Algorithmic Reward).}
        \label{fig:}
    \end{subfigure}
    \hfill
    \begin{subfigure}{0.47\linewidth}
        \centering
        \includegraphics[width=\linewidth]{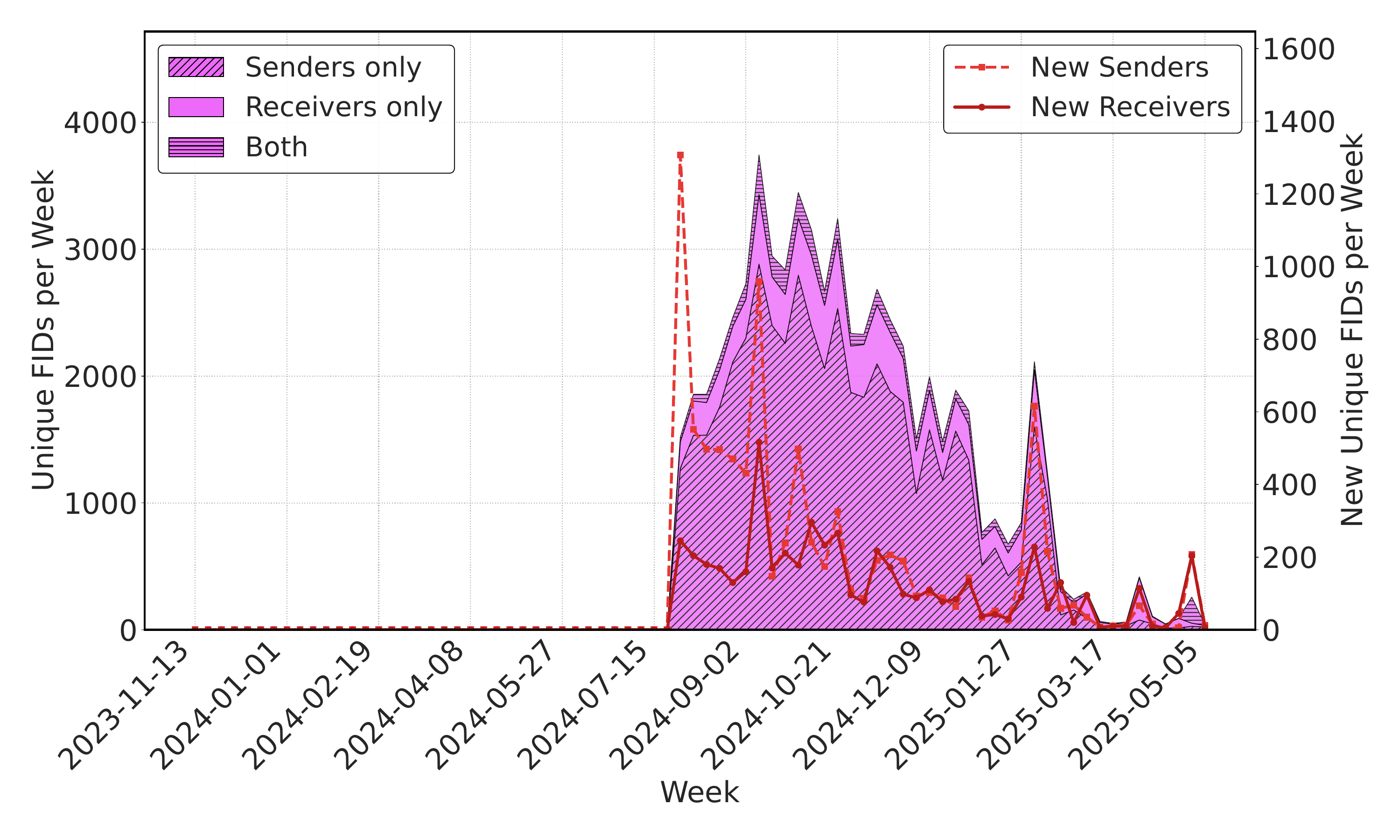}
        \caption{MOXIE (Inter-\ac{fid} Tipping).}
        \label{fig:}
    \end{subfigure}
    \hfill
    \begin{subfigure}{0.47\linewidth}
        \centering
        \includegraphics[width=\linewidth]{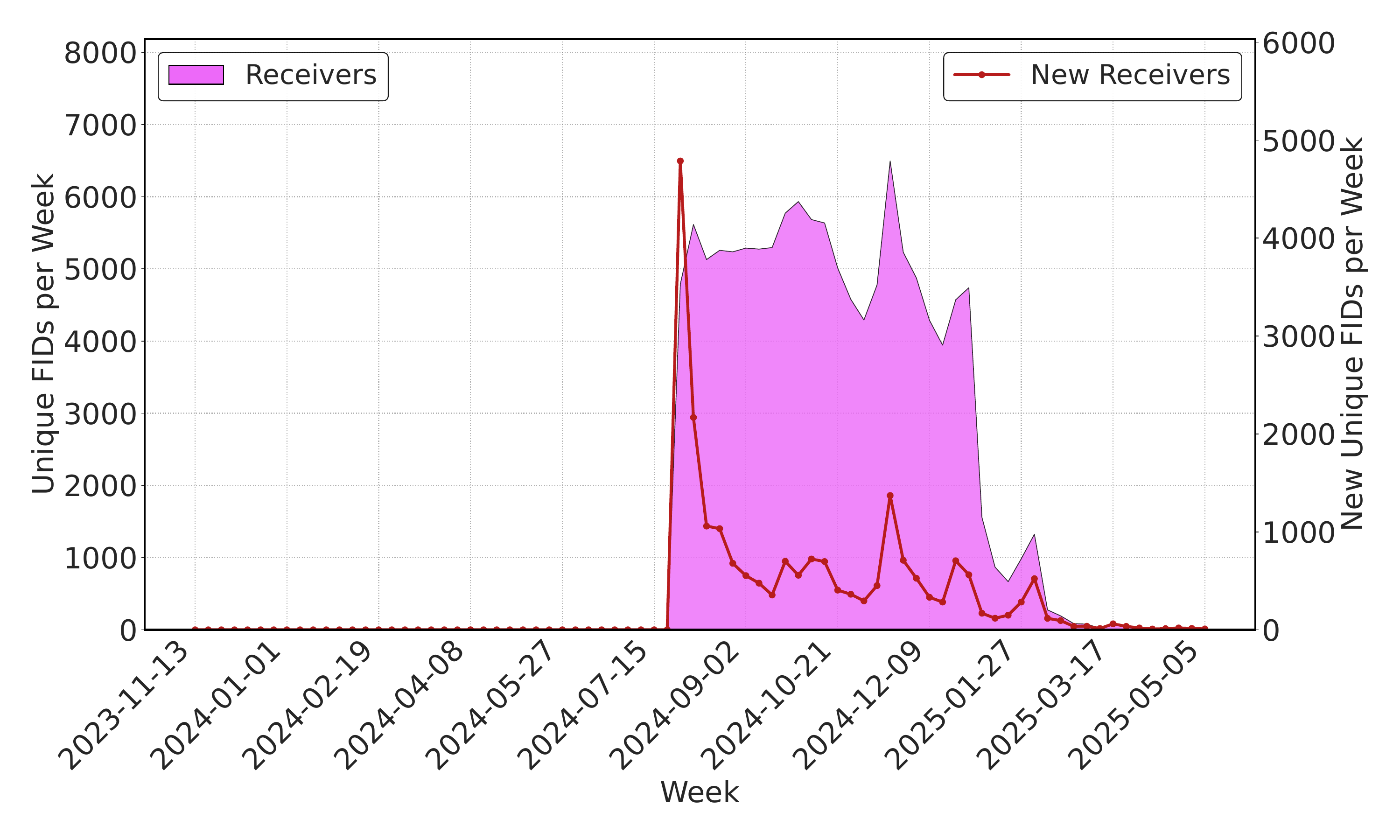}
        \caption{MOXIE (Third-party Algorithmic Reward).}
        \label{fig:}
    \end{subfigure}
    \hfill
    \begin{subfigure}{0.47\linewidth}
        \centering
        \includegraphics[width=\linewidth]{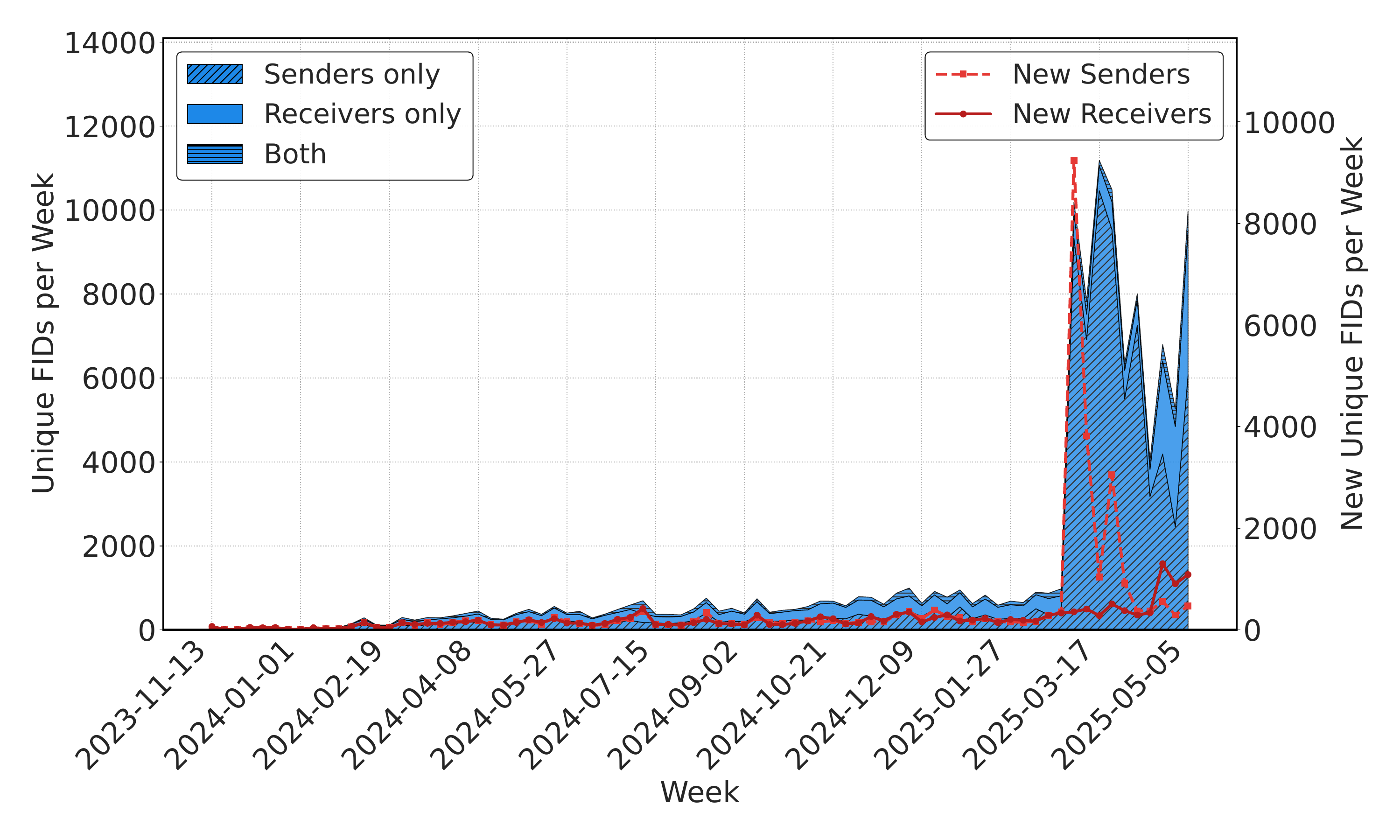}
        \caption{USDC (Inter-\ac{fid} Tipping).}
        \label{fig:}
    \end{subfigure}
    \hfill
    \begin{subfigure}{0.47\linewidth}
        \centering
        \includegraphics[width=\linewidth]{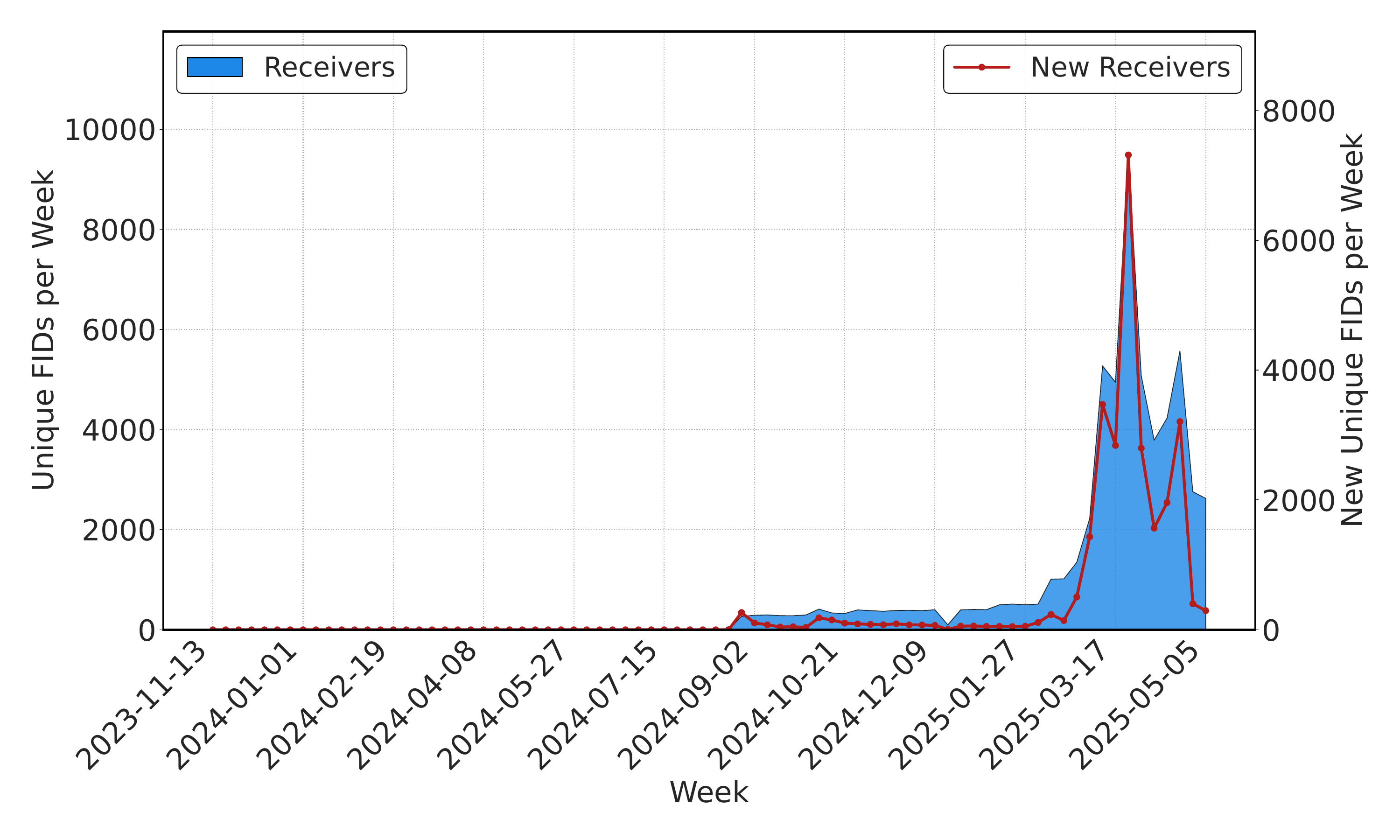}
        \caption{USDC (Official Algorithmic Reward).}
        \label{fig:}
    \end{subfigure}
    \caption{Stacked area charts with overlaid lines visualize participation for each token-mechanism pair.}
    \label{fig:stacked_and_line_each_token_abc}
    \Description{Stacked area charts with overlaid lines visualize participation for each token-mechanism pair.}
\end{figure}

\Cref{fig:stacked_and_line_each_token_abc} visualizes the participation for each token-mechanism pair in the form of stacked area charts with overlaid lines. 
The shaded areas represent the weekly number of unique participants---either senders (diagonal pattern), receivers (solid fill), or both (horizontal pattern). 
The solid line indicates the number of new receivers appearing each week, while the dashed line indicates the number of new senders appearing each week. 
This visualization captures both the cumulative engagement and the dynamics of new participant inflow over time.

%%%%%%%%%%%%%
% Table 10
\begin{table}[t]
\centering
\footnotesize
\caption{Network statistics comparison.}
\label{tab:echo_network_comparison}
\begin{tabular}{lll}
  \toprule
  Metric & Follow-only & Follow + Tip \\
  \midrule
  Nodes & 883,712 & 883,906 \\
  Edges & 159,539,953 & 159,595,800 \\
  \# Communities & [332, 359] & [334, 372] \\
  Largest SCC & 172 & [173, 174] \\
  Total SCCs & 173 & [174, 175] \\
  Avg. Deg. & 361.06 & 361.11 \\
  Max Deg. & 564,120 & 564,989 \\
  Modularity & [0.54389, 0.54580] & [0.53196, 0.54625] \\
  \bottomrule
\end{tabular}
\end{table}

%Table 11
\begin{table}[t]
\centering
\footnotesize
\caption{Community detection (follow-only network).}
\label{tab:echo_community_detection}
\setlength{\tabcolsep}{1.5pt}
\renewcommand{\arraystretch}{0.95}
\begin{tabular}{llllll}
  \toprule
  \multirow{2}{*}{Metric} & \multirow{2}{*}{Louvain} & \multicolumn{4}{c}{Infomap} \\
  \cmidrule(l){3-6}
   & & Level 0 & Level 1 & Level 2 & Level 3 \\
  \midrule
  Communities & [332, 359] & 187 & 8,301 & 330,165 & 5,122 \\
  Largest Comm. & $\approx$ 462,230 & 880,428 & 330,867 & 23,127 & 2,766 \\
  Median Comm. & 4 & 4 & 3 & 1 & 2 \\
  \midrule
  \multicolumn{6}{l}{Tip Edge Distribution:} \\
  Intra-Comm. & $\approx$ 67\% & 99.37\% & 37.5\% & 3.84\% & 0.26\% \\
  Inter-Comm. & $\approx$ 33\% & 0.63\% & 62.5\% & 96.16\% & 99.74\% \\
  \bottomrule
\end{tabular}
\end{table}

%%%%%%%%%%%%%%%%%%%%%%%%%%%%%%%%%%%%%%%%%%%%

%%%%%%%%%%%%%%%% stacked area & line %%%%%%%%%%%%%%

\FloatBarrier
\section{Effectiveness of Token Incentives on Social Activities}
\label{app:causality_effect}
%%%%%%%%%%%%%%%%%%%%%%%%%%%%%%%%%%%%%%

%%%%%%%%%%%% Balance Table (reception T=0) %%%%%%%%%%%%
\begin{figure}
    \centering
    \begin{subfigure}{0.32\linewidth}
        \centering
        \includegraphics[width=\linewidth]{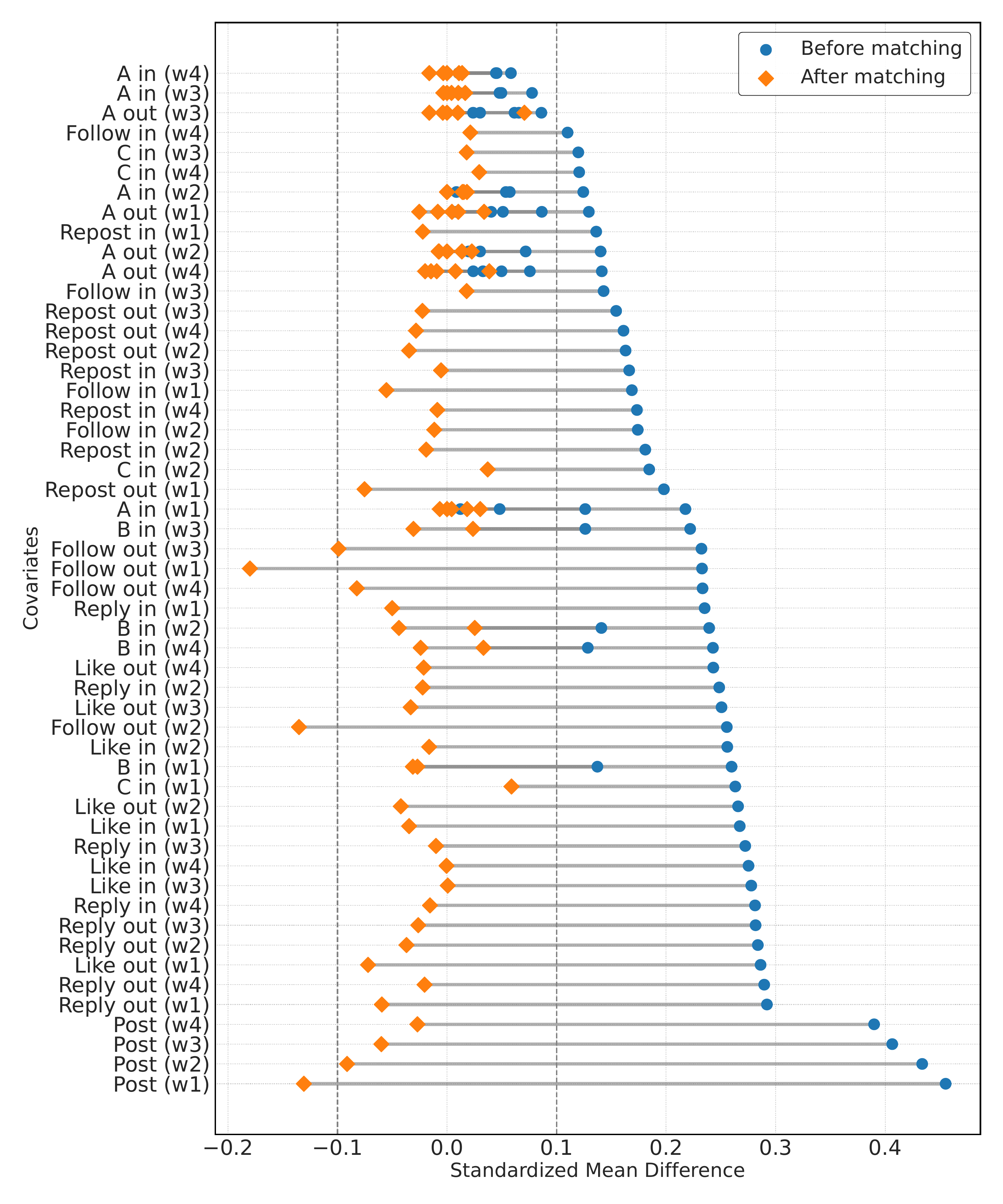}
        \caption{DEGEN (Inter\ac{fid} Tipping).}
        \label{fig:psm_balance_table_degen_A_fixedBreakpoint}
    \end{subfigure}
    \hfill
    \begin{subfigure}{0.32\linewidth}
        \centering
        \includegraphics[width=\linewidth]{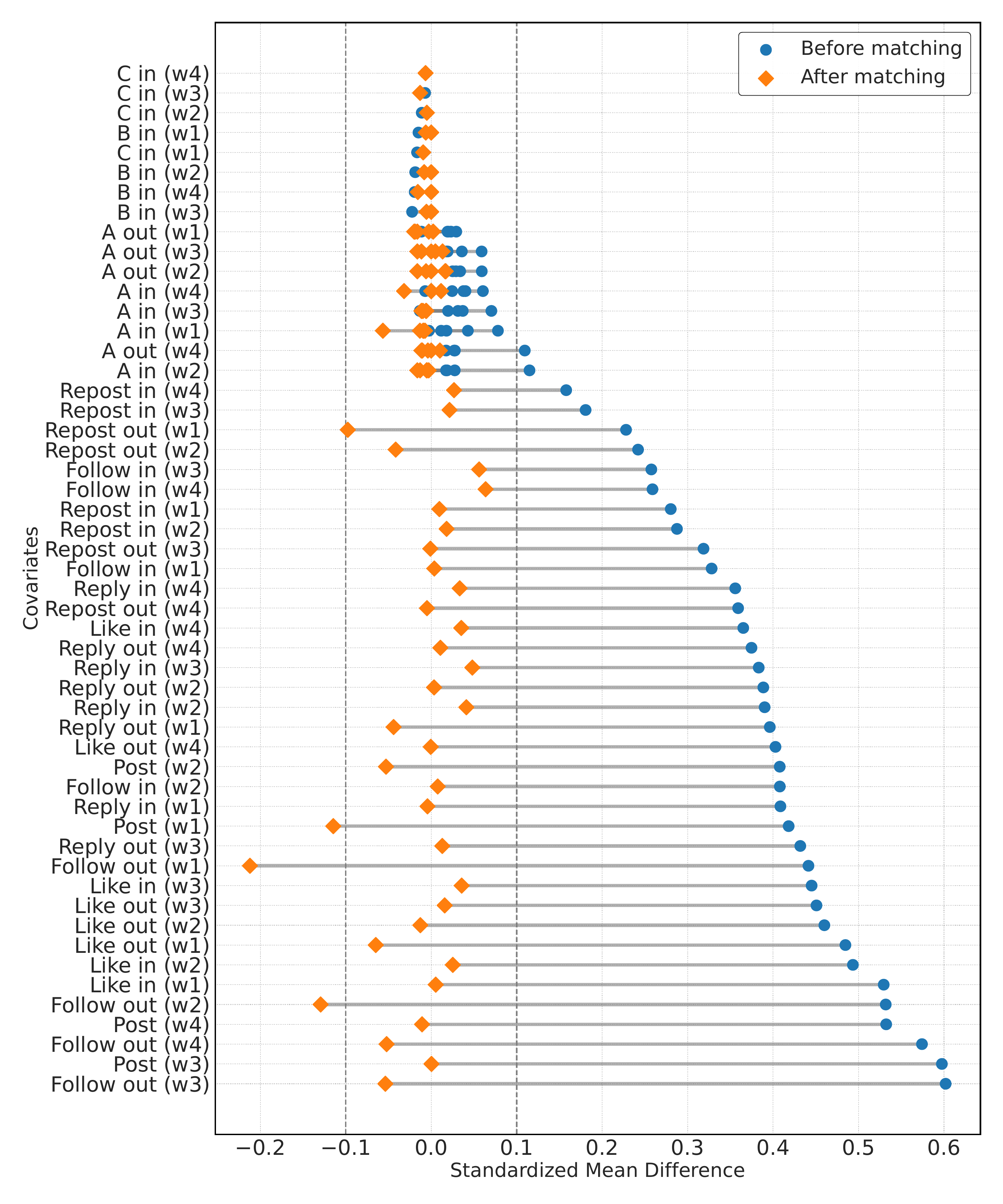}
        \caption{DEGEN (3rd. Algo. Reward).}
        \label{fig:psm_balance_table_degen_B_fixedBreakpoint}
    \end{subfigure}
    \hfill
    \begin{subfigure}{0.32\linewidth}
        \centering
        \includegraphics[width=\linewidth]{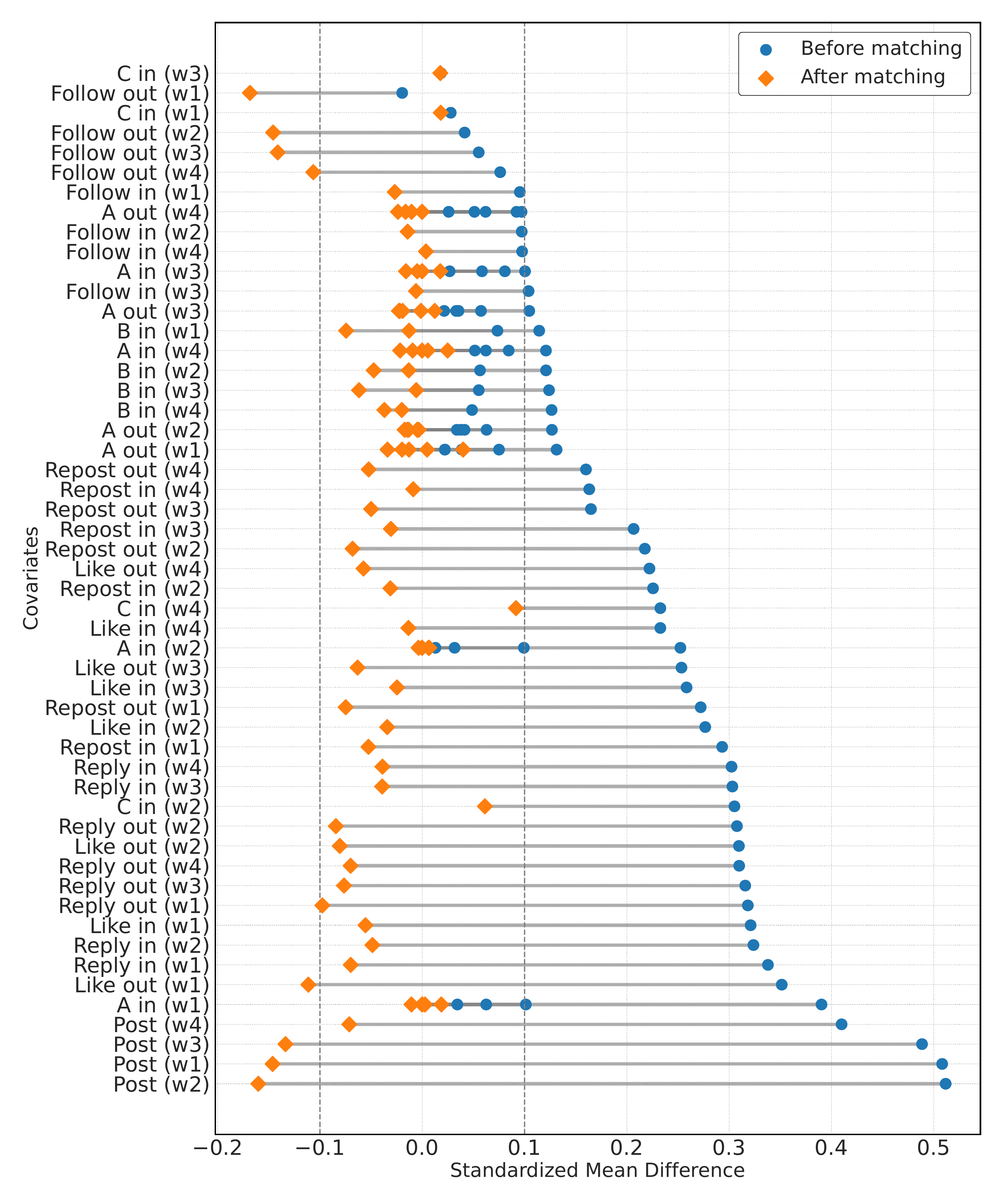}
        \caption{HIGHER (Inter\ac{fid} Tipping).}
        \label{fig:psm_balance_table_higher_A_fixedBreakpoint}
    \end{subfigure}
    \hfill
    \begin{subfigure}{0.32\linewidth}
        \centering
        \includegraphics[width=\linewidth]{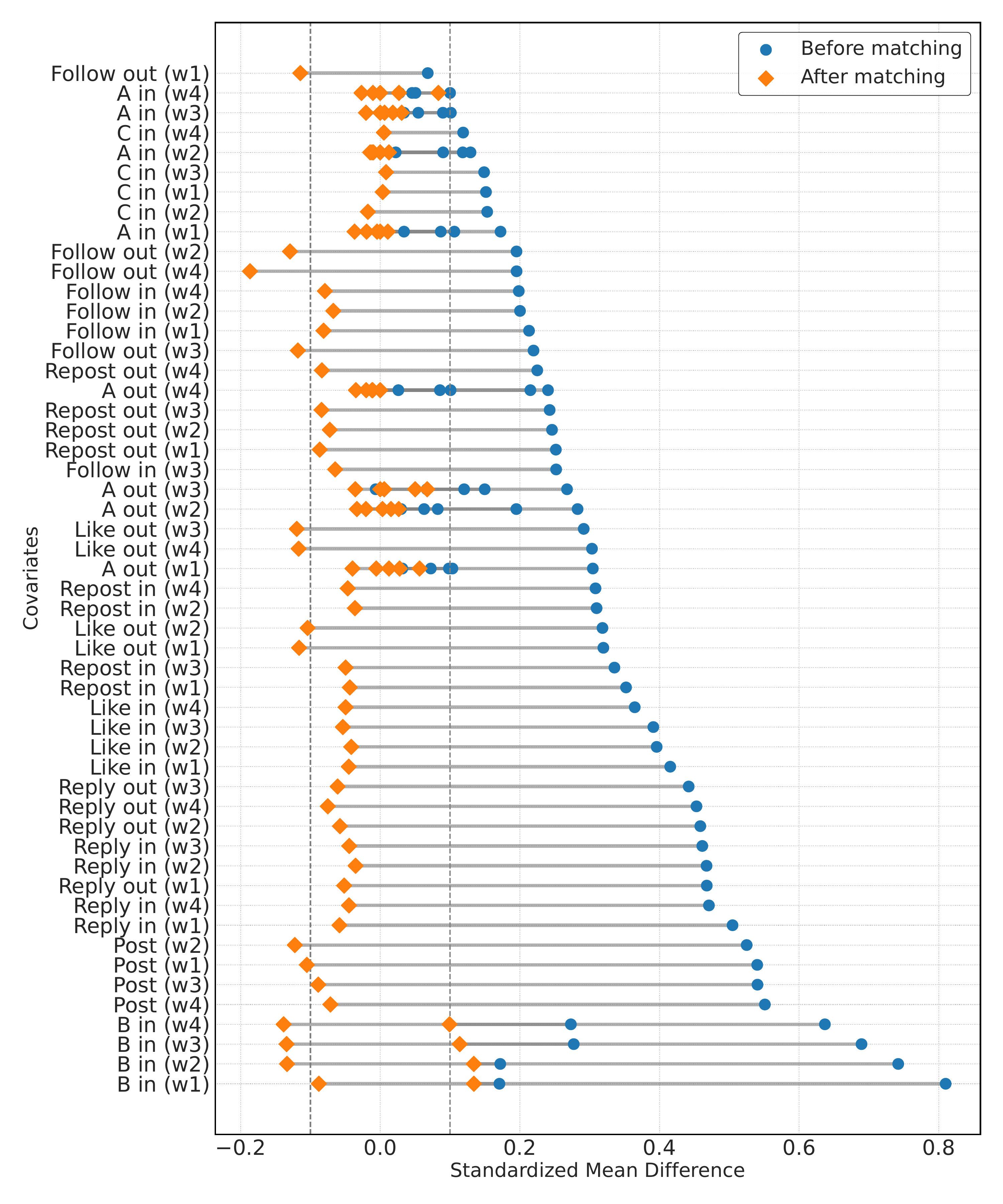}
        \caption{MOXIE (Inter\ac{fid} Tipping).}
        \label{fig:psm_balance_table_moxie_A_fixedBreakpoint}
    \end{subfigure}
    \hfill
    \begin{subfigure}{0.32\linewidth}
        \centering
        \includegraphics[width=\linewidth]{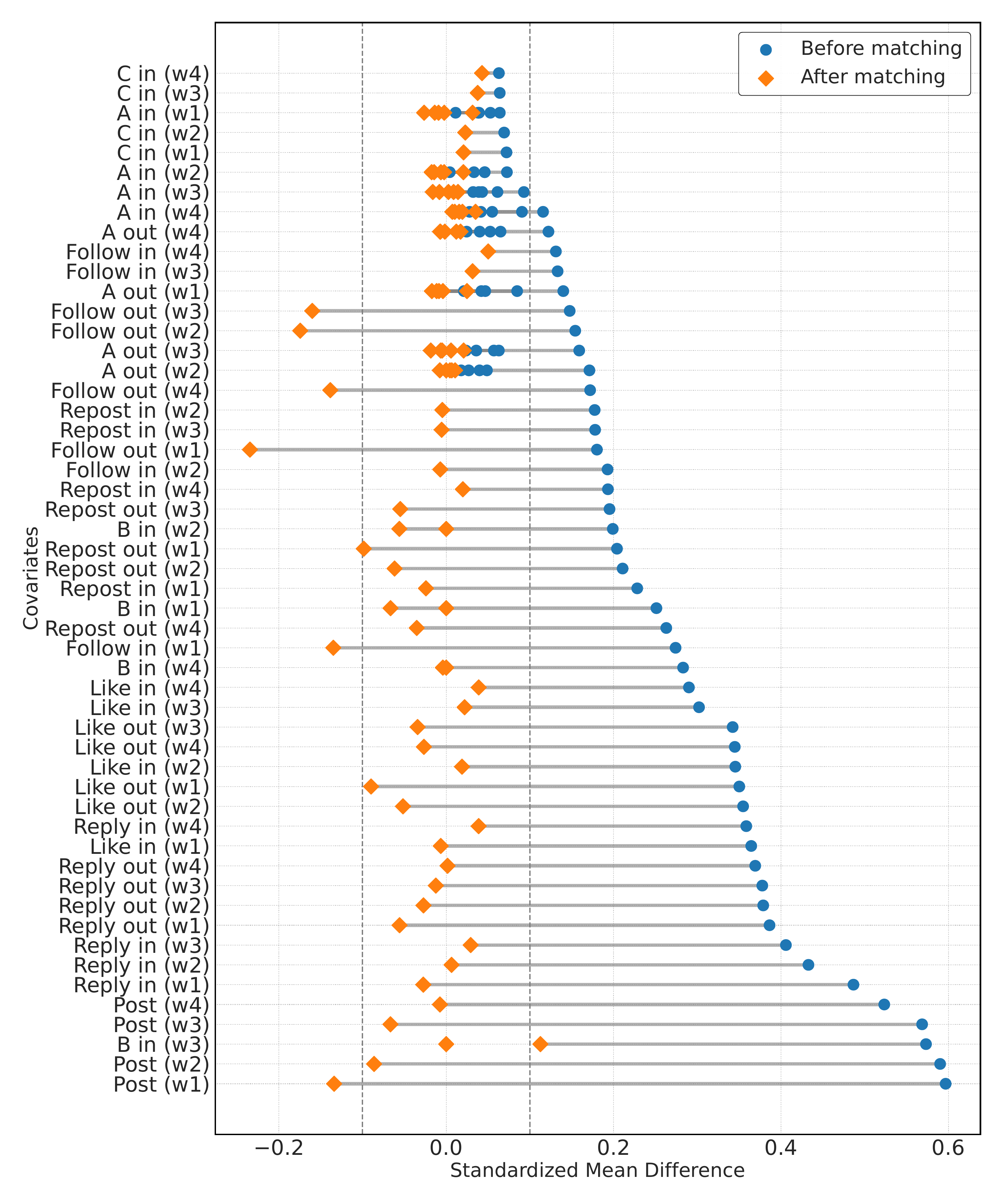}
        \caption{MOXIE (3rd. Algo. Reward).}
        \label{fig:psm_balance_table_moxie_B_fixedBreakpoint}
    \end{subfigure}
    \hfill
    \begin{subfigure}{0.32\linewidth}
        \centering
        \includegraphics[width=\linewidth]{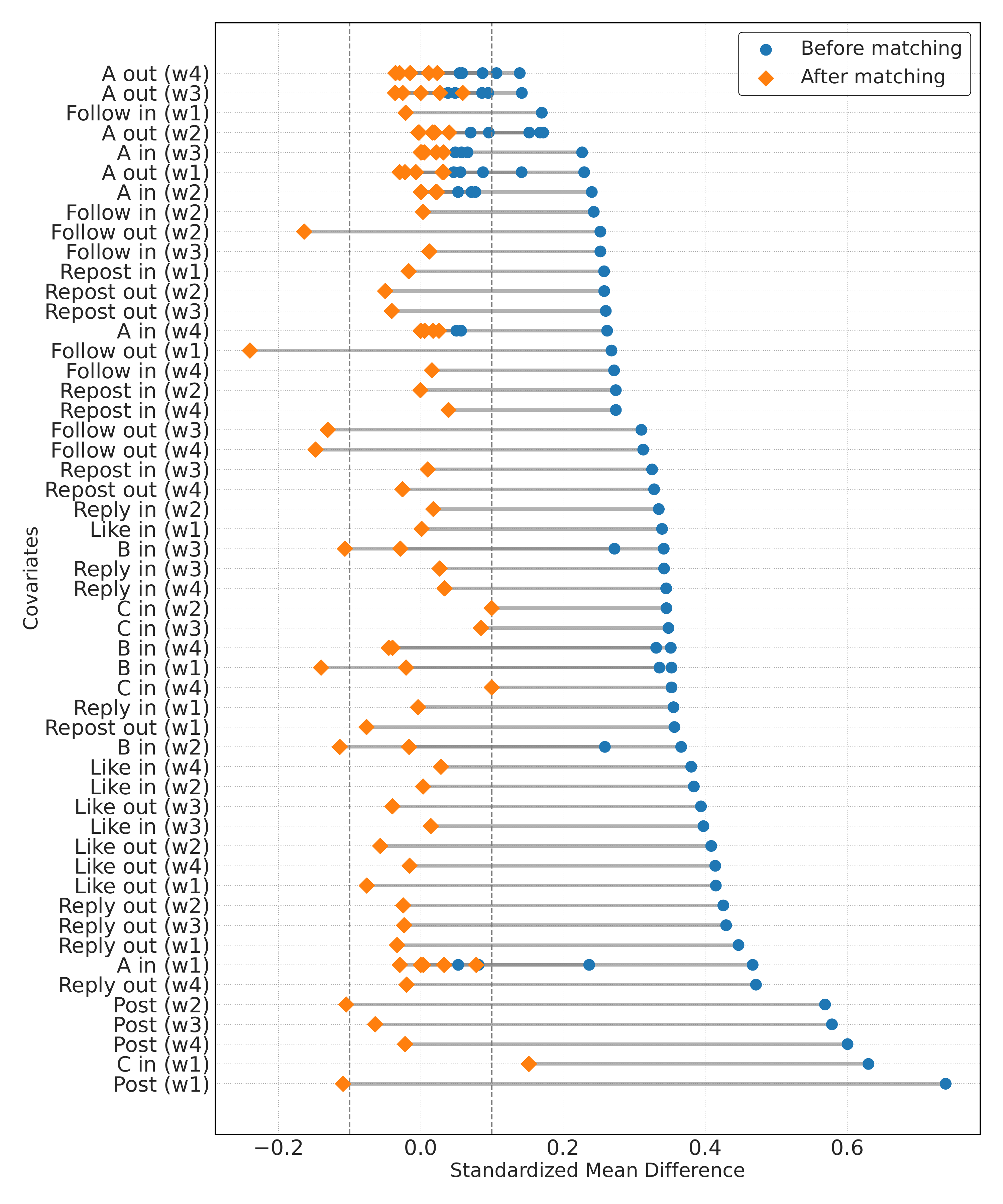}
        \caption{TN100X (Inter\ac{fid} Tipping).}
        \label{fig:psm_balance_table_tn100x_A_fixedBreakpoint}
    \end{subfigure}
    \hfill
    \begin{subfigure}{0.32\linewidth}
        \centering
        \includegraphics[width=\linewidth]{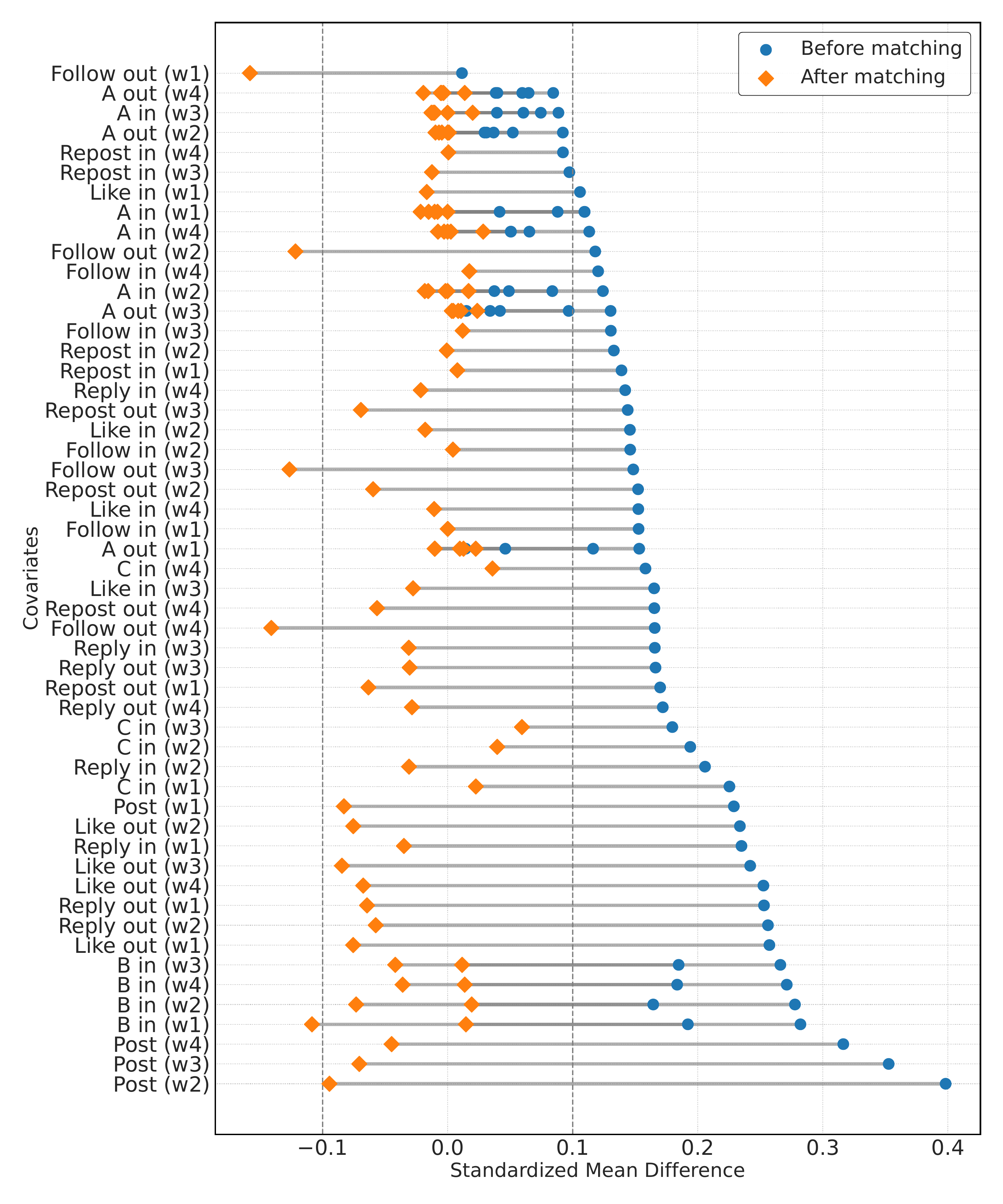}
        \caption{USDC (Inter\ac{fid} Tipping).}
        \label{fig:psm_balance_table_usdc_A_fixedBreakpoint}
    \end{subfigure}
    \quad
    \begin{subfigure}{0.32\linewidth}
        \centering
        \includegraphics[width=\linewidth]{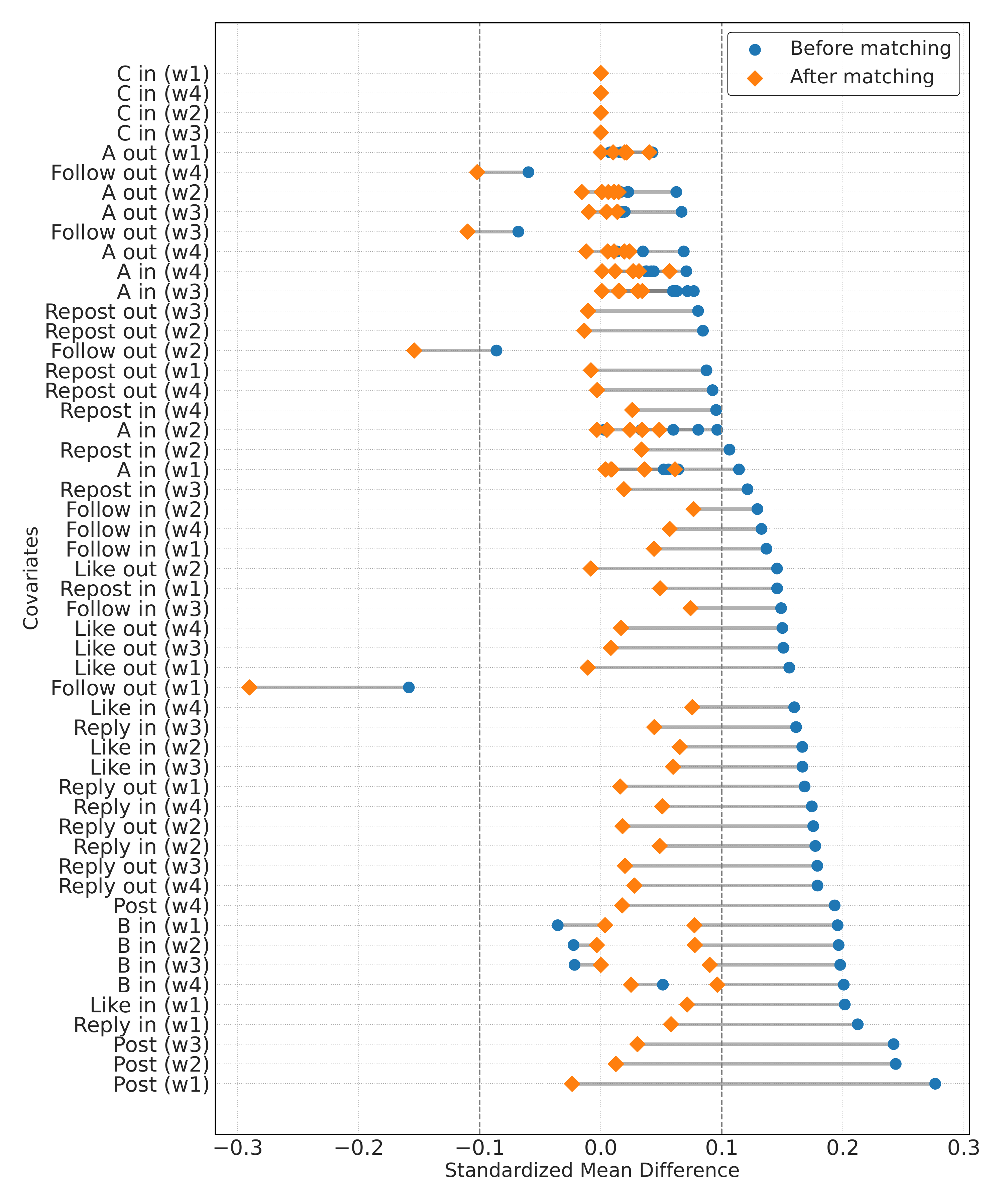}
        \caption{USDC (Off. Algo. Reward).}
        \label{fig:psm_balance_table_usdc_C_fixedBreakpoint}
    \end{subfigure}
    \caption{Balance diagnostics for propensity score matching (PSM).}
    \label{fig:psm_balance_tables_approach_1_appendix}
    \Description{Balance tables for propensity score matching (PSM). While most covariates achieve balance ($SMD<0.1$), notable imbalances are observed in outbound follows.
SMD denotes standardized mean difference.}
\end{figure}

\Cref{fig:psm_balance_tables_approach_1_appendix} demonstrates the balance tables for propensity score matching (PSM) analysis across token incentive mechanisms (see \Cref{sec:binary_psm_did}). 
While most covariates achieve balance ($SMD<0.1$), notable imbalances are observed in outbound follows. $SMD$ denotes standardized mean difference.
Token-related covariates on the Y-axis are defined as follows: \enquote{A in/out}: inbound/outbound Inter-FID Tipping; \enquote{B in}: inbound \emph{Third-party Algorithmic Rewards} (3rd Algo. Reward); \enquote{C in}: inbound \emph{Official Algorithmic Rewards} (Off. Algo. Reward).

\Cref{tab:intensity_effect_olsReg_summary_complete} illustrates a more detailed result table for the regression analysis in \Cref{sec:continuous_ols}, including standard errors ($SE$) and $R^2$.
Standard errors ($SE$) are reported within parentheses ( ), and the coefficient of determination $R^2$ is enclosed in square brackets [ ].

%%%%%%%%%%%%% regression reward intensity_complete_version %%%%%%%%%%%%
\begin{table}
\centering
\caption{
Detailed regression summary of continuous treatment intensity with social activities.
}
\label{tab:intensity_effect_olsReg_summary_complete}
\footnotesize
\setlength{\tabcolsep}{1.6pt}
\renewcommand{\arraystretch}{0.99}
\begin{tabular}{l*{8}{c}}
    \toprule
    \multicolumn{9}{c}{\textbf{Temporal Alignment: First Reward Reception Date as T+0}} \\
    \midrule
    & \multicolumn{5}{c}{\textbf{Inter-\ac{fid} Tipping}} 
    & \multicolumn{3}{c}{\textbf{Algorithmic Reward}} \\
    \cmidrule(lr){2-6} \cmidrule(lr){7-9}
    \textbf{Action} 
    & \textbf{DEGEN} 
    & \textbf{TN100X} 
    & \textbf{HIGHER} 
    & \textbf{MOXIE} 
    & \textbf{USDC} 
    & \textbf{DEGEN} 
    & \textbf{MOXIE} 
    & \textbf{USDC} \\
    \midrule
post 
    & \textcolor{+green}{\textbf{1.5703***}} & 5.0001 & \textcolor{+green}{\textbf{1.9701***}} & -0.0517 & -0.0805 & \textcolor{+green}{\textbf{12.7407***}} & \textbf{15.2254***} & \textbf{8.2309***} \\
    & (0.60)\ [0.16] & (3.08)\ [0.32] & (0.65)\ [0.28] & (0.92)\ [0.04] & (0.43)\ [0.01] & (2.93)\ [0.22] & (0.74)\ [0.12] & (1.20)\ [0.18] \\
reply\_out   
    & -4.4867 & 8.6610 & -3.8922 & 10.6198 & -0.3703 & 9.4023 & \textbf{59.2029***} & \textbf{-19.1090***} \\
    & (3.50)\ [0.56] & (30.62)\ [0.64] & (3.07)\ [0.54] & (2.57)\ [0.42] & (2.75)\ [0.08] & (17.01)\ [0.69] & (4.39)\ [0.27] & (5.25)\ [0.52] \\
reply\_in     
    & 1.7207 & -35.1615 & -2.4438 & 10.4600 & \textbf{-10.2936**} & -4.8940 & \textcolor{+green}{\textbf{57.1112***}} & 5.1841 \\
    & (3.31)\ [0.78] & (29.81)\ [0.87] & (2.75)\ [0.83] & (3.87)\ [0.59] & (3.57)\ [0.24] & (12.67)\ [0.80] & (3.24)\ [0.56] & (5.53)\ [0.74] \\
like\_out      
    & -1.4999 & 6.4651 & 1.4243 & -4.1602 & \textbf{7.5337**} & -9.7807 & \textbf{22.8327***} & \textcolor{+green}{\textbf{32.0449***}} \\
    & (4.81)\ [0.50] & (32.23)\ [0.50] & (4.24)\ [0.66] & (3.73)\ [0.26] & (3.54)\ [0.09] & (33.93)\ [0.57] & (3.02)\ [0.42] & (6.20)\ [0.60] \\
like\_in       
    & -1.2148 & 25.2958 & \textbf{17.9290***} & \textbf{-27.5476***} & 0.5102 & \textbf{-77.0576***} & \textbf{7.7791**} & \textcolor{+green}{\textbf{73.9531***}} \\
    & (5.34)\ [0.80] & (23.53)\ [0.96] & (3.89)\ [0.91] & (6.55)\ [0.74] & (3.98)\ [0.09] & (24.70)\ [0.75] & (3.50)\ [0.70] & (6.19)\ [0.90] \\
repost\_out     
    & 0.4519 & 7.1819 & \textbf{2.4716*} & -0.9295 & -0.0732 & -5.6276 & -0.1039 & \textcolor{-red}{\textbf{-9.1389***}} \\
    & (1.52)\ [0.44] & (7.93)\ [0.43] & (1.36)\ [0.56] & (1.28)\ [0.23] & (1.08)\ [0.05] & (10.63)\ [0.45] & (0.70)\ [0.34] & (2.27)\ [0.46] \\
repost\_in      
    & 1.7003 & -5.8862 & \textbf{-5.2017***} & \textbf{5.6276**} & \textbf{8.7852*} & 5.0918 & \textcolor{-red}{\textbf{-6.2899***}} & \textcolor{-red}{\textbf{-30.2670***}} \\
    & (2.00)\ [0.76] & (8.61)\ [0.94] & (1.47)\ [0.80] & (2.05)\ [0.25] & (2.18)\ [0.06] & (8.19)\ [0.64] & (1.16)\ [0.57] & (2.19)\ [0.81] \\
follow\_out     
    & 0.2749 & -0.9579 & 0.0211 & \textbf{14.5753***} & 0.6082 & 0.9927 & \textbf{20.3066***} & \textbf{-54.1936***} \\
    & (3.29)\ [0.14] & (12.09)\ [0.17] & (1.77)\ [0.10] & (2.98)\ [0.45] & (1.46)\ [0.04] & (15.90)\ [0.16] & (2.40)\ [0.45] & (3.17)\ [0.15] \\
follow\_in    
    & -12.4111 & -57.8221 & \textbf{-19.9274***} & \textbf{9.9655***} & 15.6006 & \textcolor{+green}{\textbf{196.8301***}} & \textbf{6.3755**} & \textcolor{+green}{\textbf{242.8866***}} \\
    & (8.22)\ [0.21] & (45.63)\ [0.56] & (6.43)\ [0.46] & (2.68)\ [0.40] & (5.43)\ [0.06] & (34.16)\ [0.10] & (3.21)\ [0.47] & (5.15)\ [0.56] \\
    \midrule
    pop.\_size & 40836 & 3459 & 15849 & 5252  & 15872 & 47748 & 21505 & 28181 \\
    sample\_size & 38532 & 3438 & 13854 & 5041 & 13879 & 47748 & 21482 & 27484 \\
    \bottomrule
\end{tabular}

\par\vspace{2mm}
\noindent\small
\textit{\textbf{Symbols:} $^{***}p<0.001$, $^{**}p<0.05$, $^{*}p<0.01$. Coefficients deemed statistically significant are presented in bold; Coefficients associated with statistically significant causal effects are highlighted in both bold and color.}
\end{table}

\end{appendices}

% \appendix

\end{document}